\documentclass[aps, prc, 11pt, tightenlines, letterpaper, amsmath, amssymb, preprintnumbers, floatfix, longbibliography, nofootinbib,superscriptaddress]{revtex4-2}
\usepackage[T1]{fontenc}
\usepackage{bm}
\usepackage{xspace}
\usepackage[dvipsnames]{xcolor}
\usepackage{graphicx}
\usepackage{tabularx}
\usepackage{enumitem}
\usepackage{braket}
\usepackage{float}
\usepackage[normalem]{ulem}
\newcolumntype{C}[1]{>{\centering\let\newline\\\arraybackslash\hspace{0pt}}m{#1}}

\usepackage{nccmath}
\usepackage{cancel}

\usepackage[hypertexnames=false]{hyperref}
\hypersetup{
    colorlinks=true,       
    linkcolor=blue,          
    citecolor=blue,        
    filecolor=blue,      
    urlcolor=blue           
}

\newcolumntype{Y}{>{\centering\arraybackslash}X}

\usepackage{mathtools}
\usepackage{orcidlink}

\def\pislash{ {\pi\hskip-0.54em /} }

\def\nopi{ {\rm EFT}(\pislash) }

\def\llra{{\relbar\joinrel\longrightarrow}}

\def\mapup#1{{\smash{\mathop{\llra}\limits^{#1}}}}

\usepackage{stackengine}
\stackMath

\makeatletter
\def\l@subsubsection#1#2{}
\makeatother

\begin{document}
\title{Reformulation of the pionless effective field theory \\ for practical few- and many-body applications}

\author{S.R.~Beane}
\email{silas@uw.edu}
\affiliation{Department of Physics, University of Washington, Seattle, WA 98195, USA}
\author{R.C.~Farrell}
\email{rolandf@caltech.edu}
\affiliation{Institute for Quantum Information and Matter (IQIM) and Department of Physics,
California Institute of Technology, CA 91125, USA}
\author{C.P.~Shotwell}
\email{cshotwel@uw.edu}
\affiliation{Department of Physics, University of Washington, Seattle, WA 98195, USA}
\author{U.~van~Kolck}
\email{vankolck@ijclab.in2p3.fr}
\affiliation{European Centre for Theoretical Studies in Nuclear Physics and Related Areas (ECT*), Fondazione Bruno Kessler, 38123 Villazzano (TN), Italy}
\affiliation{Universit\'e Paris-Saclay, CNRS-IN2P3, IJCLab, 91405 Orsay, France}
\affiliation{Department of Physics, University of Arizona, Tucson, AZ 85721, USA}

\preprint{NT@UW-26-7}
\date{\today}

\begin{abstract}
  \noindent The effective field theory (EFT) of contact forces which
  describes few-nucleon systems at distance scales larger than the
  pion Compton wavelength ($\nopi$) can be defined in terms of
  momentum-dependent potentials that contain a fixed number of simple
  poles and generate the effective-range expansion to a given fixed
  order. These potentials provide a practical generalization of the
  standard $\nopi$ power-counting scheme which, for instance,
  naturally accounts for the large effective ranges in the s-wave channels.
  Nucleon-nucleon phase shifts computed in this formulation of
  $\nopi$ are compared to modern phase-shift analyses, and
  effective-range parameters are tabulated.  The $\nopi$
  potentials, which provide a rational approximation of the S-matrix,
  are easily extended into the regime of pion physics through the
  inclusion of additional S-matrix poles and zeros, and provide an accurate
  description of data up to pion-production threshold. The
  potentials in this scheme are in an important sense weak over all
  relevant distance scales, and are therefore of utility for systematic
  calculations of many-nucleon observables.
\end{abstract}

\maketitle
\newpage{}
\tableofcontents
\thispagestyle{empty}
\newpage{}

\section{Introduction}
\pagenumbering{arabic}
\noindent

Effective-range theory provides an excellent description of
the nucleon-nucleon (NN) system, up to momentum scales where it is
expected to break down due to the onset of pion physics, and even
beyond.  Effective field theory (EFT) provides a useful paradigm for
understanding this qualitatively and quantitatively via an analysis of
the various momentum scales in the
system~\cite{Hammer:2019poc}.  Due to
strong NN interactions, a new dynamical infrared (IR) momentum scale,
denoted $\aleph$, is generated, which is small as compared to the
ultraviolet (UV) scale ${\cal M}\sim M_\pi=\ell_\pi^{-1}$, where
$M_\pi$ is the pion mass and $\ell_\pi$ is the range of the
interaction~\cite{vanKolck:1997ut,Kaplan:1998tg,Kaplan:1998we,vanKolck:1998bw}. In
the NN system, the s-wave scattering lengths, $a_0\sim \aleph^{-1}$, are large as compared to $\ell_\pi$.  If the
effective ranges and shape parameters are taken to scale
as (powers of) $\ell_\pi$, then, in its most economical form, the
pionless EFT ($\nopi$), which probes momenta $k\sim \aleph \ll {\cal
  M}$, treats $k\,a_0\sim 1$ effects to all orders and all
effective-range and shape-parameter effects, which scale as positive
powers of $k\,\ell_\pi\ll 1$, in perturbation
theory~\cite{vanKolck:1997ut,Kaplan:1998tg,Kaplan:1998we,vanKolck:1998bw}.  This
formulation of $\nopi$ has several advantages: order-by-order
renormalization is manifest, it can be extended to more-nucleon systems, and the coupling of the EFT to external
currents is straightforward~\cite{Hammer:2019poc}. 
However, in both s-wave channels the
effective ranges, $r_0$, are somewhat larger than natural size while
the shape parameters, $v_{2,0}$ etc., 
appear to be smaller than naive scaling
would suggest. This experimental observation suggests formulating
$\nopi$ in a manner that can treat $k\,r_0$ effects to all orders,
thereby matching onto unexpanded effective-range theory at effective-range
order, while allowing shape-parameter effects to be treated in perturbation theory or summed to all orders. This was achieved long ago by making use of
field redefinitions and energy-dependent interactions, and it indeed
leads to a useful formulation of $\nopi$ in the NN system
including processes coupled to external
probes~\cite{Kaplan:1999qa,Beane:2000fi}. 

\begin{figure*}
\begin{center}
  \includegraphics[width=\textwidth,height=\textheight,keepaspectratio]{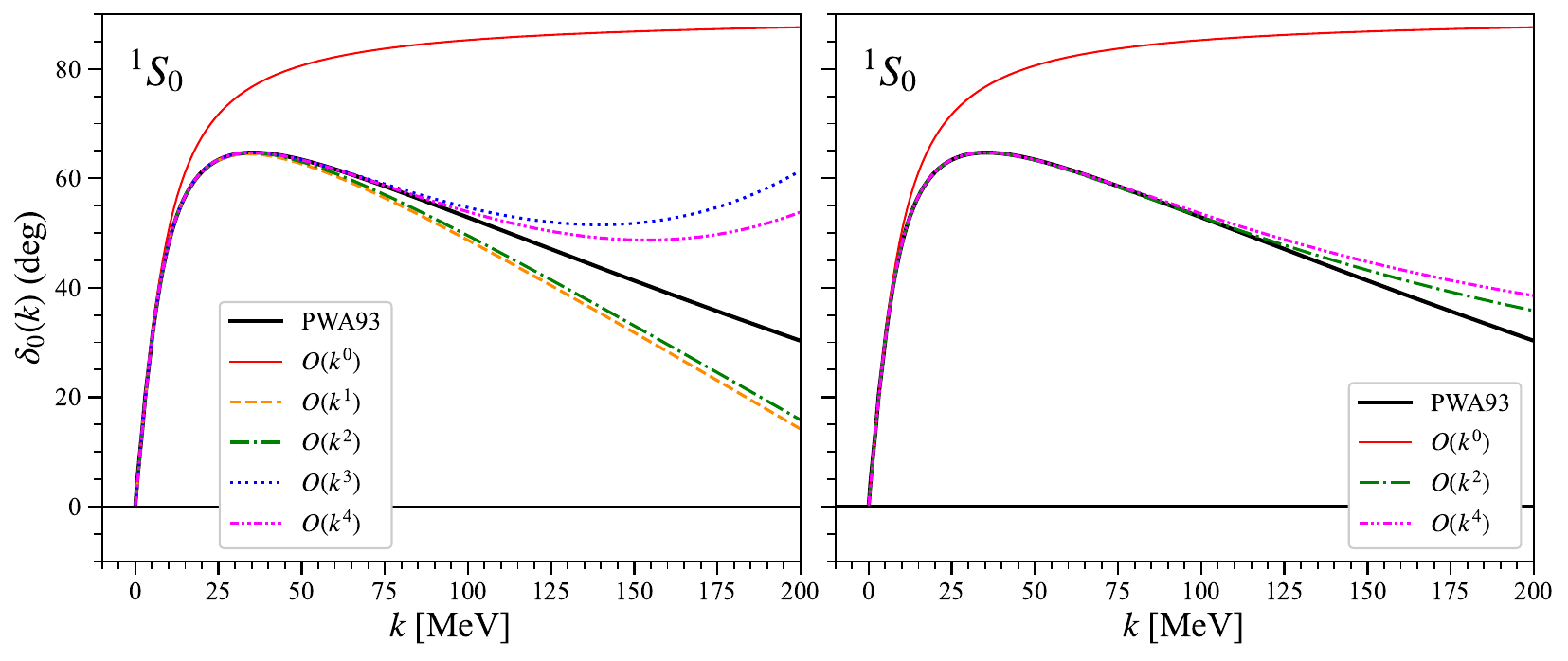}
  \includegraphics[width=\textwidth,height=\textheight,keepaspectratio]{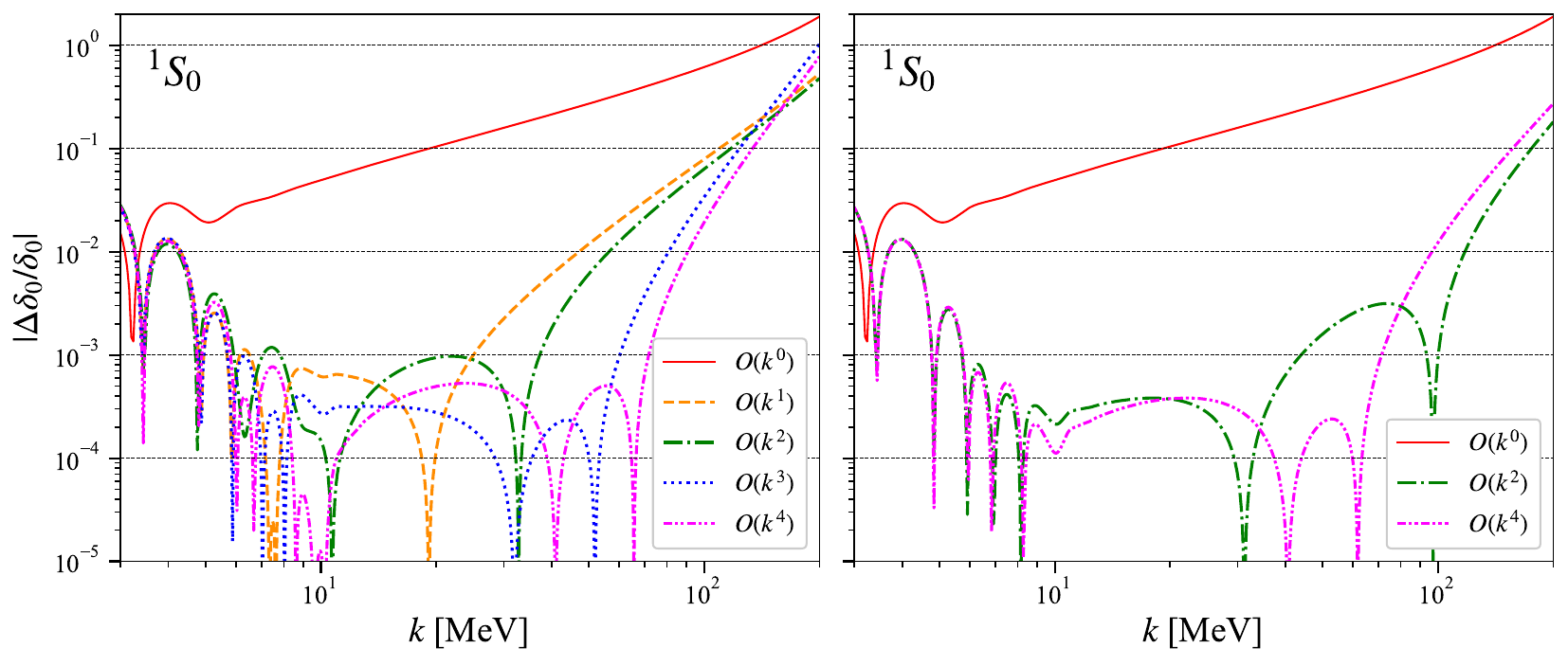}
  \caption{\small The phase shifts in the neutron-proton $^1S_0$ channel in 
  $\nopi$ (upper left) and the effective-range expansion (upper right) are plotted as functions of the relative momentum in the center-of-mass frame. The relative differences between the predicted phase shifts and the Nijmegen partial-wave analysis~\protect\cite{NNonline} are graphed in the lower two plots, again as functions of the relative momentum.  In $\nopi$, the scattering amplitude is expanded in powers of $k$, while in effective-range theory, $k\cot\delta_0$ or, equivalently, the inverse amplitude is expanded in powers of $k$.
  \label{fig:Propaganda}}
\end{center}
\end{figure*}

Figure~\ref{fig:Propaganda} compares the phase shifts obtained in the neutron-proton $^1S_0$ channel by treating $r_0$, $v_{2,0}$, etc. in perturbation theory with those obtained by summing them to all orders.\footnote{ 
The effective-range parameters used in the right set of plots are obtained from the fits in Section~\ref{sec:NNcomptodata}, in which the truncated effective-range expansion is fit to the Nijmegen partial-wave analysis~\cite{NNonline} 
up to some maximum 
momentum $k_{\max}$. The parameters used in the left set of plots are
obtained similarly, except that the prediction of $\nopi$ is fit to phase-shift data. 
The shape parameters are particularly sensitive to the fitting procedure, and there is little consensus in the literature on their numerical values. Different approaches, like refitting the effective-range parameters at each order, will produce different shape parameters and different Lepage plots; however, the qualitative conclusions of this paragraph are insensitive to these differences. }
The description of the 
$^1S_0$ phase shifts improves order by order in $\nopi$, as can be seen in the Lepage plot~\cite{Lepage:1997cs} for the deviation from empirical values~\cite{NNonline}. In fact, the convergence radius of the theory in the NN system has been estimated to be at or above $M_\pi$~\cite{Ekstrom:2024dqr}, despite the first branch cut starting at $M_\pi/2$. 
Still, re-summing the effective range to all orders clearly outperforms treating it in perturbation theory. 
By contrast, re-summing the first shape parameter makes little difference compared to treating it in perturbation theory, and both treatments actually worsen the results compared to re-summing only the effective range.
Improvement of the lowest orders can be obtained by re-summing only a fraction of $r_0$~\cite{Contessi:2024vae}, but it is, of course, maximal with the re-summation of the full effective range. The importance of the effective range has been emphasized in a similar context in Refs.~\cite{Kievsky:2023xgy,Tumino:2024qpk} (and references therein).

A large (and negative) effective range represented by an energy-dependent interaction can be incorporated in calculations of the three-body system~\cite{Griesshammer:2023scn} based on the Skornyakov--Ter-Martirosian equation, but energy-dependent interactions are difficult to handle in many-body ($N>3$) calculations based on the Schr\"odinger equation. 
These
difficulties
have recently motivated efforts to
reformulate $\nopi$ with momentum-dependent
interactions which treat
effective-range effects to all orders~\cite{
Beane:2021xrk,Beane:2021dab,Timoteo:2023dan}\footnote{For the same idea applied to the short-range interactions in chiral EFT, see Ref.~\cite{Peng:2021pvo}.}. Here it is shown that $\nopi$
can be formulated in terms of momentum-dependent interactions
whose power counting adapts naturally to large effective-range
effects.  In the standard formulation of $\nopi$, the scattering
amplitude is computed to a fixed order in the momentum expansion;
i.e., local operators
are considered up to a fixed order
in the number of derivatives. However, this formulation becomes
problematic when momentum-dependent interactions become strong, unless the effective range is negative~\cite{Wigner:1955zz,Phillips:1996ae,Hammer:2009zh,Hammer:2010fw}. 
The strongest two-body interactions generate the unitarity limit, where the S-matrix is $-1$ and $k\cot \delta_0=0$. Thus, for strong interactions, it is
more natural and convenient to work at fixed order in
the inverse scattering amplitude. Of course, this method---effective-range theory---implies an infinite re-summation of local operators
at each order, and one would like a procedure for achieving this which
leads to unique, systematic observable predictions.

In the EFT of contact operators, the momentum-dependent interactions
are non-local\footnote{In this paper, (non-)local is used in two distinct ways:
  a local operator is a product of field operators evaluated at the same spacetime
  point, whereas a local interaction or potential is diagonal in the
  coordinate-space representation, i.e., $V(r,r') = V(r)\delta(r-r')/r^2$ where $r$ is the relative separation.} and separable to any finite order in the momentum
expansion. It is therefore well motivated to assume that separability
survives re-summation. Moreover, for efficient bookkeeping it is
convenient to define separable potentials that match precisely onto
the effective-range expansion to a given order. The resulting
potentials give rise to a definition of $\nopi$---here
referred to as Effective Pole Theory (EPT)---in
terms of
momentum-dependent, separable and non-local
potentials which lead to
rational S-matrix elements; that is, ratios of polynomials of finite
degree in the momenta. 
This implies that the S-matrix
is specified by a finite number of 
simple poles and zeros, and the unitarity of the
S-matrix then implies an exact invariance with respect to permutation
of the pole positions.
This symmetry has several useful consequences:
it greatly facilitates the inverse problem of finding potentials which
give rise to a specific rational S-matrix, and in certain special
cases it implies symmetry-protected (two-body) phase-equivalent
potentials, which prove useful both in gauging the relevance of three- and
higher-body forces, and in finding exactly phase-equivalent potentials with
distinct off-shell behavior.

In addition to providing a definition of $\nopi$ in terms of
momentum-dependent potentials, EPT
allows a
simple extension to modeling momenta that probe pion and other
short-distance physics. An EFT, by definition, contains all of the
singularities of the propagating degrees of freedom that are
consistent with unitarity, and therefore, for instance, when probing
momenta of order $M_\pi/2$, pions should be included as degrees of
freedom and one is in the realm of chiral EFT.  This should be done in
such a way that singular interactions are properly renormalized~\cite{Beane:2000wh,Long:2007vp}. Such a scheme exists (see Ref.~\cite{vanKolck:2020llt} for a review), but details are still being worked out (see Refs.~\cite{Gantenberg:2026xzy,Thim:2026mzm} for recent developments).
The
uncontrolled approximations~\cite{Hammer:2019poc}
that one makes in using chiral EFT without
consistent renormalization  may be as severe as taking the simple
procedure of approximating pion branch cuts by simple poles via a
re-summation of the effective-range expansion using
EPT---a re-summation guided by the existence of zeros of the scattering amplitude~\cite{SanchezSanchez:2017tws,Behzadmoghaddam:2020pqr}\footnote{The same re-summation of short-range interactions can be profitably done also in chiral EFT~\cite{Kaplan:1996nv,SanchezSanchez:2017tws,SanchezSanchez:2020kbx,Yang:2020pgi,SanchezSanchez:2021dki,Behzadmoghaddam:2023gsh} and in halo/cluster EFT~\cite{Rupak:2018gnc}.}.
In any event,
EPT provides a model of the NN interaction which fits the NN phase
shifts to arbitrary accuracy up to pion-production threshold, and
matches onto the theory of contact forces in the infrared.

A major advantage of EPT is that the potentials are typically soft
over phenomenologically interesting momentum scales. Indeed, an
important motivation for this work is the development of accurate
momentum-dependent potentials which are weak over the momentum scales
of interest. Specifically, what is desired is a perturbative expansion
in the quantum-mechanical two-body potential, $V({\bf k},{\bf k}')$,
which is a function of the relative momenta, ${\bf k}$ and ${\bf k}'$,
of the incoming and outgoing particles.  Consider, for simplicity, a
neutron gas. For a dilute gas, the on-shell potential may be
approximated by $V(k,k)=4\pi a_0/M$, where $M$ is the nucleon mass,
$a_0$ is taken to be the singlet (i.e., neutron-neutron) s-wave scattering length, and the
effective expansion parameter is $|k_F a_0|\ll 1$. Given the large
scattering length in the neutron-neutron system, this perturbative
expansion is useful only at extremely low densities that are not of
particular phenomenological interest.  When generalizing to the momentum-dependent separable potentials of EPT, the above naively suggests the appropriate effective expansion parameter will be ${M k}V(k,k)/4\pi$. 
This is, in fact, the estimate for an iteration of the potential in the power counting of $\nopi$~\cite{Hammer:2019poc}.
In this work it is shown that this expansion parameter is generally
small over the entire range of momenta up to pion-production
threshold, or densities up to nuclear saturation density.
However, perturbativeness is not a property of $V(k,k)$ alone, but of the full Lippmann-Schwinger kernel. Weinberg eigenvalues~\cite{Weinberg:1963zza,Bogner:2005sn} provide a direct diagnostic of this kernel: even when the on-shell potential is weak, its iteration through intermediate states can generate large eigenvalues, signaling nonperturbative behavior. 
For the rank-one separable interactions considered in this work, this information is compressed into a single complex eigenvalue whose imaginary part is tied to the on-shell strength $V(k,k)$, and whose real part involves the product $V(k,k) k\cot\delta(k)$.

This paper is organized as follows. Section~\ref{sec:RSM} provides a review of
basic properties of the rational form of the S-matrix. Section~\ref{sec:pot def} focuses on the inverse problem of generating momentum-dependent potentials which give rise to 
S-matrices with a specific number of poles. In Section~\ref{sec:bsprop}, properties of the deuteron bound state are derived for the separable potentials used in EPT. Section~\ref{sec:powcount} develops a power-counting scheme 
which is equivalent to $\nopi$.
In Section~\ref{sec:NNcomptodata}, an order-by-order comparison with the Nijmegen 
partial-wave analysis is given.
An extension of EPT to center-of-mass momenta up to pion-production threshold is provided in Section~\ref{sec:NNtopiprod}. Section~\ref{sec:conc} summarizes the results and conclusions.

\section{Properties of the rational S-matrix}
\label{sec:RSM}
\noindent
In this section, a class of S-matrix elements that are defined by their simple poles $\{\gamma\}$ in the complex momentum plane is considered.
Unitarity implies an ${\cal S}_d$ permutation symmetry that acts on the $\{\gamma\}$, and all scattering observables are invariant with respect to ${\cal S}_d$.

\subsection{The scattering amplitude defined}

\noindent The partial-wave expansion of the NN scattering amplitude below inelastic threshold, in the absence of mixing, may be expressed as

\begin{equation}
T(k,\cos\theta)\ =\   \sum_{\ell\geq 0}\left(2\ell +1\right)T_\ell(k)P_\ell(\cos\theta) \ = \  -\frac{4 \pi}{M} \sum_{\ell\geq 0}   \frac{2\ell +1}{k \cot\delta_\ell(k) - i k }\, P_\ell(\cos\theta) 
\label{eq:reampgen}
\end{equation}
 in terms of the relative momentum in the center-of-mass frame $k=\sqrt{2ME}$
 and the scattering angle $\theta$. The  S-matrix element for a given partial wave is

\begin{equation}
  S_\ell(k) \ =\ e^{2i\delta_\ell(k)} \ =\ 1 \ -\, i\frac{M k}{2\pi}\,T_\ell(k)\ .
\label{eq:smatpwexp}
\end{equation}
Here it is assumed that the underlying interaction is given by strictly contact forces. This assumption may be made precise
in the EFT context as will be seen below. Given this assumption, one expects an effective-range expansion (ERE) to $\mathcal{O}(k^{2n})$ of the form

\begin{equation}
k^{2\ell +1} \cot\delta_\ell(k)\ =\  -\frac{1}{a_\ell} + \frac{1}{2} r_\ell {k^2} +  \sum_{n'=2}^n v_{n',\ell}\, k^{2n'} \ ,
\label{eq:eregen}
\end{equation}
where $a_\ell$ is the scattering ``length'',
$r_\ell$ the effective ``range'', and $v_{n,\ell}$ the shape parameters.
Therefore, the singularities of the scattering amplitude are determined by the zeros, $\kappa=i \gamma_j$, of the polynomial

\begin{equation}
  0 \ = \ -\frac{1}{a_\ell} +
\frac{1}{2} r_\ell {\kappa^2} + \sum_{n'=2}^n v_{n',\ell}\, \kappa^{2n'}
- i \kappa^{2\ell+1}\ .
\label{eq:eregenbssol}
\end{equation}
The $\gamma_j$ are generally complex numbers\footnote{Unitarity
  requires that complex poles $\{\gamma\}$ come in complex-conjugate pairs.}. For nonredundant physical poles, real, (negative) positive $\gamma_j$ correspond to (virtual)
bound states. It is clear that the polynomial in Eq. \eqref{eq:eregenbssol} has degree $d$, where
\begin{equation}
    d=\max\{d_{\min}, 2n\}\ , \ \ \ d_{\min}\equiv2\ell+1\ .
\end{equation}
Thus, the minimal number of poles of 
the S-matrix in the $\ell$th partial wave is $d_{\min}$.

\subsection{The S-matrix and symmetric polynomials}

\noindent Generally, in the EFT of contact forces in three spatial
dimensions, the S-matrix is a ratio of polynomials in $k$. Consider
the special class of forces that give rise to an ERE of finite
degree $2n$. 
Suppressing the partial-wave index for simplicity, the T-matrix element is
\begin{align}
    T(k) \ = \ -\frac{4\pi}{M}\frac{\lambda k^{2\ell}}{\prod_{j=1}^d(\gamma_j+ik)}\ ,
\end{align}

where $\lambda\in\mathbb{R}$ is chosen such that the coefficient on the $k^{2\ell+1}$ term in $\lambda^{-1}{\prod_{j=1}^d(\gamma_j+ik)}$ is $-i$. The S-matrix element can then be expressed as a ratio of two polynomials,
\begin{equation}
  S(k) \ =\ e^{2i\delta(k)} \ =\ \prod_{j=1}^d \frac{(\gamma_j-ik)}{(\gamma_j+ik)} 
  \ =\ (-1)^d\prod_{j=1}^d \frac{(k+i\gamma_j)}{(k-i\gamma_j)} \   .
\label{eq:smatgen}
\end{equation}
This class of forces constitutes a special subset of interactions that generate an S-matrix that is a ratio of polynomials in $k$ of finite degree. 
Due to unitarity, $S(k)$ is the ratio of a polynomial and its complex conjugate for real $k$.
Therefore, it
is manifestly invariant with respect to permutations of the pole positions; that is, the group ${\cal S}_d$ of
the permutations of the $\gamma_j$.

This symmetry implies that all scattering parameters can be expressed in terms of elementary symmetric polynomials (ESPs) of the pole positions. These ESPs are defined
as

\begin{equation}
{\bf {\large e}}_k^{(d)} \equiv\  {\bf {\large e}}_k (\gamma_1,\ldots,\gamma_d)\ =\ \sum_{1\leq j_1<j_2<\cdots<j_k\leq d} \gamma_{j_1} \cdots\gamma_{j_k} \ .
\label{eq:esp1}
\end{equation}
The ESPs are compact building blocks that greatly facilitate the inverse problem of generating potentials that give rise to rational
S-matrices. For instance, consider the T-matrix from above. It can be recast into the form

\begin{equation}
  T(k) \ =\ -\frac{4\pi}{M}\frac{\lambda k^{2\ell}}{\sum_{j=0}^d (i k)^j {\bf { e}}_{d-j}^{(d)}} \ .
\label{eq:smatgensps}
\end{equation}
It is straightforward to express the effective-range parameters in terms of
the ESPs. One finds 

\begin{equation}
  a_\ell\ =\ (-1)^\ell\frac{{\bf { e}}_{d-(2\ell+1)}^{(d)}}{{\bf {\large e}}_d^{(d)}} \  ,\ \ \  r_\ell\ =\ 2(-1)^\ell\frac{{\bf {\large e}}_{d-2}^{(d)}}{{\bf {\large e}}_{d-(2\ell+1)}^{(d)}}
\ ,\ \ \ v_{n',\ell}\ =\ (-1)^{n'+\ell+1}\frac{{\bf {\large e}}_{d-2n'}^{(d)}}{{\bf { e}}_{d-(2\ell+1)}^{(d)}}\, (n'=2,\dots,n) \ ,
\label{eq:esp2}
\end{equation}
subject to the constraints

\begin{equation}
0 \ =\ {\bf { e}}_{d-(2\ell'+1)}^{(d)} \; ,\; \forall \; \ell'\neq \ell \ .
\label{eq:constesp2}
\end{equation}
These constraints ensure that the only odd power of $k$ that appears in the inverse amplitude is $ik^{2\ell+1}$, as required by unitarity.

\section{A potential definition of \texorpdfstring{$\nopi$}{pionless EFT}}
\label{sec:pot def}
\noindent
This section considers the inverse problem: the structure of the S-matrices described in the previous section constrains the potentials that can generate them. 
In particular, on-shell potentials can be chosen to possess some of the same poles as the target S-matrix. The remaining poles are necessarily generated dynamically through the Lippmann-Schwinger equation. However, the ${\cal S}_d$ symmetry implies that this distinction between poles that are explicit in the potential and those that are generated dynamically is arbitrary. Therefore, there is often a family of phase-equivalent potentials that are generated by permuting explicit and dynamically generated poles~\cite{RevModPhys.21.488}.
Many of these potentials have UV divergences that require regularization and renormalization. Dimensional regularization with the power-divergence subtraction (PDS) scheme~\cite{Kaplan:1998tg,Kaplan:1998we,Phillips:1998uy} is particularly informative as poles of the coupling $g(\mu)$ in the renormalization-group (RG) scale $\mu$ are also ESPs of the S-matrix pole positions $\{\gamma\}$. A similar connection between the symmetries of the S-matrix and properties of the PDS RG scale was made in Ref.~\cite{Beane:2021xrk}.

\subsection{The Lippmann-Schwinger equation}
\label{sec:rgflow}

\noindent The Lippmann-Schwinger (LS) equation for the scattering amplitude $T$, projected onto states with orbital angular momentum $\ell$, is
\begin{equation}
    T_{\ell}(p,p';k) \ =\  V_{\ell}(p, p') \ +\ M \int \frac{d^3\mathbf{q}}{(2\pi)^3}\, V_{\ell}(p, q)\frac{1}{k^2-q^2+i\epsilon}T_{\ell}(q, p';k) \ ,
\end{equation}
where the spin and total angular momentum labels are left implicit.

Consider one-term (rank-$1$) separable potentials of the form
\begin{align}
    V_{\ell}(p, p') = g_{\ell}\mathcal{G}_\ell(p)\mathcal{G}_{\ell}(p')\ ,
\label{r1sepv}
\end{align}
where the constant $g_{\ell}$ and the function $\mathcal{G}_\ell(p)$ are to be determined.
Using the ansatz for the scattering amplitude,
\begin{align}
    T_{\ell}(p, p'; k) = N_{\ell}(-ik)\mathcal{G}_\ell(p)\mathcal{G}_{\ell}(p')\ ,
\end{align}
the LS integral equation gives
\begin{align}
    N_{\ell}(-ik) = \left(\frac{1}{g_\ell}-I_\ell(-ik)\right)^{-1}\ , \ \ \ I_\ell(-ik) = M\int\frac{d^3\mathbf{q}}{(2\pi)^3}\frac{\mathcal{G}_\ell(q)^2}{k^2-q^2+i\epsilon} \ .
\end{align}
For single-channel scattering, the on-shell scattering amplitude is
\begin{align}
    T_{\ell}(k) = \mathcal{G}_\ell(k)^2\left(\frac{1}{g_\ell}-I_\ell(-ik)\right)^{-1}\ .
    \label{scscatt}
\end{align}
All divergent integrals are evaluated in dimensional regularization with PDS and renormalized at the RG scale $\mu/2$
via 
\begin{equation}
\left(\frac{\mu}{2}\right)^{3-D}  M\int \frac{d^D{\bf q}}{(2 \pi)^{D}} \, \frac{q^{2n}}
{{k^2}- {q^2}+i\epsilon}\ \  \mapup{\rm PDS}\ \ -\frac{M}{4\pi} k^{2n} \left(\mu +i k \right) \ .
\label{eq:IEdef}
\end{equation}

\subsection{The minimal potential and fixed points of the RG}
\label{sec:rgflowbcLO}

\noindent This subsection and the next two will focus on the inverse problem of finding momentum-dependent potentials that
generate the ERE to a given fixed order. Taking the rational approximation for the S-matrix and requiring
that the underlying interaction generates a fixed-order polynomial allows
a general way of analyzing the fixed-point structure of $\nopi$.
As noted above, the minimum number of S-matrix poles in the $\ell$th partial
wave is $d_{\min}=2\ell+1$. In the s-wave, the minimal potential one can write down is
\begin{equation}
  V_0^{0}\left(p,p'\right)\ = \ g_0(\mu) \ =\ - \frac{4\pi}{M} \frac{1}{-\gamma_1+\mu} \ = \ - \frac{4\pi}{M} \frac{1}{-{\bf {\large e}}_{1}^{(1)}+\mu} \ .
\label{eq:LOswave}
\end{equation}
The resulting S-matrix has a sole pole at $\gamma_1=1/a_0$, which is generated dynamically by solving the LS equation. The ERE is given by
\begin{equation}
k\cot\delta_0^0(k) \ = \  -\frac{1}{a_0}\ .
\label{eq:Tslonopi}
\end{equation}
This s-wave potential is the coefficient of the local four-nucleon
operator with no derivatives. It exhibits, as will be shown in detail
below, the unitary (trivial) fixed point at $\gamma_1=0$
($\gamma_1=\infty$).

For the p-wave, the minimum number of S-matrix poles is
$d_{\min}=3$. However, the potential
\begin{equation}
  V_1^0\left(p,p'\right)\ \propto \ p\, p' \ ,
\label{eq:noexistLOpwave}
\end{equation}
which arises from a local four-nucleon operator with two
derivatives, does not exist; i.e., it yields a trivial
S-matrix~\cite{Bertulani:2002sz}. This can be understood as a consequence of causality, as discussed below.
Therefore, the simplest such p-wave potential
is given by~\cite{Li:2023hwe}
\begin{equation}
  V_1^{0}\left(p,p'\right)\ = \  g_1(\mu)
  \frac{p\, p' }{\sqrt{{\gamma_1^2}+{p^2}}\sqrt{{\gamma_1^2}+{p^{\prime 2}}}   } \ .
\label{eq:LOpwave}
\end{equation}
The Hermiticity of the potential requires that $\gamma_1$ be real. 
The solution of the Lippmann-Schwinger equation is, via Eq.~\eqref{scscatt},
\begin{equation}
k^3\cot\delta_1^0(k) \ = \  -\frac{1}{a_1} \; +\; \frac{1}{2} r_1 {k^2} \ ,
\label{eq:Tplonopi}
\end{equation}
with the constraint
\begin{equation}
  \frac{1}{a_1}\ =\ -\gamma_1^2 \left(\frac{r_1}{2}+|\gamma_1|\right) \ ,
\label{eq:pwconstraint}
\end{equation}
which requires $a_1<0$ if $r_1\ge 0$. The on-shell potential $V_1^0(k,k)$ has two poles $\pm i|\gamma_1|$ but only $i|\gamma_1|$ survives as a pole of the scattering amplitude. This potential construction works if at least one S-matrix pole is on the positive imaginary axis. With $\gamma_1>0$, in PDS 
\begin{equation}
  g_1(\mu)= - \frac{4\pi}{M}\frac{1}{ -{\bf {\large e}}_{1}^{(3)}+\mu}\ ,
\end{equation}
as well as
\begin{equation}
  a_1\ =\ -\frac{1}{\gamma_1 \gamma_2 \gamma_3} \ =\ -\frac{1}{{\bf {\large e}}_3^{(3)}} \  , \ \ \
  r_1\ =\ -2 \left( \gamma_1+\gamma_2+\gamma_3\right) \ =\ -2{\bf {\large e}}_{1}^{(3)}  \ ,
\label{eq:pwlopars}
\end{equation}
and
\begin{equation}
0 \ =\  \gamma_1\gamma_2+\gamma_1\gamma_3+\gamma_2\gamma_3 \ =\ {\bf {\large e}}_{2}^{(3)}\ ,
\label{eq:pwloparsb}
\end{equation}
in agreement with Eq.~\eqref{eq:esp2} and Eq.~\eqref{eq:constesp2}. 

To generate the ERE to a finite order while avoiding causality constraints, it is necessary to consider a potential that is a special re-summation of the EFT of contact operators. Furthermore, unlike in the s-wave case, only two of the three S-matrix poles are generated dynamically. The third pole corresponds to a nonanalyticity in the potential and is thus a redundant pole. For higher partial waves and higher orders in the ERE, it will be necessary to encode increasing numbers of these redundant poles into non-local potentials. Hence, the ${\cal S}_d$
permutation symmetry of $S_\ell(k)$ will be generated dynamically by
solving the LS equation.

Similarly, one finds that d-wave potentials with zero or one redundant pole do not exist. The simplest potential has two redundant poles, generating an S-matrix with
$d_{\min}=5$ poles, and so on. In general, for single-channel scattering, the simplest potential that generates the ERE to a fixed order is
\begin{equation}
  V_\ell^{0}\left(p,p'\right)\ = g_\ell(\mu) \mathcal{G}_{\ell}(p)  \mathcal{G}_{\ell}(p')\ ,
\label{eq:LOellwaveb}
\end{equation}
where
\begin{equation}
  g_\ell(\mu) \ =\ - \frac{4\pi}{M} \frac{1}{-{\bf {\large e}}_{1}^{(2\ell+1)}+\mu} \ , \ \ \  \mathcal{G}_{\ell}(p)\ =\  p^\ell\prod_{j=1}^{\ell} \frac{1}{\sqrt{\gamma_j^2+{p^2}}} \ ,
\label{eq:LOellwavec}
\end{equation}
with $\Re\gamma_j>0$ for all redundant poles $\gamma_j$ appearing in the potential. For $\ell\geq1$, the solution is
\begin{equation}
k^{2\ell +1} \cot\delta^0_\ell(k)\ =\  -\frac{1}{a_\ell} \ +\ \frac{1}{2} r_\ell {k^2} \ +\  \sum_{n'=2}^\ell v_{n',\ell}\, k^{2n'} \ ,
\label{eq:eregenLOsol}
\end{equation}
with effective-range parameters given as ESPs of the pole positions in
Eq.~\eqref{eq:esp2}.
Notice that the potential is singular and 
has a pole at the RG scale $\mu={\bf {\large e}}_{1}^{(2\ell+1)}$, which for the s- and p-waves corresponds to the inverse scattering length and the negative of half the effective range, respectively.

From the perspective of the momentum expansion, the fact that this family of potentials generates the ERE to $\mathcal{O}(k^{2\ell})$ is natural
since the branch cut in the inverse amplitude appears at
$\mathcal{O}(k^{2\ell+1})$.  Formally, one can view this as a
consequence of Wigner bounds; i.e., causality constraints on
non-relativistic scattering, which have implications for the analytic
structure of the
S-matrix~\cite{PhysRev.74.131,PhysRev.83.249,PhysRev.91.1267,Beane:2021bpr} and,
for systems arising from finite-range forces, for the spread of
allowed values of the effective-range
parameters~\cite{Wigner:1955zz,Phillips:1996ae,Hammer:2009zh,Hammer:2010fw}.
For instance, the Wigner bound drives the p-wave parameter $r_1$ to $-\infty$
as the range of the interaction is taken to zero, which is precisely what
renders the scattering amplitude obtained from the potential, Eq.~\eqref{eq:noexistLOpwave},
trivial. Hence,
the potentials 
constructed above serve as a natural starting point in each partial wave.

Note that with this family of potentials, one has the parametric freedom to (attempt to) take the limit that any or all ERE parameters, save for $a_\ell$, vanish. In the case where all other ERE parameters vanish, the inverse scattering amplitude is 
\begin{align}
   T_\ell(k)^{-1} \propto -\frac{1}{a_\ell}-ik^{2\ell+1}\ .
\end{align}
The S-matrix poles are then proportional to the $(2\ell+1)$th roots of unity. In particular\footnote{The pole permutation symmetry ${\cal S}_d$ makes the labeling of the $\gamma_j$ arbitrary.}, 

\begin{align}
    \gamma_{j}=\frac{(-1)^\ell}{(a_\ell)^{1/(2\ell+1)}}\exp \left(\frac{2\pi i j}{2\ell+1}\right) \ , \ \ \ j=1,...,2\ell+1\ .
\end{align} 
Note that this limit is not guaranteed to exist since, for example, for $\ell$ odd and $a_\ell>0$, the $\ell+1$ poles with $\Re\gamma_j>0$ are all complex and a Hermitian potential with $\ell$ poles cannot be written down. An example is provided by Eq. \eqref{eq:pwconstraint} for the p-wave, where $r_1=0$ does not allow $a_1>0$.

Taking $g_\ell^\star$ to be the coupling with ${\bf {\large e}}_{1}^{(2\ell+1)}=0$ and defining
${\hat g_\ell} \equiv g_\ell/g_\ell^\star$, the beta-functions for the rescaled couplings
are 
\begin{equation}
  {\hat\beta({\hat g}_\ell)}\ =\  \mu \frac{d}{d\mu} {\hat g}_\ell(\mu) \ =\ -{\hat g}_\ell(\mu)\left({\hat g}_\ell(\mu)-1\right) \ ,
   \label{eq:c0betafn}
\end{equation}
which have nominal fixed points at ${\hat g}_\ell=0$ (trivial) and $1$ (nontrivial) in correspondence with ${\bf {\large e}}_{1}^{(2\ell+1)}=\infty$  and $0$, respectively, at fixed $\mu$.
At the nontrivial fixed point, scale invariance requires $\gamma_j=0\,\forall\, j$ and the potential takes the form
\begin{equation}
  V(\cos\theta)\ \supset \  (2\ell+1)g_\ell^\star P_\ell(\cos\theta) \ .
\label{eq:LOellwaventfp}
\end{equation}
However, note that this potential is represented by a local
operator in EFT only for the s-wave. For all $\ell\neq 0$, the
potential cannot be represented by a local operator and therefore does not represent a physical system described by short-range interactions\footnote{This conclusion  appears to be consistent with
  Refs.~\cite{Nishida:2011np,Braaten:2011vf} which, for the p-wave,
  find that in the nontrivial, scale-invariant limit there are low-energy states
  with negative probability.}.

Viewing EFT expansions as perturbations around fixed points of the RG,
the above analysis implies that, while the s-wave accommodates
expansions about the trivial or unitarity fixed points, all higher
waves are expansions about trivial fixed points. Hence, the RG
analysis of the EPT potentials recovers standard $\nopi$ power
counting of Refs.~\cite{vanKolck:1997ut,Kaplan:1998tg,Kaplan:1998we,vanKolck:1998bw}.
It is important to note that, for the s-wave, this power counting is not strictly an expansion
about the unitarity fixed point\footnote{
  An expansion about the unitarity fixed point has been considered for few-nucleon observables in Refs.~\cite{Konig:2016utl,Konig:2016iny,vanKolck:2017jon,Dawkins:2019vcr,Konig:2019xxk,Teng:2024exc,Griesshammer:2025els} and, recently, for
  correlation functions at fixed baryon-number charge in Refs.~\cite{Chowdhury:2023oas,Beane:2024kld,Beane:2025tum}.}.
It is an expansion about a fixed point, not of the non-relativistic conformal symmetry, but rather of
a different symmetry~\cite{Beane:2020wjl,Beane:2021xrk,Beane:2021dab}.

\subsection{Corrections to the minimal potential}
\label{sec:rgflowsub}

\noindent The minimal potential introduced above can be regarded as the first in a family of potentials, $V^m_\ell$, that generate the ERE to successively higher powers of $k$. For $m\geq1$, 
\begin{equation}
V_\ell^{m}\left(p,p'\right)\ =\ g_{m,\ell} \mathcal{G}_{m,\ell}(p)  \mathcal{G}_{m,\ell}(p')
\label{eq:NpLOellwaveb}
\end{equation}
with
\begin{equation}
  g_{m,\ell} \ =\ (-1)^m \frac{4\pi}{M}\, {\bf {\large e}}_{2m-1}^{(2m+2\ell)}\ \ \ , \ \ \  \mathcal{G}_{m,\ell}(p)\ =\  p^\ell
  \prod_{j=1}^{m+\ell} \frac{1}{\sqrt{\gamma_j^2+{p^2}}} \ ,
\label{eq:NpLOellwavec}
\end{equation}
where again $\Re\gamma_j>0$ for $j=1, ..., m+\ell$.
For $m\geq 1$, this potential is regular at the origin. For $\ell,m\geq 0$ the solution is
\begin{equation}
k^{2\ell +1} \cot\delta^{m}_\ell(k)\ =\  -\frac{1}{a_\ell} \ +\ \frac{1}{2} r_\ell {k^2} \ +\  \sum_{n=2}^{m+\ell} v_{n,\ell}\, k^{2n} \ .
\label{eq:eregenNpLOsol}
\end{equation}
That is, the potential $V_\ell^m$ generates the ERE truncated at $\mathcal{O}(k^{2m+2\ell})$. In particular, in the s-wave the $V_0^1$ potential produces two S-matrix poles, one being a redundant pole on the positive imaginary axis. An alternative way~\cite{RevModPhys.21.488,Tumino:2024qpk} to produce the same on-shell amplitude employs the local Eckart potential~\cite{Eckart:1930zza}.

As one goes to higher order in the EFT expansion, the interaction is more singular and shorter distances are probed.
On the other hand, as one goes to higher order in the EPT expansion, the interaction is more rapidly convergent and
less short-distance physics is probed. This makes sense as adding poles allows increasing amounts of short-distance physics
to be removed.

\subsection{Obstructions to solving the inverse problem}
\label{constraints}

\noindent For an effective-range expansion of finite order $2m+2\ell$ to have a corresponding separable potential with $m+\ell$
redundant poles, it is clear that (i) 
the polynomial in Eq. \eqref{eq:eregenbssol} must admit $m+\ell$ zeros $k=i\gamma_j$ with 
$\Re \gamma_j>0$ 
 so that a potential can be written down, and (ii) those poles $\{\gamma\}$ must be either real or come in complex-conjugate pairs so that the potential is Hermitian. While these constraints seem abstract, they can be better understood by analyzing the bound states generated by the above separable potentials.

For S-matrices of the form in Eq.~\eqref{eq:smatgen}, the phase shift is given by 

\begin{equation}
    \delta(k)-\delta(\infty) = \sum_{j=1}^d \arg \left(k+i\gamma_j\right) \  ,
\end{equation}
where the $\arg$ function has the range $[-\pi, \pi)$.
At $k=0$, each pole's contribution to the phase shift is $\arg(i\gamma_j)$. If $\gamma_j\in \mathbb{R}$ and $\gamma_j>0$, its contribution to the phase shift at $k=0$ is $\pi/2$. If $\gamma_j$ is complex, but $\Re\gamma_j>0$, $\gamma_j$ and its complex conjugate will contribute a combined $\pi$ to the phase shift. Likewise, poles with $\Re\gamma_j<0$ contribute an average of $-\pi/2$ to the phase shift at $k=0$.
Therefore, 

\begin{align}
    \delta(0)-\delta(\infty) = \frac{\pi}{2}(n_+-n_-) \ ,
\end{align}
where $n_+$ ($n_-$) is the number of poles with $\Re\gamma_j>0$ ($\Re\gamma_j<0$).

For one-term separable potentials, the Fredholm determinant of the Lippmann-Schwinger kernel is easily computed~\cite{Newton1982}. For the nonsingular potentials in Eq.~\eqref{eq:NpLOellwavec}, Levinson's theorem takes the form

\begin{align}
    \delta(0)-\delta(\infty) 
    = \pi n_B \ ,
\end{align} 
where $n_B$ is the number of bound states. Since $n_B \geq 0$, $n_+\geq n_-$ and $n_+\geq m+\ell$. In other words, the constraint that $n_+ \geq m+\ell$ comes from constraining the asymptotic behavior of the phase shift to be compatible with Levinson's theorem.

Any T-matrix that corresponds to a finite-order ERE cannot cross $0$ for any finite, real, and nonzero value of $k$. Thus $\Delta\delta = 0$ or $\pm\pi$. The case of $\Delta\delta = -\pi$ is excluded by Levinson's theorem. The sickness in such an ERE can be attributed to the highest-order nonvanishing effective-range parameter giving the phase shift the wrong asymptotic behavior. Clearly, such an ERE can be fixed by adding a shape parameter at the next order with the appropriate sign to give the phase shift the correct asymptotic behavior. The corresponding potential is well-defined in the one-sided limit as that shape parameter is sent to 0. It just cannot be defined when that shape parameter is equal to 0.

In the case where the potential is singular as in Eq.~\eqref{eq:LOellwavec},
\begin{equation}
    \delta(0)-\delta(\infty) 
    = \pi \left(n_B-\frac{1}{2}\right) \ .
\end{equation}
$n_B\geq0$ requires that $n_+\geq n_--1$ and $n_+ \geq \ell$. Since the phase shift cannot cross zero, it is clear that $\Delta\delta = \pm\pi/2$, meaning $n_+\geq n_--1$ is guaranteed for all such EREs.

It is also required that the $m+\ell$
poles with $\Re\gamma_j>0$ are either real or come in complex-conjugate pairs. This may not be true if $n_+$ is even, $m+\ell$
is odd, and all poles $\gamma_j$ are complex. In this case, $n_B = 1$, but there exists no zero of the polynomial in Eq. \eqref{eq:eregenbssol}
corresponding to a bound state with a real energy $\gamma_j\in\mathbb{R}$ and $\gamma_j>0$. Clearly, such an ERE is sick. Again, this problem is solved by adding a higher-order effective-range parameter so that either (i) $n_-$ increases to equal $n_+$ and the bound state vanishes or (ii) $n_+$ increases by 1 and there is guaranteed to be a real pole $\gamma_j$. Again, such a potential is well-defined in the limit that this new effective-range parameter is sent to 0, just not at 0.

While the obstructions to writing down a potential become less transparent in the coupled-channel case, in practice, they can also be circumvented in the same way, by adding a small shape parameter at the next order.

\subsection{The coupled Lippmann-Schwinger equation}

 \noindent The coupled 
 LS integral equations for the scattering amplitude $T$, projected onto states with orbital angular momentum $\ell$, are
 \begin{equation}
     T_{\ell,\ell'}(p,p';k) \ =\  V_{\ell,\ell'}(p, p') \ +\ M \sum_{\ell''}\int_0^\infty \frac{dq\, q^2}{2\pi^2}\, V_{\ell,\ell''}(p, q)\frac{1}{k^2-q^2+i\epsilon}T_{\ell'',\ell'}(q, p';k) \ ,
 \end{equation}
 where the spin and total angular momentum labels are left implicit. The relation between the on-shell scattering amplitude and the S-matrix is
 \begin{equation}
  S_{\ell,\ell'}(k) \ =\  \delta_{\ell,\ell'}\ -\, i\frac{M k}{2\pi}\,T_{\ell,\ell'}(k)\ .
\label{eq:smatpwexpcc}
\end{equation}

 Consider one-term (rank-$1$) separable potentials of the form
 \begin{align}
     V_{\ell,\ell'}(p, p') = g_{\ell,\ell'}\mathcal{G}_\ell(p)\mathcal{G}_{\ell'}(p')\ ,
 \label{r1sepv-coupled}
 \end{align}
 where the constants $g_{\ell,\ell'}$ and the functions $\mathcal{G}_\ell(p)$ are to be determined.
 Using the ansatz for the scattering amplitude,
 \begin{align}
     T_{\ell,\ell'}(p, p'; k) = \tau_{\ell,\ell'}(-ik)\mathcal{G}_\ell(p)\mathcal{G}_{\ell'}(p')\ ,
 \end{align}
 the LS integral equations become a set of coupled algebraic equations of the form

 \begin{align}
     \tau_{\ell,\ell'}(-ik) = g_{\ell,\ell'}+ M\sum_{\ell''}g_{\ell,\ell''}\int_0^\infty\frac{dq\,q^2}{2\pi^2}\frac{\mathcal{G}_{\ell''}(q)^2}{k^2-q^2+i\epsilon}\tau_{\ell'',\ell'}(-ik)\ .
 \end{align}
This is solved by

 \begin{align}
 \tau_{\ell, \ell'}(-ik)  = \sum_{\ell''}\left[\left(\hat{\mathbf{1}}-\hat{\mathbf{g}}\hat{ \mathbf{I}}(-ik)\right)^{-1}\right]_{\ell,\ell''}\left[\hat{\mathbf{g}}\right]_{\ell'',\ell'}\ ,
 \end{align}
 where $\left[\hat{\mathbf{g}}\right]_{\ell,\ell'} = g_{\ell,\ell'}$, $\left[ \hat{ \mathbf{I}}(-ik)\right]_{\ell, \ell'} = I_\ell(-ik)\delta_{\ell,\ell'}$, and 
 
 \begin{align}
    I_\ell(-ik) = M\int_0^\infty\frac{dq\,q^2}{2\pi^2}\frac{\mathcal{G}_{\ell}(q)^2}{k^2-q^2+i\epsilon}\ . 
    \label{intDef}
 \end{align}

 The real M-matrix is defined in relation to the on-shell T-matrix by
 \begin{align}
     \hat{\mathbf{T}} = -\frac{4\pi}{Mk}\left(\hat{\mathbf{M}}-i\hat{\mathbf{1}}\right)^{-1}\ .
 \end{align}
 The M-matrix obeys the coupled-channel effective-range expansion
 \begin{align}
     k^{\ell+\ell'+1}M_{\ell,\ell'} = -\left[\hat{\mathbf{a}}^{-1}\right]_{\ell,\ell'} + \frac{1}{2}\left[\hat{\mathbf{r}}\right]_{\ell,\ell'}k^2+\sum_{n'=2}^n\left[\hat{\mathbf{v}}_{n'}\right]_{\ell,\ell'}k^{2n'}+\mathcal{O}\left(k^{2n+2}\right)\ .
 \end{align}
 In the single-channel case, the M-matrix is a one-by-one matrix with $M_\ell = \cot \delta_\ell$.
 For a one-term separable potential, 
 \begin{align}
     M_{\ell,\ell'} = -\frac{4\pi}{Mk}\frac{1}{\mathcal{G}_{\ell}(k)\mathcal{G}_{\ell'}(k)}\left[\hat{\mathbf{g}}^{-1}-\hat{\mathbf{I}}(-ik)\right]_{\ell,\ell'}+i\delta_{\ell, \ell'} \ .
\end{align}

It is well known that two spin triplet channels with the same total angular momentum $J$, but differing orbital angular momenta $\ell = J-1$ and $\ell = J+1$ can mix. In this case, the $S$-matrix can be represented as a 2-by-2 matrix using the Stapp parametrization

\begin{align}
    S = \begin{pmatrix}
        \text{cos}\,2\epsilon_J \ e^{2i\delta_{J-1}} & i\,\text{sin}\,2\epsilon_J \ e^{i(\delta_{J-1}+\delta_{J+1})} \\
        i\,\text{sin}\,2\epsilon_J \ e^{i(\delta_{J-1}+\delta_{J+1})} & \text{cos}\,2\epsilon_J\ e^{2i\delta_{J+1}}
    \end{pmatrix}\ , \label{Stapp}
\end{align}
where $\epsilon_J$ is the mixing angle. The scattering phase shifts and mixing angle can be calculated from the M-matrix elements:

\begin{align}
    \delta_{J-1} &= \frac12 \arg S_{J-1,J-1} \\
    &= \frac12 \arctan \left(
    \frac{
    2M_{J-1,J-1}\left(M_{J+1,J+1}^2+1\right)
    -2M_{J+1,J-1}^2M_{J+1,J+1}
    }{
    \left(M_{J+1,J+1}^2+1\right)
    \left(M_{J-1,J-1}^2-1\right)
    -2M_{J+1,J-1}^2M_{J+1,J+1}M_{J-1,J-1}
    +M_{J+1,J-1}^4
    }
    \right) \ ,
\end{align}
 while $\delta_{J+1}$ is obtained by swapping $J-1 \leftrightarrow J+1$ in the above formula, and

\begin{align}
    \tan^22\epsilon_J &= -\frac{S_{J-1,J+1}^2}{S_{J+1,J+1}S_{J-1,J-1}}\\
    \Rightarrow\epsilon_J&= \pm\frac12 \arctan\left(\frac{2M_{J+1,J-1}}{\sqrt{\left(\det\hat{\mathbf{M}}+1\right)^2+\left(M_{J+1,J+1}-M_{J-1,J-1}\right)^2}}\right)\ ,
\end{align}
where the appropriate sign is determined by matching to the threshold behavior of $S_{J-1, J+1}$.
\subsection{The coupled-channel potential} 

\noindent Consider specializing the one-term separable potential from Eq.~\eqref{r1sepv-coupled} to use the form factors from Eq.~\eqref{eq:LOellwavec} and Eq.~\eqref{eq:NpLOellwavec}. Since the diagonal part of the M-matrix is precisely what one would find in the single-channel case, save for replacing $1/g_{\ell}$ with $\left[\hat{\mathbf{g}}^{-1}\right]_{\ell,\ell}$, this separable form factor again produces the ERE to fixed order in the diagonal part of the M-matrix. The only qualification is that $\left[\hat{\mathbf{g}}^{-1}\right]_{\ell,\ell}$ must be renormalized rather than $1/g_{\ell}$. 

Here, the $\{\gamma\}$ are no longer interpreted as S-matrix poles, but they are still zeros of $k^{2\ell+1}M_{\ell,\ell}-ik^{2\ell+1}$. However, the redundant $\{\gamma\}$ that are encoded into the potential are still singularities of the scattering amplitude because they are zeros of the determinant $ \det\left(\hat{\mathbf{M}}-i\hat{\mathbf{1}}\right)$, which the scattering amplitude, T, is inversely proportional to. In particular, the $\{\gamma\}$ in $\mathcal{G_\ell}$ are still poles of $T_{\ell,\ell}(k)$.

The off-diagonal M-matrix element is given by

\begin{align}
    M_{J-1,J+1} = -\frac{4\pi}{Mk}\frac{1}{\mathcal{G}_{J-1}(k)\mathcal{G}_{J+1}(k)}\left[\hat{\mathbf{g}}^{-1}\right]_{J-1,J+1}\ .
\end{align}
Note that since the $\mathcal{G}_{\ell}(k)$'s do not depend on the renormalization scale $\mu/2$, $\left[\hat{\mathbf{g}}^{-1}\right]_{J+1,J-1}$ is $\mu$-independent. 
Since the free parameters in the $\mathcal{G}_{\ell}(k)$'s are fixed by the on-diagonal piece of the ERE, these potentials only provide direct parametric control over the off-diagonal scattering length through $\left[\hat{\mathbf{g}}^{-1}\right]_{J+1,J-1}$, where

\begin{align}
    \left[\hat{\mathbf{a}}^{-1}\right]_{J+1,J-1} = \left.\frac{4\pi}{M}\frac{k^{2J}}{\mathcal{G}_{J-1}(k)\mathcal{G}_{J+1}(k)}\left[\hat{\mathbf{g}}^{-1}\right]_{J+1,J-1}\right|_{k=0}\ .
\end{align}
For this choice of $\mathcal{G}_\ell$, an infinite number of correlated, higher-order effective-range parameters are generated that depend nontrivially on the on-diagonal effective-range parameters and $\left[\hat{\mathbf{a}}^{-1}\right]_{J+1,J-1}$.

\subsection{Weinberg eigenvalues}
\label{sec:weigen}

\noindent A more general diagnostic of the softness of momentum-space potentials than that considered in
the introduction is provided by the Weinberg eigenvalues~\cite{Weinberg:1963zza,Bogner:2005sn}. The
LS equation in operator form is
\begin{equation}
   \hat T (z)\ =\  \hat V \ +\ \hat V \hat G_0(z) \hat T(z) \ ,
\end{equation}
where $z$ is a complex energy, with $\hat V$ an energy-independent potential, and $\hat G_0(z)$ the free two-particle
Green's function. The convergence of the Born series may be studied by defining the eigenvalue system,
\begin{equation}
\hat V \hat G_0(z) \ket{\Psi_\nu(z)} \ =\ \eta_\nu(z) \ket{\Psi_\nu(z)} \ ,
\end{equation}
where $\eta_\nu(z)$ are the Weinberg eigenvalues. 
The magnitude of the largest Weinberg eigenvalue is the effective expansion parameter of the Born series and 
informs the error induced by truncating the series at a finite order.

For the general coupled-channel case, 
\begin{equation}
\sum_{\ell'}M \int_0^\infty\frac{dq\,q^2}{2\pi^2}\frac{V_{\ell,\ell'}(p,q)}{k^2-q^2+i\epsilon} \langle{q, \ell'}\ket{\Psi_\nu(k)} \ =\ \eta_\nu(k) \langle{p,\ell}\ket{\Psi_\nu(k)} \ .
\end{equation}
Substituting a separable ansatz of the form $\langle{p,\ell}\ket{\Psi_\nu(k)} = D_\ell(-ik)\mathcal{G}_\ell(p)$, this set of integral equations reduces to the algebraic equations
\begin{align}
    \left(\eta\hat{\mathbf{1}}-\hat{\mathbf{g}}\hat{\mathbf{I}}(-ik)\right)\hat{\mathbf{D}} = 0\ .
\end{align}
The Weinberg eigenvalues are the eigenvalues of the matrix $\hat{\mathbf{g}}\hat{\mathbf{I}}(-ik)$. It can be further shown that this matrix has the same eigenspectrum as 
\begin{align}
    \hat{\mathbf{1}}-\hat{\mathbf{V}}(k, k)\hat{\mathbf{T}}^{-1}(k)\ .
\end{align}
This matrix is usually easier to compute in practice. This form makes it clear that the Weinberg eigenvalues are explicitly dependent on the regulator through the on-shell potential $\hat{\mathbf{V}}(k, k)$.

For the single-channel case, this eigenvalue problem gives a single Weinberg eigenvalue of 
\begin{equation}
\eta(k) \ =\ 1 \ +\ \frac{M k}{4\pi}V(k,k)\left( \cot\delta(k)\;-\;i \right) \ .
\end{equation}
The imaginary part of $\eta(k)$ is the naive effective expansion
parameter postulated in the introduction. However, returning to the potential $V=4\pi a_0/M$, one can check that it has a Weinberg eigenvalue whose real part precisely vanishes. The effective expansion parameter $ka_0$ can be reinterpreted as not just the strength of the potential (or the imaginary part of the Weinberg eigenvalue), but as the full Weinberg eigenvalue, up to a phase.

Note that at least one $\eta(k)=1$ at every non-redundant pole of the S-matrix. This reflects the fact that iterating the potential to all orders is necessary to generate the non-redundant poles of the scattering amplitude, including bound states. In this sense, channels supporting low-lying bound states, such as the deuteron in the neutron-proton system, are inherently nonperturbative.

\section{Bound-state properties}

\label{sec:bsprop}

\noindent 
The existence of bound states constrains the parameters of the potentials that can sustain them.
Properties of the deuteron, the bound state in the $^3S_1$-$^3D_1$ channel in the neutron-proton system, are useful metrics for assessing the performance of the coupled-channel potentials. In particular, the asymptotic D/S ratio of the deuteron is examined below.

In the $^3S_1$-$^3D_1$ coupled-channel system, the bound-state LS equations are
\begin{equation}
\psi_\ell(p)= -\frac{M}{p^2+\gamma_d^2}\sum_{\ell'}
\int_0^\infty \frac{q^2\,dq}{2\pi^2}\; V_{\ell,\ell'}(p,q)\,\psi_{\ell'}(q) \ ,
\label{eq:LSbound}
\end{equation}
where the wavefunctions, $\psi_\ell(p)$, are evaluated at external 
momentum $p$, and $\gamma_d>0$ is the deuteron binding momentum.

Assuming the interaction of Eq.~\eqref{r1sepv-coupled}, one can define
\begin{equation}
C_\ell \equiv -M\int_0^\infty \frac{q^2\,dq}{2\pi^2}\,\mathcal{G}_\ell(q)\,\psi_\ell(q) \ .
\label{eq:Cdef}
\end{equation}
In matrix notation, the condition for a bound-state solution is
\begin{equation}
\big[\hat{\mathbf 1}-\,\hat{\mathbf I}(\gamma_d)\,\hat{\mathbf g}\big]\hat{\mathbf C}=0\; .
\label{eq:algCcond}
\end{equation}
Therefore, a nontrivial solution exists if and only if
\begin{equation}
\det\!\big[\hat{\mathbf 1}-\,\hat{\mathbf I}(\gamma_d)\,\hat{\mathbf g}\big]=0\; ,
\label{eq:det_condition}
\end{equation}
or, explicitly,
\begin{equation}
(1-I_0 g_{00})(1-I_2 g_{22}) - I_0 I_2\, g_{02}g_{20}=0\; .
\label{eq:detexpl}
\end{equation}

To obtain the wavefunctions, let $r\equiv C_2/C_0$ and define
\begin{equation}
\tilde v_0 \equiv g_{00}+g_{02}r,\qquad
\tilde v_2 \equiv g_{20}+g_{22}r.
\end{equation}
Then the wavefunctions are
\begin{equation}
\psi_0(p)= C_0\,\tilde v_0\,\frac{\mathcal{G}_0(p)}{p^2+\gamma_d^2},\qquad
\psi_2(p)= C_0\,\tilde v_2\,\frac{\mathcal{G}_2(p)}{p^2+\gamma_d^2}\; .
\label{eq:psiexplicit}
\end{equation}
From (either row of) Eq.~\eqref{eq:algCcond}, one finds
\begin{equation}
r= \frac{1-I_0(\gamma_d) g_{00}}{I_0(\gamma_d) g_{02}}
= \frac{I_2(\gamma_d) g_{20}}{1-I_2(\gamma_d) g_{22}}\; .
\end{equation}

The final step is to obtain $C_0$. Imposing the momentum-space normalization
\begin{equation}
1=\sum_{\ell}\int_0^\infty \frac{p^2\,dp}{2\pi^2}\,|\psi_\ell(p)|^2 
\label{eq:nmom}
\end{equation}
and defining
\begin{equation}
N_\ell(\gamma_d)\equiv \int_0^\infty \frac{p^2\,dp}{2\pi^2}\,\frac{\mathcal{G}_\ell(p)^2}{(p^2+\gamma_d^2)^2}\ ,
\label{eq:Ndef}
\end{equation}
gives
\begin{equation}
C_0=\frac{1}{\sqrt{|\tilde v_0|^2N_0(\gamma_d)+|\tilde v_2|^2N_2(\gamma_d)}}\; ,
\label{eq:C0norm}
\end{equation}
where $C_0$ has been chosen to be real and positive.

The deuteron coordinate-space wavefunctions are available via the Fourier transform
\begin{equation}
u(r)=\frac{2}{\pi}\,r\int_0^\infty dp\,p^2\,j_0(pr)\,\psi_0(p),\qquad
w(r)=\frac{2}{\pi}\,r\int_0^\infty dp\,p^2\,j_2(pr)\,\psi_2(p),
\label{eq:coordspwfs}
\end{equation}
where the $j_\ell$ are spherical Bessel functions. For large $r$, 
\begin{equation}
u(r)\xrightarrow[r\to\infty]{} A_S\,e^{-\gamma_d r},
\qquad
w(r)\xrightarrow[r\to\infty]{} A_D\,e^{-\gamma_d r}\left(1+\frac{3}{\gamma_d r}+\frac{3}{(\gamma_d r)^2}\right), 
\label{eq:asymptoticcoordforms}
\end{equation}
and the asymptotic D/S ratio is
\begin{equation}
\xi \equiv \frac{A_D}{A_S} \;=\;-\frac{\tilde v_2\,\mathcal{G}_2(i\gamma_d)}{\tilde v_0\,\mathcal{G}_0(i\gamma_d)}\; ,
\end{equation}
up to an overall phase, where 
\begin{align}
    \frac{\tilde v_2}{\tilde v_0} = \frac{g_{02}I_0(\gamma_d)}{1-g_{22}I_2(\gamma_d)} = \frac{1-g_{00}I_0(\gamma_d)}{g_{02}I_2(\gamma_d)}\ .
\end{align}
The asymptotic D/S ratio is a long-distance bound-state property accessible to EPT.

\section{Power counting}

\label{sec:powcount}

\noindent When computing few- and many-body observables, it is important to have a power-counting scheme that provides a consistent ordering of the potentials across all partial waves. 

For this reformulation of $\nopi$, all effective-range parameters are treated as naturally sized and scale as powers of the UV scale $\mathcal{M}$: 
\begin{itemize}[nosep]
    \item $a_\ell\sim\mathcal{M}^{-2\ell-1}$
    \item $r_\ell\sim\mathcal{M}^{2\ell-1}$
    \item $v_{n, \ell} \sim \mathcal{M}^{2\ell-2n+1}$
\end{itemize}
At N$^p$LO, the ERE is expanded to $\mathcal{O}(k^{2p-2\ell})$ in the $\ell$th partial wave, so that each successive order adds two additional powers of the small momentum $k$. In cases where a potential cannot be written down in EPT that generates the desired ERE, small, formally subleading terms are added to the ERE to remedy the pathology. For example, the first few orders introduce the following parameters:

\begin{itemize}[nosep]
    \item LO: $a_0$
    \item NLO: $a_1$, $r_0$
    \item 
    NNLO: $a_2$, $r_1$, $v_{2,0}$
    \item 
    N$^3$LO: $a_3$, $r_2$, $v_{2,1}$, $v_{3, 0}$
\end{itemize}

The ordering is chosen so that, if all effective-range parameters were treated in perturbation theory, this would be equivalent to expanding $\delta_\ell$, and likewise $\epsilon_\ell$, to $\mathcal{O}(k^{2p+1})$. In EPT, however, the scattering amplitude and all observables are computed exactly from the (re-summed) EPT potentials. None of the effective-range parameters are treated in perturbation theory.  Consequently, unnaturally large effective-range parameters that would otherwise need to be promoted and re-summed in $\nopi$ are naturally accommodated.

\subsection{Power counting in a single channel}

\noindent The manual for constructing the N$^p$LO potential in the $\ell$th partial wave is as follows:

\begin{itemize}[nosep]
\item Truncate the ERE at $\mathcal{O}(k^{2p-2\ell})$. This ERE is generated by the potential $V^m_\ell$ with $m+\ell = \max\{\ell, p-\ell\}$ poles.
\item If an EPT potential cannot be written down for the desired ERE, instead use the potential $V^{m+1}_\ell$, hence increasing the number of redundant poles by $1$, to $m+\ell+1$ and thus extending the ERE by one term. Set the $\mathcal{O}(k^{2m+2\ell+2})$ parameter to be small or naturally sized with the sign needed to generate the appropriate asymptotic behavior. Then fit this ERE to data with all lower-order effective-range parameters as free parameters. 
\end{itemize}

In the above scheme, the lowest-order ERE in the $\ell$th partial wave begins at N$^\ell$LO with $a_\ell$. The first subleading correction to the ERE comes from $r_{\ell}$ at N$^{\ell+1}$LO, and so on. The first $\ell+1$ nontrivial orders, up to N$^{2\ell}$LO, in $\nopi$ can be constructed using the $V_\ell^0$ potential, which generates the ERE up to $\mathcal{O}(k^{2\ell})$. (Recall that these potentials often, but not always, have the parametric freedom to set unwanted effective-range parameters to zero.) Each subsequent order after N$^{2\ell}$LO uses the next subleading potential.

If, at any order, such an ERE does not have a corresponding EPT potential, one can add a subleading effective-range parameter of small or natural size to cure its sickness without significantly changing the behavior of the scattering amplitude at momentum scales of interest.

Consider, for example, the following channels, which start at LO and NLO. 

\subsubsection{\texorpdfstring{$^1S_0$}{1S0}}

\noindent The LO potential is $V_0^0$, or $g_0(\mu)$, which generates the scattering length $a_0$ to all orders. The NLO potential is $V_0^1$
and generates the first two terms in the ERE, $a_0$ and $r_0$, to all orders.

The NNLO potential should generate $a_0$, $r_0$, and $v_{2,0}$ to all orders; however, $v_{2,0}$ is found to be negative. This gives the phase shift the wrong asymptotic behavior, $\delta(0)-\delta(\infty) = -\pi$, and a $V_0^2$ potential cannot be written down. As suggested in Section~\ref{constraints}, this S-matrix is fixed by inserting a shape parameter with the correct sign, $v_{3,0} = +0.1$ fm$^5$, so that $\delta(0)-\delta(\infty) = 0$ and a potential can be written down. This necessitates using the potential with three redundant poles, namely the $V_0^3$ potential with this fake but subleading shape parameter. At 
N$^3$LO, there is no obstruction to employing the $V_0^3$ potential with three redundant poles that generates the physical $a_0$, $r_0$, $v_{2,0}$, and $v_{3,0}$.

\subsubsection{\texorpdfstring{$^3P_0$}{3P0}}

\noindent At NLO, which is the lowest nontrivial order in the ERE, the coupling 
$g_1(\mu)$ and the pole $\gamma_1$ of the $V_1^0$ potential are correlated to reproduce $a_1$ while enforcing $r_1 = 0$. 
Since $a_1<0$, one redundant pole on the positive imaginary axis is sufficient to accomplish this. At NNLO the same potential can be re-fitted to yield an $r_1>0$. At N$^3$LO, one employs the $V_1^1$ potential with two redundant poles to further accommodate $v_{2,1}>0$.

\subsubsection{\texorpdfstring{$^1P_1$, $^3P_1$}{1P1, 3P1}}

\noindent The construction of the potential is qualitatively the same in the $^1P_1$ and $^3P_1$ channels. Unlike $^3P_0$, however, in these channels $a_1>0$.
As described in Section~\ref{sec:rgflowbcLO}, 
the polynomial in Eq. \eqref{eq:eregenbssol}
has only complex zeros $\gamma_j$ with $\Re\gamma_j>0$ and a Hermitian $V_1^0$ potential with one real pole $\gamma_1$ cannot be written down. As suggested in Section~\ref{constraints}, this ERE is fixed by instead considering the potential with two redundant poles. The coupling 
$g_1$ and the two poles $\gamma_1$ and $\gamma_2$ are correlated such that $a_1$ takes its physical value, $r_1 =0$, and $v_{2, 1} = -2$ fm is fixed at a constant value. A Hermitian potential can be written down using the two complex-conjugate poles with $\Re\gamma_j>0$, and this potential, the $V_1^1$, produces the desired ERE up to natural-sized, subleading corrections from $v_{2, 1}$. 

At NNLO, the $V_1^0$ potential is calibrated 
to produce an ERE with $a_1$ and $r_1$ at their physical values. There is no obstruction because $r_1$ is sufficiently negative. N$^3$LO uses the $V_1^1$ potential with two redundant poles, and so on.

\subsection{Power counting in a coupled channel}

\noindent The manual for constructing the N$^p$LO potential in the $J$th mixed channel is somewhat more complicated than in the single-channel case:

\begin{itemize}[nosep]
\item Truncate the ERE at $\mathcal{O}(k^{2p-2J+2})$ and $\mathcal{O}(k^{2p-2J-2})$ for $M_{J-1, J-1}$ and $M_{J+1, J+1}$, respectively. 
\item Add a total of $r=\max\{0, p-J\}$ auxiliary shape parameters to the two on-diagonal EREs, one for each off-diagonal effective-range parameter required at N$^p$LO that cannot be controlled directly using the parameters of the EPT potential. The potential that generates this ERE has $m_{J-1} = \max\{J-1, p-J+1+r_{J-1}\}$ and $m_{J+1} = \max\{J+1, p-J-1+r_{J+1}\}$ poles in $\mathcal{G}_{J-1}$ and $\mathcal{G}_{J+1}$, respectively. The numbers of auxiliary shape parameters assigned to each ERE are denoted by $r_{J-1}$ and $r_{J+1}$ where $r_{J-1}+r_{J+1} = r$.
\item Any extra auxiliary degrees of freedom should be appropriately constrained prior to fitting data. If $a_{J-1, J+1}$ is subleading, set it equal to zero. If $a_{J+1, J+1}$ is subleading, fix it close to the physical value, but do not fit it with the other parameters. Other extra auxiliary effective-range parameters can be set to zero.
\item As in the single-channel case, if a $\mathcal{G}_\ell$ cannot be written down in EPT corresponding to the desired on-diagonal ERE, increase the number of redundant poles by $1$ to $m_{\ell}+1$ and thus extend the ERE order by $\mathcal{O}(k^2)$. Set the $\mathcal{O}(k^{2m_{\ell}+2})$ parameter to be small or naturally sized with the sign needed to generate the appropriate asymptotic behavior. Then fit this ERE to data with all lower-order effective-range parameters as free parameters.
\end{itemize}

As the most important examples, consider the two coupled channels $^3S_1$-$^3D_1$ and $^3P_2$-$^3F_2$, which start at LO and NLO, respectively. For simplicity, the following examples will not consider sick cases where the desired ERE does not have a corresponding EPT potential and the prescription in the last bullet above is necessary. 

\subsubsection{\texorpdfstring{$^3S_1$-$^3D_1$}{3S1-3D1}}

\noindent
In the $^3S_1$-$^3D_1$ channel, the LO ERE consists of only $a_{00}$. At this order, the $^3S_1$ and $^3D_1$ channels are decoupled and the mixing angle $\epsilon_1$ is precisely 0. The form factor $\mathcal{G}_0$ is $1$, which corresponds to the $V_0^0$ potential, while the form factor $\mathcal{G}_2$ can be freely set to 0.

At NLO, the desired effective-range parameters are $a_{00}$, $r_{00}$, and $a_{02}$. The s-wave $\mathcal{G}_0$ is picked to have one redundant pole so that it generates precisely $a_{00}$ and $r_{00}$, but no further terms in the ERE for $M_{00}$---this $\mathcal{G}_0$ is picked from the $V_0^1$ potential. Since there is nonzero mixing, it is necessary to specify a nontrivial form factor, $\mathcal{G}_2$, for the d-wave. The simplest choice is to take $\mathcal{G}_2$ from the minimal potential $V_2^0$ that has 2 redundant poles. In summary, the form factors are

\begin{align}
    \mathcal{G}_0(k) = \frac{1}{\sqrt{\gamma_1^2+k^2}}\ , \ \ \ \mathcal{G}_2(k) = \frac{k^2}{\sqrt{\beta_1^2+k^2}\sqrt{\beta_2^2+k^2}}\ .
\end{align}

This potential has a total of six degrees of freedom, three embedded in the form factors in addition to the three couplings, which is far more than the three necessary to specify $a_{00}$, $r_{00}$, and $a_{02}$. The three degrees of freedom introduced by the d-wave potential are auxiliary since no prediction is made for $\delta_2$ at this order. In principle, these three parameters can be set freely. For simplicity, $g_{22}$ and the two poles, $\beta_1$ and $\beta_2$, are constrained such that $r_{22}$ = $v_{2, 22}$ = 0 and $a_{22}$ is near its physical value of $\approx 6.5$ fm$^{5}$. Then, $a_{00}$, $r_{00}$, and $a_{02}$ can be fit to data and used to determine all six parameters of the potential. In summary, the M-matrix elements fit to data are

\begin{align}
    kM_{00} &= -\left[\hat{\mathbf{a}}^{-1}\right]_{00}+\frac{1}{2}r_{00}k^2\ ,\ \ \  k^5M_{22} = -\left[\hat{\mathbf{a}}^{-1}\right]_{22}\,\\
    k^3M_{20} &= -\frac{4\pi}{M}\left[\hat{\mathbf{g}}^{-1}\right]_{20}\sqrt{\gamma_1^2+k^2}\sqrt{\beta_1^2+k^2}\sqrt{\beta_2^2+k^2}\\
    &= -\left[\hat{\mathbf{a}}^{-1}\right]_{20}\sqrt{1+\frac{k^2}{\gamma_1^2}}\sqrt{1+\frac{k^2}{\beta_1^2}}\sqrt{1+\frac{k^2}{\beta_2^2}}\ .
\end{align}
Note that with the auxiliary degrees of freedom appropriately constrained, there are only three free, independent variables.
 
NNLO should produce $a_{00}$, $r_{00}$, $v_{2,00}$, $a_{02}$, $r_{02}$, and $a_{22}$ in the ERE. This suggests using the form factors from the $V_0^2$ potential and the $V_2^0$ potential, both of which have two redundant poles. Once the two auxiliary degrees of freedom of the d-wave potential are constrained such that the subleading $r_{22}$ and $v_{2, 22}$ are zero, there are only five degrees of freedom left over to fix the six effective-range parameters at NNLO. In particular, there is only ever one independent parameter, $g_{02}$, in the mixing part of the potential, whereas two effective-range parameters, $a_{02}$ and $r_{02}$, must be reproduced in $M_{02}$.
An additional degree of freedom can be introduced by allowing the auxiliary $r_{22}$ to be nonzero. Since the $\mathcal{O}(k^2)$ and higher-order behavior of $k^3M_{02}$ is correlated with the on-diagonal elements of the M-matrix in a highly nontrivial way, there is no practical way to express $r_{02}$ in terms of the other parameters. Instead of fitting $r_{02}$ directly, $r_{22}$ is used as an additional degree of freedom so that a fit can reproduce the $\mathcal{O}(k^2)$ behavior of $k^3M_{02}$.

In summary, the form factors used at NNLO are

\begin{align}
    \mathcal{G}_0(k) = \frac{1}{\sqrt{\gamma_1^2+k^2}\sqrt{\gamma_2^2+k^2}}\ , \ \ \ \mathcal{G}_2(k) = \frac{k^2}{\sqrt{\beta_1^2+k^2}\sqrt{\beta_2^2+k^2}}\ .
\end{align}
There are a total of seven degrees of freedom, but one is constrained by setting the auxiliary $v_{2, 22}=0$. The M-matrix elements fit to data are

\begin{align}
    kM_{00} &= -\left[\hat{\mathbf{a}}^{-1}\right]_{00}+\frac{1}{2}r_{00}k^2+v_{2, 00}k^4\ ,\ \ \ k^5M_{22} = -\left[\hat{\mathbf{a}}^{-1}\right]_{22}+\frac{1}{2}r_{22}k^2\ ,\\
    k^3M_{20} &= -\left[\hat{\mathbf{a}}^{-1}\right]_{20}\sqrt{1+\frac{k^2}{\gamma_1^2}}\sqrt{1+\frac{k^2}{\gamma_2^2}}\sqrt{1+\frac{k^2}{\beta_1^2}}\sqrt{1+\frac{k^2}{\beta_2^2}}\ ,
\end{align}
where $a_{00}$, $r_{00}$, $v_{2, 00}$, $a_{02}$, $a_{22}$, and $r_{22}$ are the six fit parameters.

N$^3$LO should produce $a_{00}$, $r_{00}$, $v_{2,00}$, $v_{3, 00}$, $a_{02}$, $r_{02}$, $v_{2, 02}$, $a_{22}$, and $r_{22}$. The minimal d-wave potential, $V_2^0$, has one auxiliary degree of freedom, but two auxiliary degrees of freedom are needed to fit both the $\mathcal{O}(k^2)$ and $\mathcal{O}(k^4)$ behavior of $k^3M_{02}$. Consequently, it is necessary to introduce an additional auxiliary degree of freedom in, say, the s-wave part of the potential by increasing the number of poles, and the number of shape parameters, by one. The form factors used are 

\begin{align}
    \mathcal{G}_0(k) = \frac{1}{\sqrt{\gamma_1^2+k^2}\sqrt{\gamma_2^2+k^2}\sqrt{\gamma_3^2+k^2}\sqrt{\gamma_4^2+k^2}}\ , \ \ \ \mathcal{G}_2(k) = \frac{k^2}{\sqrt{\beta_1^2+k^2}\sqrt{\beta_2^2+k^2}}\ .
\end{align}
There are a total of nine degrees of freedom, all of which are left free. The M-matrix elements fit to data are

\begin{align}
    kM_{00} &= -\left[\hat{\mathbf{a}}^{-1}\right]_{00}+\frac{1}{2}r_{00}k^2+v_{2, 00}k^4+v_{3, 00}k^6+v_{4, 00}k^8\ ,\\
    k^5M_{22} &= -\left[\hat{\mathbf{a}}^{-1}\right]_{22}+\frac{1}{2}r_{22}k^2+v_{2, 22}k^4\ ,\\
    k^3M_{20} &= -\left[\hat{\mathbf{a}}^{-1}\right]_{20}\sqrt{1+\frac{k^2}{\gamma_1^2}}\sqrt{1+\frac{k^2}{\gamma_2^2}}\sqrt{1+\frac{k^2}{\gamma_3^2}}\sqrt{1+\frac{k^2}{\gamma_4^2}}\sqrt{1+\frac{k^2}{\beta_1^2}}\sqrt{1+\frac{k^2}{\beta_2^2}}\ ,
\end{align}
where $a_{00}$, $r_{00}$, $v_{2, 00}$, $v_{3, 00}$, $v_{4, 00}$, $a_{02}$, $a_{22}$, $r_{22}$, and $v_{2, 22}$ are the nine fit parameters. Here $v_{4, 00}$ and $v_{2, 22}$ are the auxiliary degrees of freedom used to constrain the $\mathcal{O}(k^2)$ and $\mathcal{O}(k^4)$ behavior of $k^3M_{02}$.

\subsubsection{\texorpdfstring{$^3P_2$-$^3F_2$}{3P2-3F2}}

This channel proceeds very similarly to the $^3S_1$-$^3D_1$ channel. At the first nonvanishing order, NLO, only $a_{11}$ should be produced; therefore, the f-wave form factor can be freely set to 0, and the minimal p-wave form factor from the $V_1^0$ potential is used. Since the minimal p-wave potential has two free parameters, the coupling and the pole, the auxiliary degree of freedom is constrained by requiring $r_{11}=0$. In summary, the form factors are 

\begin{align}
    \mathcal{G}_1(k) = \frac{k}{\sqrt{k^2+\gamma_1^2}}\ ,\ \ \ \mathcal{G}_3=0\ ,
\end{align}
and the M-matrix fit to the data is

\begin{align}
    k^3M_{11} = -\frac{1}{a_{11}}\ ,\ \ \ M_{13}=0 \ , \ \ \ |M_{33}|\to\infty\ .
\end{align}

NNLO should produce $a_{11}$, $r_{11}$, and $a_{13}$. The p-wave M-matrix can be reproduced again with $V_1^0$. Since there is nonzero mixing, the minimal f-wave form factor is used, and the auxiliary degrees of freedom introduced are constrained by requiring $r_{33}=v_{2,33}=v_{3,33}=0$ and $a_{33}=-1$ fm$^7$, which is on the order of the physical value. In summary, the form factors are 

\begin{align}
    \mathcal{G}_1(k) = \frac{k}{\sqrt{k^2+\gamma_1^2}}\ ,\ \ \ \mathcal{G}_3=\frac{k^3}{\sqrt{k^2+\beta_1^2}\sqrt{k^2+\beta_2^2}\sqrt{k^2+\beta_3^2}}\ ,
\end{align}
and the M-matrix is 

\begin{align}
    k^3M_{11} &= -\left[\hat{\mathbf{a}}^{-1}\right]_{11}+\frac{1}{2}r_{11}k^2\ ,\ \ \ k^7M_{33}=-\left[\hat{\mathbf{a}}^{-1}\right]_{33}\ ,\\
    k^5M_{13} &= -\left[\hat{\mathbf{a}}^{-1}\right]_{13}\sqrt{1+\frac{k^2}{\gamma_1^2}}\sqrt{1+\frac{k^2}{\beta_1^2}}\sqrt{1+\frac{k^2}{\beta_2^2}}\sqrt{1+\frac{k^2}{\beta_3^2}}\ .
\end{align}
The three free parameters, $a_{11}$, $r_{11}$, and $a_{13}$, are fit to data.

N$^3$LO should produce $a_{11}$, $r_{11}$, $v_{2,11}$, $a_{13}$, $r_{13}$, and $a_{33}$. In analogy to the treatment of $^3S_1$-$^3D_1$ at NNLO, the auxiliary effective-range parameter $r_{33}$ is left free so that the fit can reproduce the $\mathcal{O}(k^2)$ behavior of $k^5M_{13}$ while the other auxiliary effective-range parameters are set to zero. The form factors are

\begin{align}
    \mathcal{G}_1(k) = \frac{k}{\sqrt{k^2+\gamma_1^2}\sqrt{k^2+\gamma_2^2}}\ ,\ \ \ \mathcal{G}_3=\frac{k^3}{\sqrt{k^2+\beta_1^2}\sqrt{k^2+\beta_2^2}\sqrt{k^2+\beta_3^2}}\ ,
\end{align}
and the M-matrix is 

\begin{align}
    k^3M_{11} &= -\left[\hat{\mathbf{a}}^{-1}\right]_{11}+\frac{1}{2}r_{11}k^2+v_{2,11}k^4\ ,\ \ \ k^7M_{33}=-\left[\hat{\mathbf{a}}^{-1}\right]_{33}+\frac{1}{2}r_{33}k^2\ ,\\
    k^5M_{13} &= -\left[\hat{\mathbf{a}}^{-1}\right]_{13}\sqrt{1+\frac{k^2}{\gamma_1^2}}\sqrt{1+\frac{k^2}{\gamma_2^2}}\sqrt{1+\frac{k^2}{\beta_1^2}}\sqrt{1+\frac{k^2}{\beta_2^2}}\sqrt{1+\frac{k^2}{\beta_3^2}}\ .
\end{align}
The six free parameters fit to data are $a_{11}$, $r_{11}$, $v_{2,11}$, $a_{13}$, $a_{33}$, and the auxiliary parameter $r_{33}$.

A discussion of cases where the best-fit on-diagonal ERE has no corresponding potential in EPT has been omitted from these examples; however, the solution is the same as in the single-channel case. In the problematic part of the M-matrix, increase the number of poles by one and fix the new, highest-order shape parameter to be small or naturally sized with the appropriate sign to cure the sickness. This prescription is used, where necessary, in the fits to phase-shift data described below.

\section{Nucleon-nucleon phase shifts in \texorpdfstring{$\nopi$}{pionless EFT}}
\label{sec:NNcomptodata}
\noindent
In this section, the EPT potentials are fit to empirical phase shifts order by order in the power counting described above. Results are presented after the fitting procedure and uncertainty estimates are discussed.

\subsection{Fitting the ERE to phase-shift data}

\subsubsection{Single-channel fits}

\noindent The ERE parameters and, by extension, the corresponding S-matrix pole positions are determined by fitting the ERE truncated at order $m$ to the Nijmegen partial-wave analysis (PWA93)~\cite{NNonline}. At sufficiently large $k$, phase shifts receive significant contributions from the first omitted effective-range parameter, whose contributions scale as $\sim k^{m'}$ with $m'>m$; thus, only Nijmegen analysis data with $k\leq k_{\max}$ are included in the fit. To quantify the dependence of the best-fit parameters on the choice of $k_{\max}$, fits are performed for 
\begin{align}
    k_{\max} \in [k_{\max,1}, k_{\max,2}]\ ,
\end{align} and the best-fit parameters are averaged. $k_{\max,1}$ is chosen as the first point where order-$m$ contributions become significant, while order-$m'$ contributions are negligible. $k_{\max, 2}$ is chosen as the last point where order-$m'$ contributions are small. 

Since the Nijmegen analysis does not provide uncertainties, the least-squares regression is weighted using only theoretical uncertainty bars associated with truncating the ERE. The leading omitted contribution is expected to scale as 

\begin{align}
    \sigma_{k^{2\ell+1}\cot\delta_\ell} = c_\ell \mathcal{M}^{2\ell+1}\left(\frac{k}{\mathcal{M}}\right)^{m'}\ ,
\end{align}
where $\mathcal{M}\sim1$ fm$^{-1}$ and $c_\ell\sim\mathcal{O}(1)$. Linear error propagation then gives
\begin{align}
    \sigma_{\delta_\ell} = c_\ell\sin^2\delta_\ell\left(\frac{k}{\mathcal{M}}\right)^{m'-2\ell-1}\ . 
\end{align}
However, below $k_{\max, 1}$, $\sigma_{\delta_{\ell}}$ is taken to be constant since order-$m'$ contributions are expected to be parametrically suppressed. It is important to put a lower bound on $\sigma_{\delta_{\ell}}$, else $\sigma\to0$ as $k\to 0$, which would overweight small deviations in low-energy data points that should otherwise be drowned out by the unquantified uncertainty coming from the Nijmegen analysis.

\subsubsection{Coupled-channel fits}

\noindent The coupled-channel fits require special care. Since omitted effective-range parameter effects become significant at different momentum scales for different observables, each phase shift and mixing angle is assigned its own fitting window. The observables are first fit with the data truncated at their respective $k_{\max,1}$, after which the cutoffs are increased simultaneously up to the corresponding $k_{\max,2}$.

The characteristic size of effective-range parameters is not universal across all elements of the M-matrix. For example, in the $^3S_1$-$^3D_1$ channel, the subleading ERE parameters of $M_{00}$ and $M_{02}$ are typically $\lesssim\mathcal{O}(1)$, but $r_{22}$ and $v_{2,22}$ of $M_{22}$ are numerically larger. The $\mathcal{O}(1)$ coefficients $c_{\ell}$, which set the characteristic size of the leading omitted contributions, are chosen accordingly. In practice, $c_\ell=$ 1, 1, 3, 5, 10, and 20 for $M_{00}$, $M_{02}$, $M_{22}$, $M_{11}$, $M_{13}$, and $M_{33}$, respectively. Below $k_{\max,1}$, however, retaining the factor of $c_\ell$ in the lower bound on the uncertainty would unduly suppress the contribution of low-energy data in sectors with large $c_\ell$. The uncertainties are instead bounded from below by
$\sin^2\delta_\ell(k_{\max,1})\left(k_{\max,1}/\mathcal{M}\right)^{m'-2\ell-1}$ where the corresponding factor of $c_\ell$ has been divided out. 

Uncertainties are propagated using linear error propagation. In the limit that the channel decouples, $\epsilon_J\to0$, the Jacobian of the transformation relating M-matrix elements $M_{J-1,J-1}$, $M_{J-1,J+1}$, and $M_{J+1,J+1}$ to the scattering data $\delta_{J-1}$, $\epsilon_J$, and $\delta_{J+1}$ is a diagonal matrix with entries $-\sin^2\delta_{J-1}$, $-\sin\delta_{J-1}\sin\delta_{J+1}$, and $-\sin^2\delta_{J+1}$. In this limit, the propagated uncertainties are

\begin{align}
    \sigma_{\delta_{\ell}} = c_\ell\sin^2\delta_\ell\left(\frac{k}{\mathcal{M}}\right)^{m'-2\ell-1}\ ,
\end{align}
and
\begin{align}
    \sigma_{\epsilon_J} = c_J\left|\sin\delta_{J-1}\sin\delta_{J+1}\right|\left(\frac{k}{\mathcal{M}}\right)^{m'-2J-1}\ ,
\end{align}
for leading omitted contributions at $\mathcal{O}(k^{m'})$ in the ERE for $M_{\ell,\ell}$ and $M_{J+1,J-1}$, respectively. This approximation is used throughout the coupled-channel analysis. Corrections away from the decoupling limit are expected to modify the uncertainties only by $\mathcal{O}(1)$ multiplicative factors.

The deuteron binding momentum $\gamma_d = 45.7022$ MeV~\cite{Mohr:2024kco} is also included as a data point in the $^3S_1$-$^3D_1$ channel. The theoretical uncertainty on $\gamma_d$ is estimated by examining the shift in the condition for a bound state due to the leading omitted contribution. At LO, the condition for a bound state is

\begin{align}
    0 = -\frac{1}{a_0}+\gamma_d\ .
\end{align}

\noindent In the decoupling limit, the condition at NLO is

\begin{align}
    0 = -\frac{1}{a_0}+(\gamma_d+\delta\gamma_d)-\frac{1}{2}r_0(\gamma_d+\delta\gamma_d)^2\ .
\end{align}
Expanding to leading order in $\delta\gamma_d$ gives

\begin{align}
    \left|\delta\gamma_d\right| \approx \frac{1}{2} \left|\frac{r_0\gamma_d^2}{1-r_0\gamma_d}\right|\approx 10\text{ MeV}\ ,
\end{align}
where $r_0$ is taken to be $\approx 2$ fm and the LO value $\gamma_d = 1/a_0 \approx 36$ MeV is used. This value is used as the theoretical uncertainty in $\gamma_d$ at LO.

At NLO, the uncertainty is estimated by adding the shape parameter $v_{2,0}$ to the ERE. The bound state shifts by
\begin{align}
    \left|\delta\gamma_d\right| \approx\left|\frac{v_{2,0}\gamma_d^4}{1-r_0\gamma_d}\right|\approx 1 \text{ MeV}\ .
\end{align}
Here $v_{2,0}\gamma_d^4$ is treated as $\mathcal{O}(\delta\gamma_d)$. For natural $v_{2,0}$, this term is suppressed relative to $r_0\gamma_d^2$ by two powers of the dimensionless ratio $\gamma_d/\mathcal{M}$. $\gamma_d$ is taken to be close to its physical value $\approx 45.7$ MeV. This procedure generalizes nicely to higher orders.

Gaussian priors centered at zero with width $c_{\ell}$ fm$^p$ are imposed on effective-range parameters beyond $a_{\ell,\ell'}$ to prevent the fit from wandering to unnaturally large values in poorly constrained directions of parameter space. Priors are unnecessary in the single-channel case since the effective-range parameters are more directly constrained by the data.

Because $M_{J+1,J-1}$ depends nonlinearly on the free effective-range parameters, averaging the best-fit parameters significantly worsens the quality of the fit. Therefore, the quoted best-fit parameters
are taken from the set of best-fit parameters closest to the average. Uncertainties are computed relative to the quoted best-fit parameters.

\subsection{Phase-shift uncertainty bands}

\subsubsection{Single-channel bands}

\noindent At each order, the uncertainty on the predicted phase shift is estimated from the variation in the sets of best-fit effective-range parameters together with contributions from the first few omitted effective-range parameters. Complete sets of effective-range parameters through $\mathcal{O}(k^m)$ are sampled from the ensemble of fits obtained by varying $k_{\max}$. Omitted effective-range parameters are sampled from uncorrelated Gaussian distributions centered at 0 with natural-sized widths. The default width is taken to be 1 fm$^p$; however, if a parameter first appearing at order $\mathcal{O}(k^n)$ is found to be numerically large when fit explicitly, then at lower orders, where that parameter is omitted, it is sampled using a width consistent with its fitted size. Furthermore, if effective-range parameters in a given channel exhibit a characteristic size larger than $\mathcal{O}(1)$, the same scaling is assumed for higher-order effective-range parameters when assigning them a width.

This ensemble of effective-range parameters is used to compute an ensemble of phase shifts. The lower and upper edges of the uncertainty band are defined by the 2.5th and 97.5th percentiles of the phase shift at each center-of-mass momentum $k$.

\subsubsection{Coupled-channel bands}

\noindent Again, a complete set of effective-range parameters is sampled from the ensemble of best fits. However, independently varying all omitted shape parameters is likely to overestimate the uncertainty in the phase shifts because it incorrectly treats ERE parameters from different M-matrix elements as uncorrelated. Furthermore, inserting subleading ERE parameters in one M-matrix element can contaminate the lower-order behavior of another phase shift or mixing angle. For example, the $\mathcal{O}(k^3)$ behavior of $\delta_0$ depends on $r_{22}$, which does not enter until $\mathcal{O}(k^7)$.

Estimating uncertainties is more transparent in the decoupling limit. Here, the $\mathcal{O}(k^{2n+1})$ uncertainty in $\delta_0$ is estimated by inserting an effective-range parameter at $\mathcal{O}(k^{2n})$ in $M_{00}$. Even when the channels are coupled, varying $M_{00}$ in this way still estimates the $\mathcal{O}(k^{2n+1})$ uncertainty in $\delta_0$ without over-counting uncertainties. Accordingly, when constructing the uncertainty band for a given phase shift or mixing angle, omitted effective-range parameters are inserted in only the corresponding M-matrix element: $M_{00}$ for $\delta_0$, $M_{02}$ for $\epsilon_1$, $M_{22}$ for $\delta_2$, etc.

As mentioned above, at LO in the $^3S_1$-$^3D_1$ channel, the mixing is set to zero. Since mixing first appears at NLO, its leading contribution must be included when estimating the truncation uncertainty at LO. Consequently, values of $a_{02}$ are sampled from a Gaussian centered on 0 with width $1$ fm$^3$. Nonzero mixing requires specifying a d-wave form factor. As discussed above, one convenient choice is to constrain the auxiliary degrees of freedom of the d-wave potential such that $a_{22}=6.5$ fm$^5$ is near the physical value and $r_{22}=v_{2,22}=0$. Accordingly, for the LO and NLO uncertainty bands, values of $a_{22}$ are sampled from a Gaussian centered at $6.5$ fm$^5$ with width $\sim 0.5$ fm$^{5}$. At NNLO, a nontrivial ERE is specified for all M-matrix elements, and the uncertainty determination proceeds straightforwardly. The $^3P_2$-$^3F_2$ channel is treated similarly.

\subsection{Results}

\begin{figure*}
\begin{center}
  \includegraphics[width=\textwidth,height=\textheight,keepaspectratio]{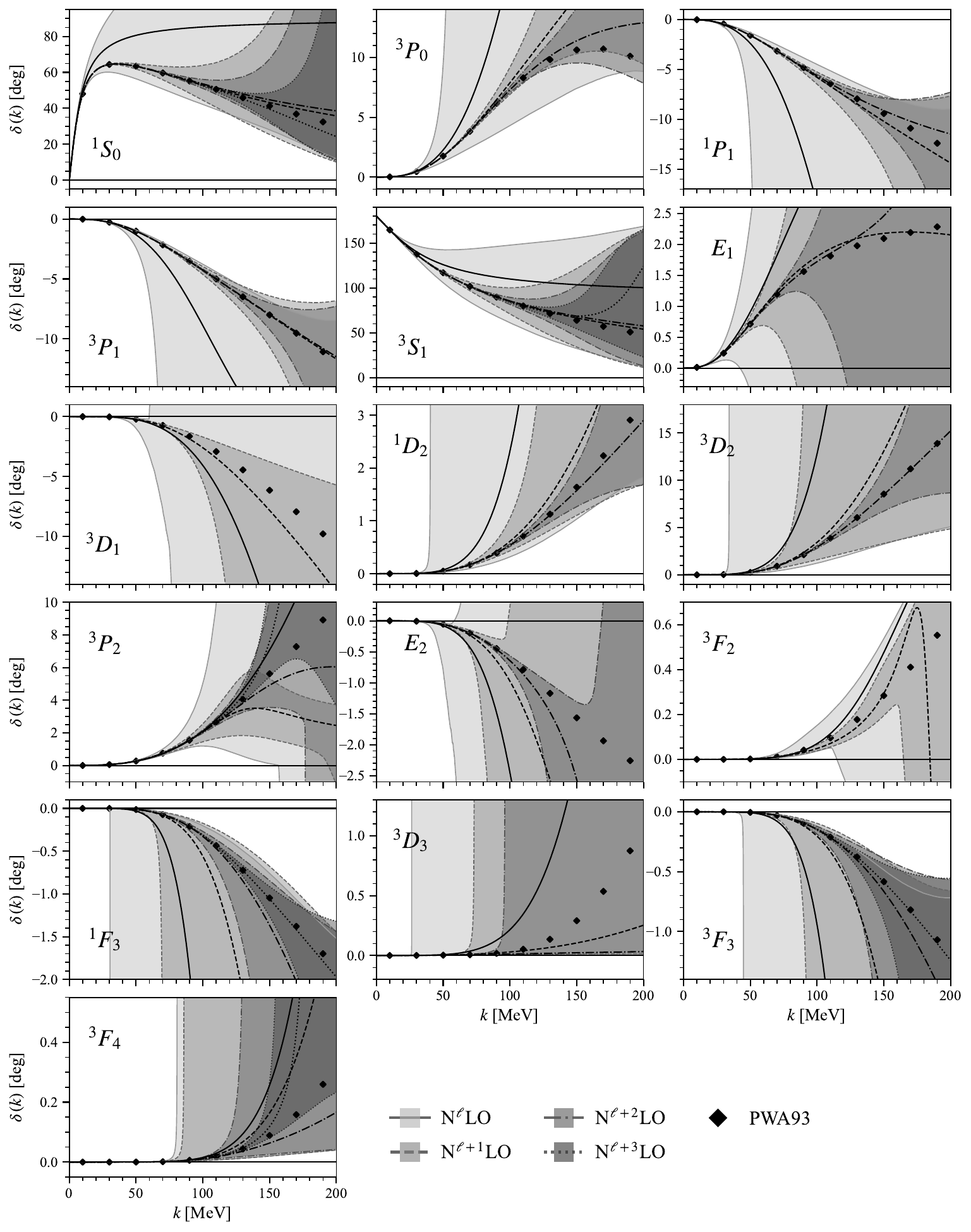}
  \caption{\small Phase shifts for all of the 
  EPT potentials fit in $\nopi$ and the corresponding uncertainty bands,
  as functions of the relative momentum.
  \label{fig:PiSlashPhaseShifts}}
\end{center}
\end{figure*}

\noindent The best-fit phase shifts in all channels through the f-waves are plotted as functions of the relative momentum in Fig.~\ref{fig:PiSlashPhaseShifts} alongside their uncertainty bands. The first three nontrivial orders are plotted in each partial wave, with a fourth included in the $^1S_0$ and mixed channels, where the convergence of the ERE is of particular interest, and in the uncoupled f-wave channels to exhaust the parametric freedom of the minimal potential. Since higher partial waves are neglected, the $^3D_3$ and $^3F_4$ channels are treated as uncoupled. A full treatment would require including mixing at N$^3$LO and N$^4$LO, respectively. 

\begin{figure*}
\begin{center}
  \includegraphics[width=\textwidth,height=\textheight,keepaspectratio]{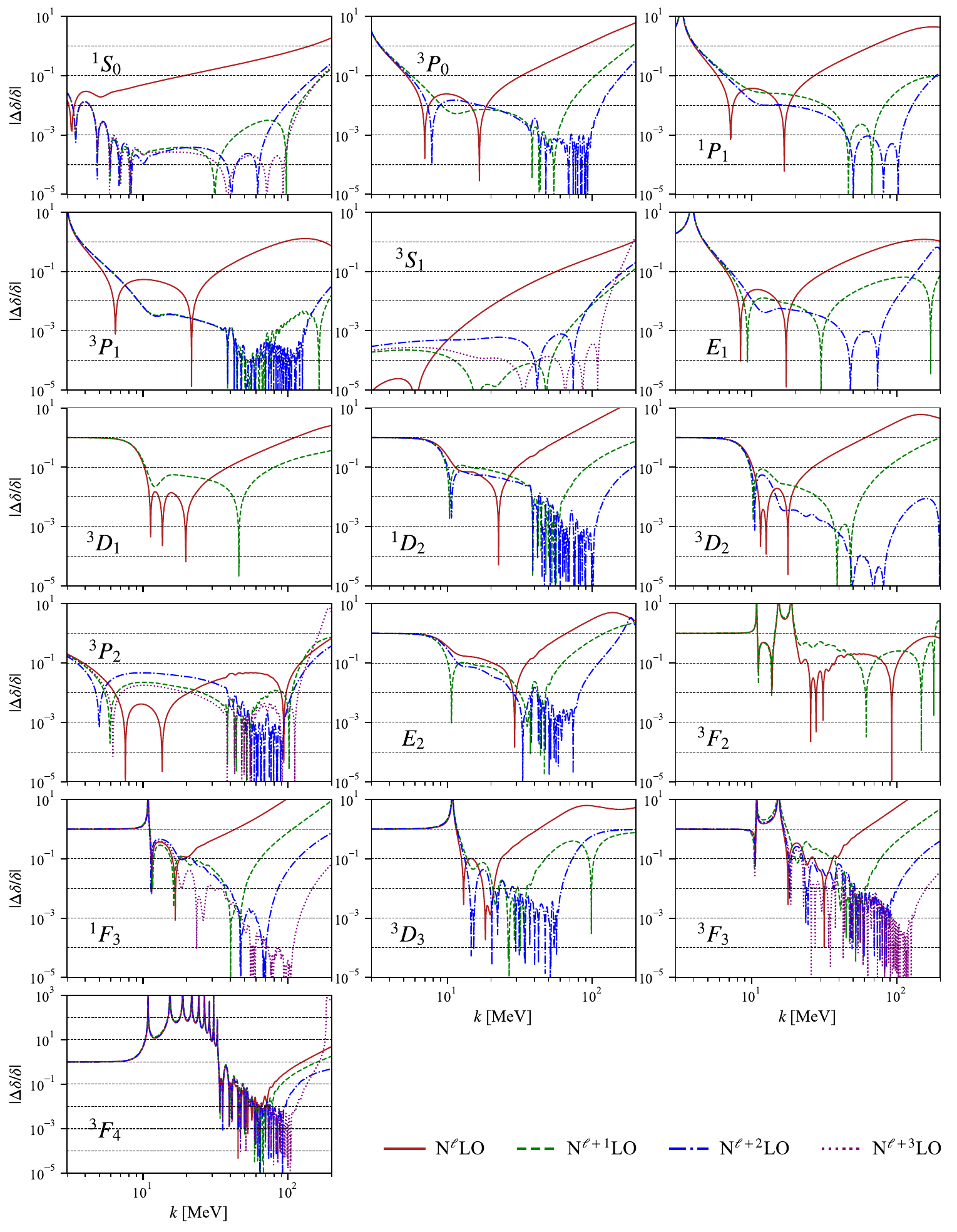}
  \caption{\small 
  Fractional errors in the phase shifts from the 
  EPT potentials fit in $\nopi$
  relative to the Nijmegen PWA, as functions of the relative momentum. The phase shifts themselves are plotted in Fig.~\ref{fig:PiSlashPhaseShifts}.
  \label{fig:PiSlashLepage}}
\end{center}
\end{figure*}

The nested, shrinking bands reflect the hierarchy of ERE parameters. The central values generally approach the empirical phase shifts as the order increases. The fractional errors in the phase shifts~\cite{Lepage:1997cs}, plotted in Fig.~\ref{fig:PiSlashLepage}, provide a more quantitative measure of order-by-order convergence. 
Most channels exhibit order-by-order convergence at momenta comparable to the pion mass. Convergence is particularly clear in the $^3P_0$, $^1P_1$, $^3D_1$, $^1D_2$, $^3D_2$, $^1F_3$, and $^3F_3$ channels. At low momenta, the fractional errors exhibit erratic behavior as the predicted phase shifts repeatedly cross the interpolated Nijmegen PWA. Some higher partial waves also exhibit spikes in the fractional error at low energies where fluctuations in the empirical phase shift bring it very near zero. Neither effect precludes order-by-order convergence at higher momenta.

Some channels exhibit less uniform convergence. For the $^1S_0$ channel, the LO, NLO, and NNLO curves are, respectively, those labeled ${\cal O}(k^0)$, ${\cal O}(k^2)$, and ${\cal O}(k^4)$ in the bottom-right panel of Fig.~\ref{fig:Propaganda}. While NNLO has a larger error than NLO, the N$^3$LO error is smaller than the NLO error. Similar behavior is observed in $^3S_1$. In $E_1$, the convergence pattern is obscured only by an accidental crossing of the NNLO and empirical mixing angle close to 200 MeV. The mixed channels exhibit less systematic order-by-order improvement at higher orders; however, this behavior is most likely caused by the difficulty of performing fits with nearly flat directions in parameter space and highly correlated parameters, both of which become increasingly problematic as the number of parameters grows.

The values of the effective-range parameters tend to be stable from order to order, except in the higher waves where the phase shifts are smaller and the fits are correspondingly less constrained. The best-fit 
parameters are listed in Table~\ref{tab:ERE_EFT} of Appendix~\ref{sec:parPiSlash}. The quoted uncertainties are the standard deviations of the best-fit parameters obtained by varying the fitting window. They do not include uncertainties in the Nijmegen analysis itself or uncertainties associated with the choice to model the data by a truncated ERE. 
As a validation of the extraction procedure, the highest-order EPT effective-range parameters are compared to those extracted~\cite{PavonValderrama:2005ku} for the Nijmegen group's NijmII potential~\cite{Stoks:1994wp}. The first two effective-range parameters in almost all channels agree to at least one significant digit, and often more. Progressively poorer agreement is observed in higher-order effective-range parameters. This behavior is consistent with the strong correlations among effective-range parameters, so small differences in the values of the first few parameters can be compensated for by much larger differences in higher-order ones. Consequently, two EREs that are approximately phase equivalent can have substantially different higher-order effective-range parameters.

The degree of agreement varies considerably by channel. The strongest agreement is in the $^1S_0$ channel where $a_0$ and $r_0$ agree to four and three digits, respectively; however, agreement worsens for higher-order parameters, with $v_{2,0}$ being off by a factor of two. In the $^3D_3$ and $^3F_4$ channels, where only $a_\ell$ can be compared, $a_2$ agrees to one digit, while $a_3$ does not. In the $^3S_1$-$^3D_1$ channel, $a_{00}$, $r_{00}$, $a_{02}$, $a_{22}$, and $r_{22}$ agree to three, two, two, one, and one digit, respectively. The $^3P_2$-$^3F_2$ channel exhibits poorer agreement, with only the scattering length parameters $a_{\ell\ell'}$ agreeing to at least one digit.

An additional comparison can be made in the $^3S_1$ channel using the coefficients in the power series expansion of $k\cot\delta_{^3S_1}$, which were also extracted by the Nijmegen group for various potential models~\cite{deSwart:1995ui}. The values obtained in EPT for these coefficients at N$^3$LO are $a_0= 5.4110(18)$ fm, $r_0=1.7714(5)$ fm, $v_2=0.01(2)$ fm$^3$, and $v_3=-1.20(12)$ fm$^5$. The first two values agree to three and two significant digits, respectively, with the Nijmegen values. The latter two values agree in order of magnitude, but $v_3$ has the opposite sign, consistent with the expected deterioration in agreement for higher-order effective-range parameters discussed above.

The potentials in Eqs. \eqref{eq:LOellwaveb} and \eqref{eq:NpLOellwaveb} are given in terms of the pole positions $\{\gamma\}$.
Table~\ref{tab:polesEFT} of Appendix~\ref{sec:parPiSlash} lists the $\{\gamma\}$ corresponding to the best-fit parameters.

\begin{figure*}
\begin{center}
  \includegraphics[width=\textwidth,height=\textheight,keepaspectratio]{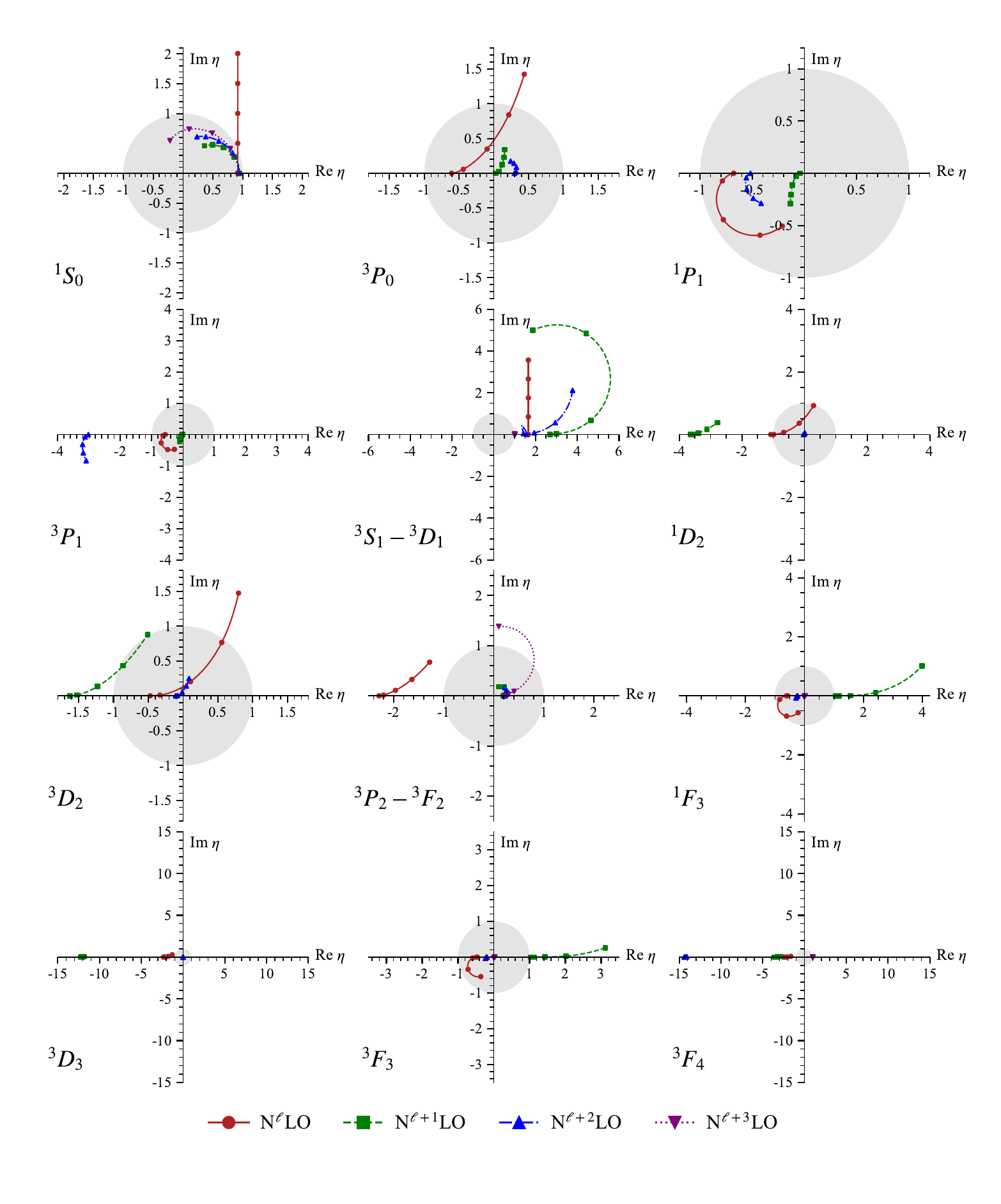}
  \caption{\small Weinberg eigenvalues for 
  EPT potentials fit in $\nopi$. These eigenvalues correspond to the phase shifts in Fig.~\ref{fig:PiSlashPhaseShifts}. For the mixed channels, which have two Weinberg eigenvalues at each momentum, the eigenvalue with larger absolute value is plotted. The points denote the Weinberg eigenvalues at $k=0, 50, 100, 150, 200$ MeV. 
  \label{fig:PiSlashWeinbergEvals}}
\end{center}
\end{figure*}

Weinberg eigenvalues for the best-fit potentials are plotted in Fig.~\ref{fig:PiSlashWeinbergEvals}. In many channels, the potential is perturbative at one or more orders in EPT. In channels with nonperturbative potentials, the magnitude of the largest Weinberg eigenvalue generally decreases as the order increases and more phase-shift data are incorporated; however, this trend is not universal. This observation suggests that provided EPT potentials incorporate sufficient amounts of low-energy scattering data, one can eventually obtain a perturbative potential that is suitable for many-body applications.

\begin{figure*}
\begin{center}
  \includegraphics[width=0.43\textwidth,height=\textheight,keepaspectratio]{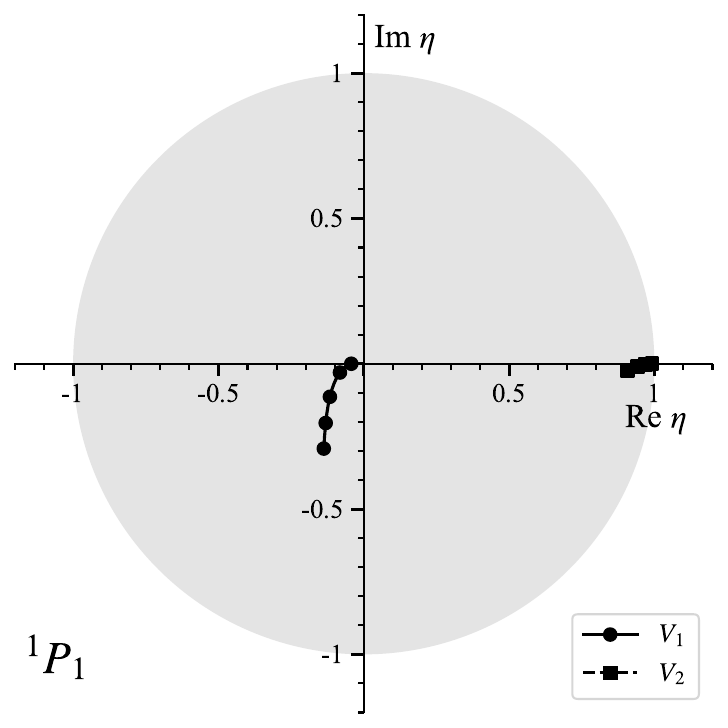}
  \includegraphics[width=0.43\textwidth,height=\textheight,keepaspectratio]{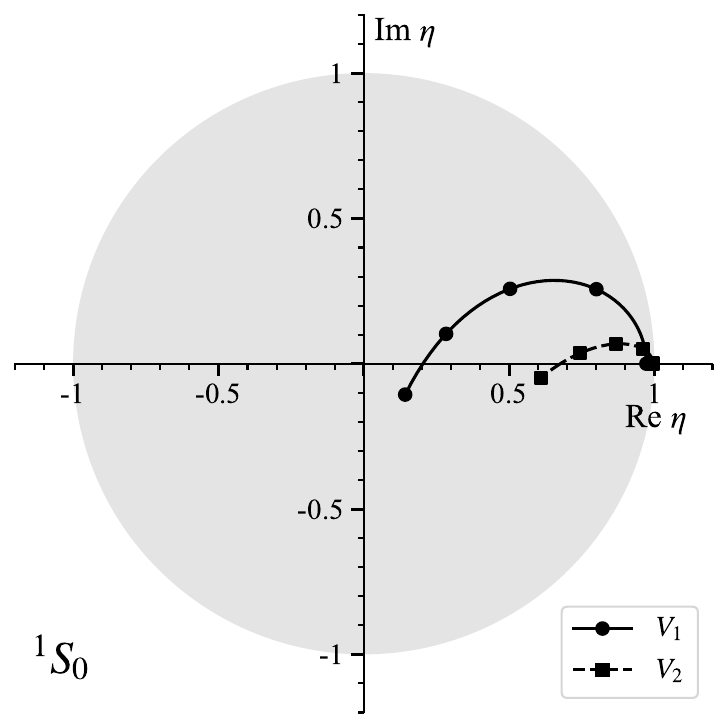}
  \caption{\small The Weinberg eigenvalues for two phase-equivalent potentials in the $^1P_1$ channel at NNLO are plotted (left). Two phase-equivalent potentials are also obtained when fitting the $V_{z,0}^0$ potential (see Section~\ref{sec:NNtopiprod}) to the $^1S_0$ 
  phase shift.
  The Weinberg eigenvalues of these potentials are also plotted (right). In both cases, the potential labeled $V_1$ has Weinberg eigenvalues with smaller absolute value, and so $V_1$ is chosen.
  The points in the left and right panels denote the Weinberg eigenvalues at $k=0, 50, 100, 150, 200$ MeV and $k=0, 100, 200, 300, 405$ MeV, respectively.
  \label{fig:phaseEquiv}}
\end{center}
\end{figure*}

When applicable, the RG scale is set to $\mu/2=\gamma_d=45.7$ MeV. At some orders, multiple phase-equivalent potentials exist because a spurious bound state pole can be exchanged with a real-valued redundant pole. While these potentials produce the same phase shifts, they have different Weinberg eigenvalues. In these cases, the potential that minimizes the magnitude of the largest Weinberg eigenvalue is chosen. Figure~\ref{fig:phaseEquiv} illustrates this choice in two cases: in the $^1P_1$ channel at NNLO and in the $^1S_0$ channel using the $V_{z,0}^0$ potential from Section~\ref{sec:NNtopiprod}. In both cases, the potential labeled $V_1$ is chosen. 

The asymptotic D/S ratio is predicted using the $^3S_1$-$^3D_1$ channel potentials through N$^3$LO as an additional test of EPT in the mixed channel. These values, along with the deuteron binding momenta, are listed in Table~\ref{tab:Deuteron}. The computed values get closer to the experimental value with increasing order, but do not exhibit clear convergence. This trend is consistent with the deterioration of the higher-order fits in the mixed channels.

\begin{table}[bt]
\caption{Deuteron parameters computed using the LO, NLO, NNLO, and N$^3$LO EPT potentials fit in $\nopi$ compared to their experimental values. The experimental value for $\gamma_d$ comes from Ref.~\cite{Mohr:2024kco} and the value for the asymptotic D/S ratio, $\xi$, comes from Ref.~\cite{PhysRevC.41.898}. \label{tab:Deuteron}\\}
\begin{minipage}{\linewidth}
\begin{center}
\begin{tabular}{cccc} \hline \hline
Order & $\sigma_{\gamma_d, \rm theory}$ (MeV) & $\gamma_{d}$ (MeV) & $\xi$\\
\hline\hline
Exp. & \text{—} & 45.7022 & 0.0256 \\
LO & 10 & 36.120(3) & 0 \\
NLO & 1 & 45.77(3) & 0.01672(9) \\
NNLO & 0.05 & 45.57(4) & 0.0245(3) \\
N$^3$LO & 0.003 & 45.70241(10) & 0.0266(2) \\
\hline\hline
\end{tabular}
\end{center}
\end{minipage}
\end{table}

\section{Nucleon-nucleon phase shifts beyond \texorpdfstring{$\nopi$}{pionless EFT}}
\label{sec:NNtopiprod}

\noindent As larger momenta are probed, and formally at $k\sim
M_\pi/2$, the pion branch-cut singularities should be included in an
EFT description with pions as fundamental degrees of
freedom. However, 
consistently maintaining regulator independence at
each order in the expansion requires a complicated combination of nonperturbative interactions and perturbation theory~\cite{vanKolck:2020llt}.  
It is therefore
of interest to push the EFT of contact interactions as far as it will
go. In this case, while regulator independence is manifest, the pion
branch-cut singularities are absent except insofar as they are
approximated by the S-matrix poles at $|k|\geq M_\pi/2$. 

In this section, it is shown how the EPT potential can be extended to the threshold for pion production, that is, momenta $|k|\sim \sqrt{M M_\pi}$, once zeros of the amplitude are included. The 
$^1S_0$, $^3S_1$ and $^3P_0$ phase shifts change sign for momenta in the range $\sim300$--$400~{\rm MeV}$. The zero in the $^1S_0$ channel is accommodated~\cite{SanchezSanchez:2017tws} by
re-summing the effective-range expansion so that it acquires a pole, and the S-matrix element has a second
trivial fixed point ($S=1$) at $k=k_*$. The zero of the $^3S_1$ channel can be reproduced similarly when the mixing with $^3D_1$ is neglected\footnote{A zero of the amplitude can be related~\cite{Krivoruchenko:2010ft} to the CDD singularities in the $N/D$ method. Separable potentials for coupled channels have been recently discussed in connection with the $N/D$ method in Ref.~\cite{Sanchez:2024xzl}.}. The procedure can be generalized to the (uncoupled) $^3P_0$ channel.

\subsection{Designer potentials and re-summed effective-range theory}

In an uncoupled s-wave, the minimal 
form of the re-summed ERE that admits a renormalizable potential is

\begin{equation}
  k \cot\delta_0(k)\ =\  -\frac{1}{a_0} + \frac{1}{2} r_0 {k^2}\frac{{k_*^2}}{{k_*^2}-{k^2}} \ .
\label{eq:eregenzero1}
\end{equation}
Noting that there are three poles in the rational S-matrix element, one finds
\begin{eqnarray}
  k_*^2 &=&  {\bf {\large e}}_2^{(3)}    \ \ , \ \   a_0 \ =\ \frac{{\bf {\large e}}_2^{(3)}}{{\bf {\large e}}_3^{(3)}}   \ \ , \ \
  r_0 \ =\ \frac{2}{{\bf {\large e}}_2^{(3)}}\left({\bf \large e}_1^{(3)}-\frac{1}{a_0}\right) \ .
\label{eq:ereparszero1}
\end{eqnarray}
A potential, $V_{z,0}^{0}$, which gives rise to this S-matrix element is
\begin{equation}
  V_{z,0}^{0}\left(p,p'\right)\ = \ \frac{4\pi}{M} \frac{1}{ -{\bf {\large e}}_{1}^{(3)}+\mu} \sqrt{\frac{{k_*^2}-{p^2}}{{\gamma_1^2}+{p^2}}}\sqrt{\frac{{k_*^2}-{p^{\prime 2}}}{{\gamma_1^2}+{p^{\prime 2}}}} \ .
\label{eq:yamazero}
\end{equation}

It is straightforward to accommodate higher-order effective-range parameters while preserving the amplitude zero by adding more poles,
\begin{equation}
    V_{z,0}^{m}(p,p') = (-1)^{m+1}\,\frac{4\pi}{M} \, 
    {\bf \large e}_{2m-1}^{(2m+2)}\sqrt{k_*^2-p^2}\sqrt{k_*^2-p'^2}\prod_{j=1}^{m+1}\frac{1}{\sqrt{(\gamma_j^2+p^2)}}\frac{1}{\sqrt{(\gamma_j^2+p'^2)}}\ .
\end{equation}
For $m\geq1$, the potential is now regular at the origin. It generates the re-summed effective-range expansion
\begin{equation}
    k\cot\delta_0 = -\frac{1}{a_0}+\frac{1}{2}r_0k^2 + \sum_{j=2}^{m}v_{j,0}k^{2j}+v_{m+1,0}k^{2m+2}\frac{k_*^2}{k_*^2-k^2}\ ,
\end{equation}
where 
\begin{eqnarray}
  k_*^2 &=&  \frac{{\bf {\large e}}_{2m+1}^{(2m+2)}}{{\bf {\large e}}_{2m-1}^{(2m+2)}}  \ \  \ \ , \ \ \ \   a_0 \ =\ \frac{{\bf {\large e}}_{2m+1}^{(2m+2)}}{{\bf {\large e}}_{2m+2}^{(2m+2)}}   \ \ \ \ , \ \ \ \
  r_0 \ =\ \frac{2{\bf\large e}_{2m-1}^{(2m+2)}}{{\bf {\large e}}_{2m+1}^{(2m+2)}}\left(\frac{{\bf \large e}_{2m}^{(2m+2)}}{{\bf\large e}_{2m-1}^{(2m+2)}}-\frac{1}{a_0}\right) \ ,  \nonumber \\
  \qquad\qquad\qquad && v_{j,0} \ =\ \frac{{\bf\large e}_{2m-1}^{(2m+2)}}{{\bf {\large e}}_{2m+1}^{(2m+2)}}\left((-1)^{j+1}\frac{{\bf \large e}_{2m-2j+2}^{(2m+2)}}{{\bf\large e}_{2m-1}^{(2m+2)}}+v_{j-1,0}\right)
  \ ,
\label{eq:ereparszero2}
\end{eqnarray}
for $2\leq j\leq m+1$ with ${\bf\large e}_{k}^{(2m+2)}=0$ for $k=1, 3, 5, ..., 2m-3$.

The amplitude zero is also accommodated in 
p-wave channels. Consider the minimal ERE,
\begin{equation}
  k^3 \cot\delta_1(k)\ =\  -\frac{1}{a_1} + \frac{1}{2} r_1 k^2 + v_{2,1} {k^4}\frac{{k_*^2}}{{k_*^2}-{k^2}} \ ,
\label{eq:eregenzero1p}
\end{equation}
where 
\begin{eqnarray}
  k_*^2 &=&  {\bf {\large e}}_2^{(5)}  \ \  \ \ , \ \ \ \   a_1 \ =\ -\frac{{\bf {\large e}}_2^{(5)}}{{\bf {\large e}}_5^{(5)}}   \ \ \ \ , \ \ \ \
  r_1 \ =\ \frac{2}{{\bf {\large e}}_2^{(5)}}\left(-{\bf \large e}_3^{(5)}-\frac{1}{a_1}\right) \ ,  \nonumber \\
  \qquad\qquad\qquad && v_{2,1} \ =\ \frac{1}{{\bf\large e}_{2}^{(5)}}\left({\bf {\large e}}_1^{(5)}+\frac{1}{2}r_1\right)
\ \ \ \ , \ \ \ \  0 \ =\ {\bf {\large e}}_4^{(5)} \ .
\label{eq:ereparszero2p}
\end{eqnarray}
A potential that gives rise to this S-matrix element is
\begin{equation}
  V_{z, 1}^{0}\left(p,p'\right)\ = \ \frac{4\pi}{M} \frac{pp^{\prime}}{ -{\bf {\large e}}_{1}^{(5)}+\mu} \sqrt{\frac{{k_*^2}-{p^2}}{({\gamma_1^2}+{p^2})({\gamma_2^2}+{p^2})}} \sqrt{\frac{{k_*^2}-{p^{\prime 2}}}{({\gamma_1^2}+{p^{\prime 2}})({\gamma_2^2}+{p^{\prime 2}})}} \ .
\label{eq:yamazerop}
\end{equation}

Again, higher-order effective-range parameters can be included by inserting more poles in the potential.

\begin{equation}
    V_{z, 1}^{m}(p,p') = (-1)^{m+1}\,\frac{4\pi}{M}\,
    {\bf \large e}_{2m-1}^{(2m+4)}pp'\sqrt{k_*^2-p^2}\sqrt{k_*^2-p'^2}\prod_{j=1}^{m+2}\frac{1}{\sqrt{(\gamma_j^2+p^2)}}\frac{1}{\sqrt{(\gamma_j^2+p'^2)}}\ .
\end{equation}
For $m\geq 1$, this potential is regular at the origin and generates the re-summed effective-range expansion

\begin{equation}
    k^3\cot\delta_1 = -\frac{1}{a_1}+\frac{1}{2}r_1k^2 + \sum_{j=2}^{m+1}v_{j,1}k^{2j}+v_{m+2,1}k^{2m+4}\frac{k_*^2}{k_*^2-k^2} \ ,
\end{equation}
where 
\begin{eqnarray}
  k_*^2 &=&  \frac{{\bf {\large e}}_{2m+1}^{(2m+4)}}{{\bf {\large e}}_{2m-1}^{(2m+4)}}  \ \  \ \ , \ \ \ \   a_1 \ =\ -\frac{{\bf {\large e}}_{2m+1}^{(2m+4)}}{{\bf {\large e}}_{2m+4}^{(2m+4)}}   \ \ \ \ , \ \ \ \
  r_1 \ =\ \frac{2{\bf\large e}_{2m-1}^{(2m+4)}}{{\bf {\large e}}_{2m+1}^{(2m+4)}}\left(-\frac{{\bf \large e}_{2m+2}^{(2m+4)}}{{\bf\large e}_{2m-1}^{(2m+4)}}-\frac{1}{a_1}\right) \ ,  \nonumber \\
  \qquad\qquad\qquad && v_{j,1} \ =\ \frac{{\bf\large e}_{2m-1}^{(2m+4)}}{{\bf {\large e}}_{2m+1}^{(2m+4)}}\left((-1)^{j}\frac{{\bf \large e}_{2m-2j+4}^{(2m+4)}}{{\bf\large e}_{2m-1}^{(2m+4)}}+v_{j-1,1}\right)
  \ ,
\end{eqnarray}
for $2\leq j\leq m+2$ with ${\bf\large e}_{k}^{(2m+4)}=0$ for $k=1, 3, 5, ...\widehat{2m-1}, \widehat{2m+1}, 2m+3$.

\subsection{Fits to nucleon-nucleon phase shifts up to 405 MeV}
\noindent All phase shifts through the f-waves are fit to the ERE beyond the radius of convergence of $\nopi$. The $^1S_0$, $^3S_1$, and $^3P_0$ channels are fit using the re-summed ERE, while the remaining channels use the standard ERE. Mixing is neglected in the fits to the coupled-channel phase shifts. Uniform uncertainties are assigned to all phase-shift data points. Since the phase shift is nearly zero where the scattering length $a_\ell$ dominates, uniform uncertainties give these low-energy data insufficient weight to make a reliable determination of $a_\ell$. The scattering length is therefore held fixed at a value obtained by fitting the phase shift near threshold. Most often, this value is taken from the $\nopi$ fits. $k_*$ is determined by interpolating the phase-shift data to locate the zero crossing and is not treated as a free fit parameter.

Fits are first performed to the minimal potential, which corresponds to an ERE with $\ell+1$ terms, or $\ell+2$ independent terms with the amplitude zero included. The fit window is then enlarged, and more effective-range parameters are included until the predicted phase shift describes the data up to center-of-mass momentum $k=405$ MeV. The only exception is the $^3F_2$ channel, where the largest $k_{\max}$ used was $375$ MeV.

\subsection{Taming pathological asymptotic behavior of the phase shifts}

Together, the $^1S_0$ and $^1P_1$ cases demonstrate the two varieties of pathological asymptotic behavior encountered. All other channels can be handled similarly.

\subsubsection{\texorpdfstring{$^1S_0$}{1S0}}

\begin{figure*}
\begin{center}
  \includegraphics[width=0.43\textwidth,height=\textheight,keepaspectratio]{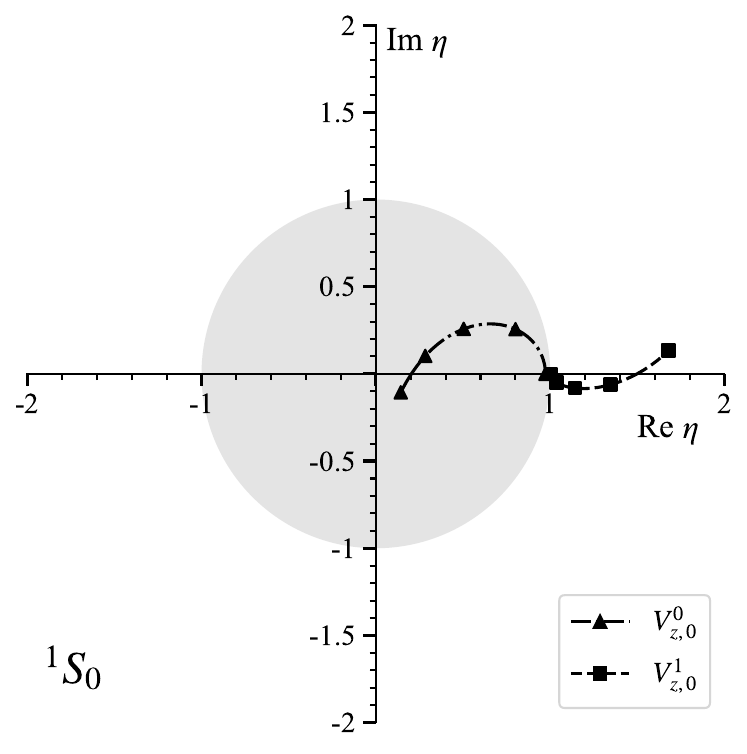}
  \includegraphics[width=0.55\textwidth,height=\textheight,keepaspectratio]{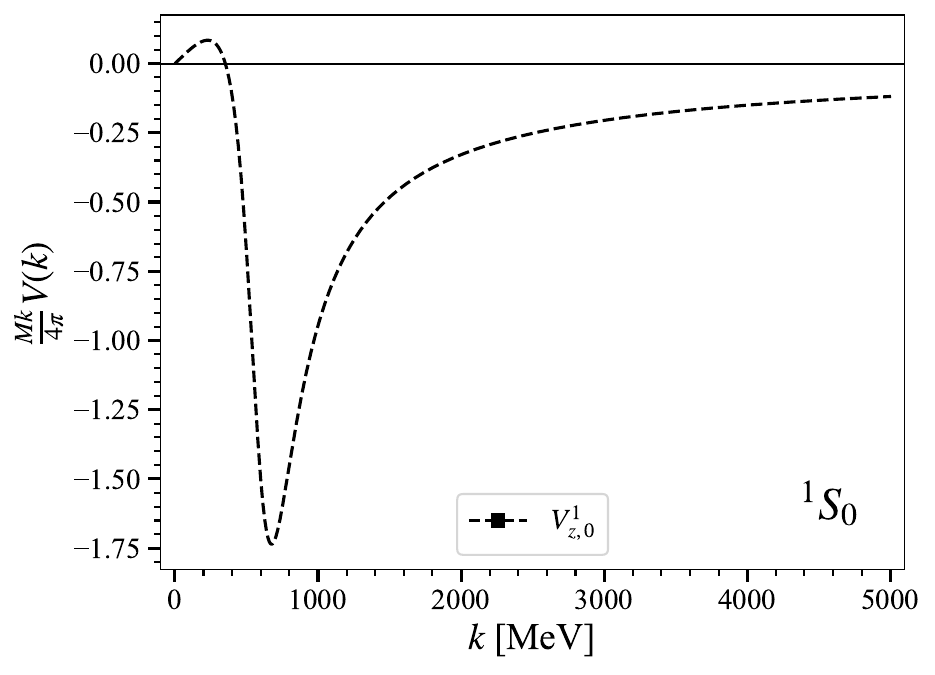}
  \caption{\small The Weinberg eigenvalues for the potentials $V_{z,0}^0$ and $V_{z,0}^1$ are plotted (left). The points correspond to the Weinberg eigenvalues at $k=0,100,200,300, 405$ MeV. The scaled potential $V_{z,0}^1$ is plotted (right) and is observed to be pathological. \label{fig:Pathological}}
\end{center}
\end{figure*}

\noindent In Fig.~\ref{fig:Pathological} (left), the Weinberg eigenvalues are plotted for the two potentials $V_{z,0}^{0}$ and $V_{z,0}^{1}$ that generate a re-summed ERE with a zero at $k_*=352.6$ MeV. $V_{z,0}^{0}$ is perturbative, but for $V_{z,0}^{1}$, the magnitude of the largest Weinberg eigenvalue exceeds 1 for almost all momenta below 405 MeV. When the potential is plotted (right), it becomes clear that this best-fit potential is pathological. The integrated potential is attractive, as expected; however, the potential is repulsive at long distances, and the proper net attractive behavior is only realized when integrating up to momentum scales well in excess of $k_*$.

To determine the source of the pathological behavior in $V_{z,0}^{1}$, it is helpful to consider the connection between the potentials that generate the re-summed ERE and the ordinary ERE. In particular, in the limit $k_*\to\infty$, $V_{z,\ell}^{m}\to V_{\ell}^{m+1}$ and the re-summed ERE becomes the ordinary ERE with the same number of independent effective-range parameters. Since the potential $V_{0}^{2}$ does not suffer from this pathology, one can bring $k_*$ in from infinity and examine when the pathological behavior arises.

\begin{figure*}
\begin{center}
  \includegraphics[width=0.43\textwidth,height=\textheight,keepaspectratio]{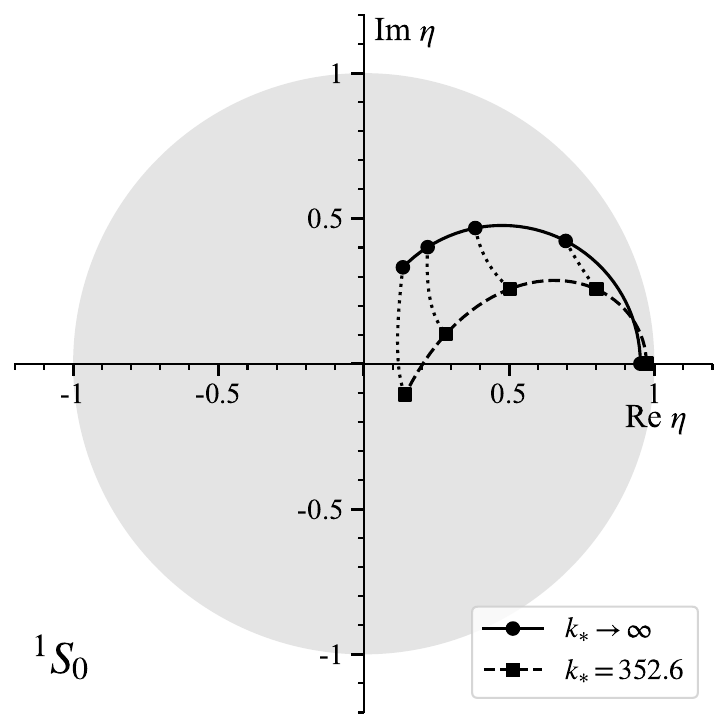}
  \includegraphics[width=0.43\textwidth,height=\textheight,keepaspectratio]{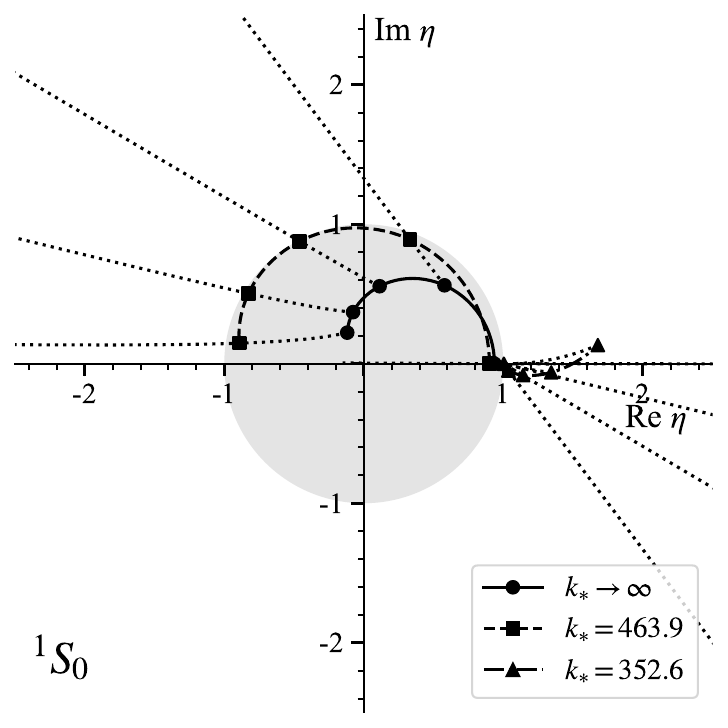}
  \caption{\small  The Weinberg eigenvalues of the $V_{z,0}^0$ potential (left) and the $V_{z,0}^1$ potential (right) as $k_*$ is brought in from infinity to its physical value. The dotted lines trace the evolution of the particular eigenvalues at $k= 0, 100, 200, 300, 405$ MeV represented by points. \label{fig:EvalEvolution}}
\end{center}
\end{figure*}

In Fig.~\ref{fig:EvalEvolution}, the Weinberg eigenvalues for $V_{z,0}^0$ and the potential itself continuously evolve as $k_*$ is brought in from infinity to its physical value. Throughout this evolution, the potential remains perturbative. For $V_{z,0}^1$, the Weinberg eigenvalues steadily grow for much of the evolution until around $k_* \lesssim 460$ MeV when the potential becomes nonperturbative. Then, the Weinberg eigenvalues diverge, come back from the opposite side of the complex plane, converge toward 1, and finally evolve to their pathological values at $k_* = 352.6$ MeV.

The sickness lies in the asymptotic behavior of the phase shift. As $k\to\infty$,

\begin{align}
    \delta_0(k) &= \arctan\left(\frac{k}{-\frac{1}{a_0}+\frac{1}{2}r_0k^2+v_{2,0}k^4\frac{k_*^2}{k_*^2-k^2}}\right)\\
    &\simeq \arctan\left(\frac{k}{\left(\frac{1}{2}r_0-v_{2,0}k_*^2\right)k^2}\right)\ .
\end{align}
If $v_{2,0}k_*^2-r_0/2 < 0$, the phase shift will traverse an additional $\Delta\delta_0 = \pi$ at large $k$ after the zero at $k_*$. This requires the potential to be non-negligible at high momentum and introduces a deep bound state. Such a potential will be nonperturbative. Even when $v_{2, 0}k_*^2 - r_0/2$ is positive but small, the appropriate asymptotic behavior of the phase shift is not realized until large momenta since $v_{2,0}k^4 k_*^2/(k_*^2-k^2)\to-v_{2,0}k_*^2k^2$ only in the regime $k^2\gg k_*^2$. Such a potential is also likely to be nonperturbative. For example, for $k_*=464$ MeV, just before the $V_{z,0}^1$ potential becomes nonperturbative, $v_{2,0}k_*^2-r_0/2 \approx 0.3$ fm. These observations motivate imposing $v_{2, 0} > 0$, so that $\delta_0(k_*) = 0$ rather than $\pi$, and $v_{2,0}k_*^2-r_0/2>\epsilon$ for an appropriate $\epsilon>0$ so the potential is perturbative.

Imposing these constraints on $V_{z,0}^1$ does restore perturbativity; however, choosing $\epsilon$ large enough to make $V_{z,0}^1$ perturbative degrades the goodness-of-fit to that of $V_{z,0}^0$ so that there is no benefit from adding $v_{2,0}$. Instead, one should proceed directly to the $V_{z,0}^2$ potential. Here, the goodness of fit can be improved drastically relative to $V_{z,0}^0$ while keeping the potential perturbative by requiring $v_{3,0}k_*^2 -v_{2,0}> \epsilon = 0.1$ fm$^3$.

\subsubsection{\texorpdfstring{$^1P_1$}{1P1}}

\noindent This lesson can also be applied to the ordinary EPT potentials introduced in Section~\ref{sec:pot def}. 
Consider the $^1P_1$ channel, which is fit to the ordinary ERE. The $V_{1}^{0}$ potential is perturbative, but the $V_{1}^{1}$ potential is nonperturbative. The culprit is, again, the asymptotic behavior of the phase shift. Since the best-fit $v_{2, 1}>0$, $\Delta\delta_1 = \pi$. This introduces an unphysical bound state and requires the potential to be non-negligible at large momenta. Perturbativity requires the last non-vanishing shape parameter to be sufficiently negative so that the potential is suppressed at large momentum. Attempting to apply this to $V_{1}^{1}$ degrades the quality of fit relative to $V_{1}^{0}$, but this constraint does produce an improved fit and a perturbative potential for $V_{1}^{2}$ with $v_{3, 1} < -\epsilon = -0.04 $ fm$^3$.

\subsection{Results}
\begin{figure*}[p]
\begin{center}
  \includegraphics[width=\textwidth,height=\textheight,keepaspectratio]{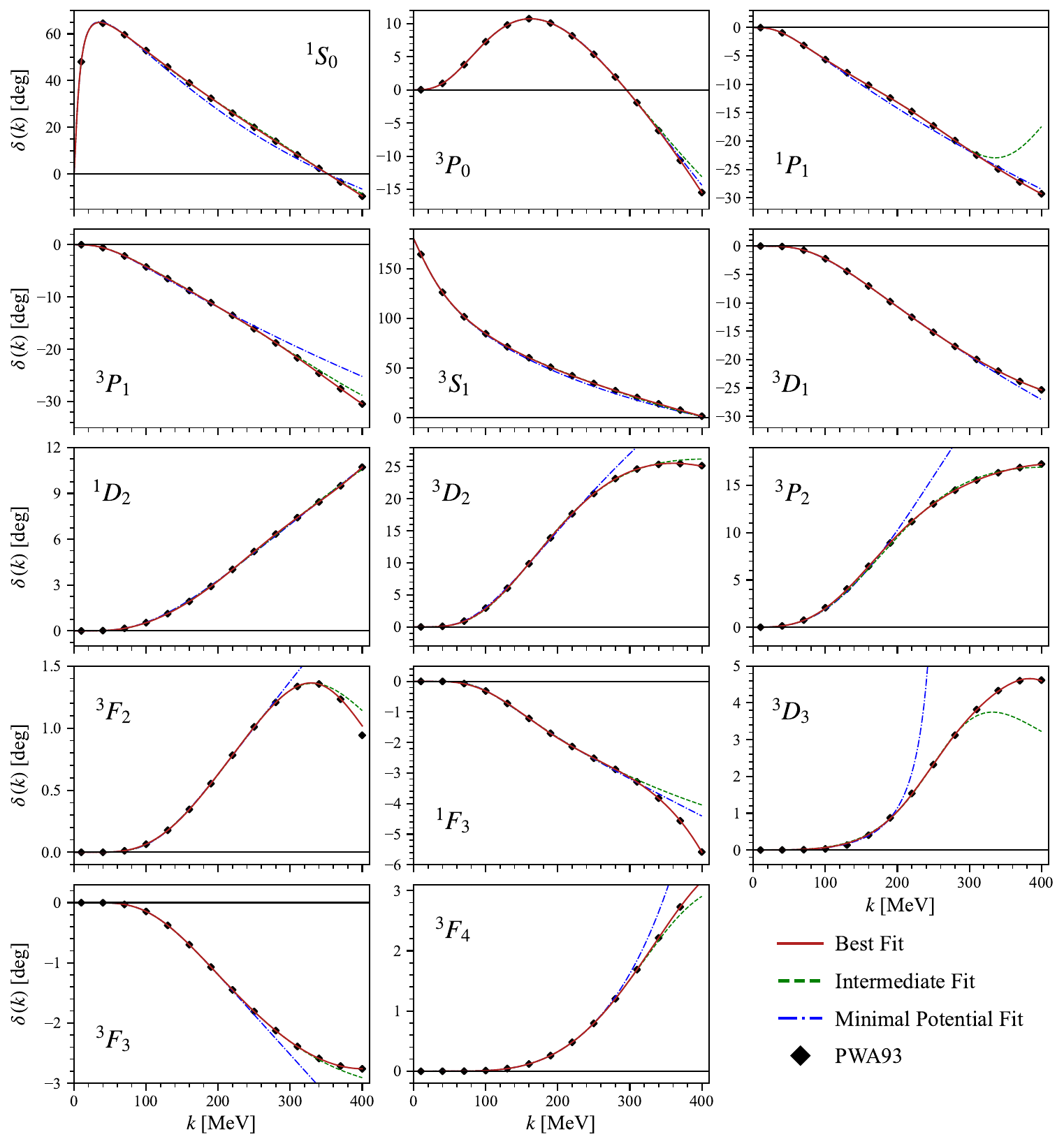}
  \caption{\small Phase shifts from EPT potentials fit up to a relative momentum of 405 MeV. The phase shifts from the minimal, intermediate, and best-fit potentials are plotted in blue, green, and red, respectively.
  \label{fig:400PhaseShifts}}
\end{center}
\end{figure*}

\begin{figure*}[p]
\begin{center}
  \includegraphics[width=\textwidth,height=\textheight,keepaspectratio]{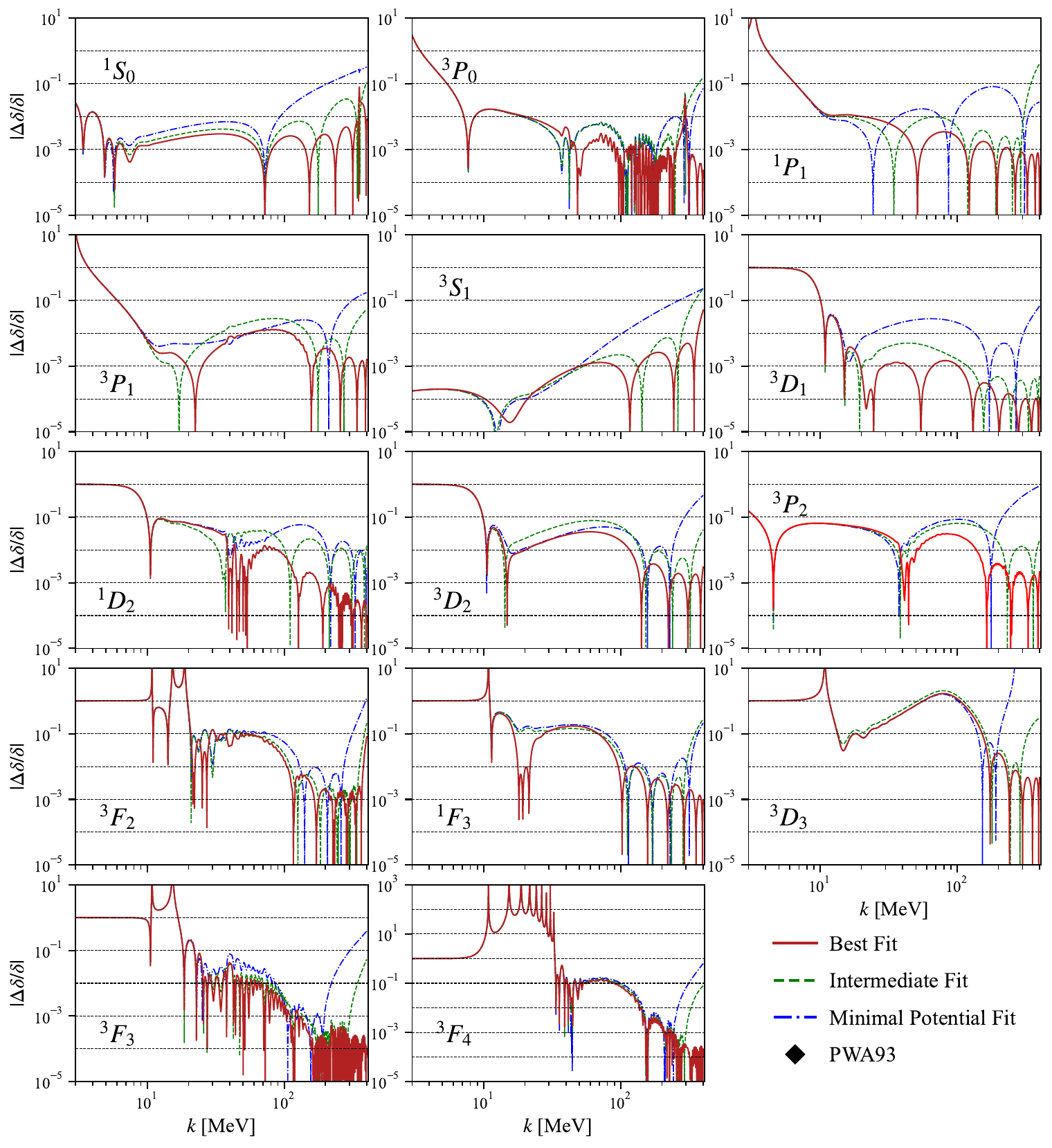}
  \caption{\small The fractional errors in the phase shifts from EPT potentials fit up to a relative momentum of 405 MeV 
  relative to the Nijmegen PWA. The phase shifts themselves are plotted in Fig.~\ref{fig:400PhaseShifts}. The minimal potential fit, an intermediate fit, and the best-fit fractional errors are plotted in blue, green, and red, respectively. 
  \label{fig:400Lepage}}
\end{center}
\end{figure*}

\noindent 
The best-fit phase shifts are plotted in Fig.~\ref{fig:400PhaseShifts}, along with predictions obtained using fewer terms in the ERE---the best-fit minimal potential and a third potential with an intermediate number of effective-range parameters---which display the systematic improvement as more shape parameters are added to the ERE.
The fractional errors in the best-fit phase shifts are plotted in Fig.~\ref{fig:400Lepage}. For potentials with a zero, the fractional error diverges where the phase shift crosses zero. To improve readability, data points within 1 MeV of the zero are omitted. 

\begin{figure*}
\begin{center}
   \resizebox{!}{0.90\textheight}{ \includegraphics[width=\textwidth,height=\textheight,keepaspectratio]{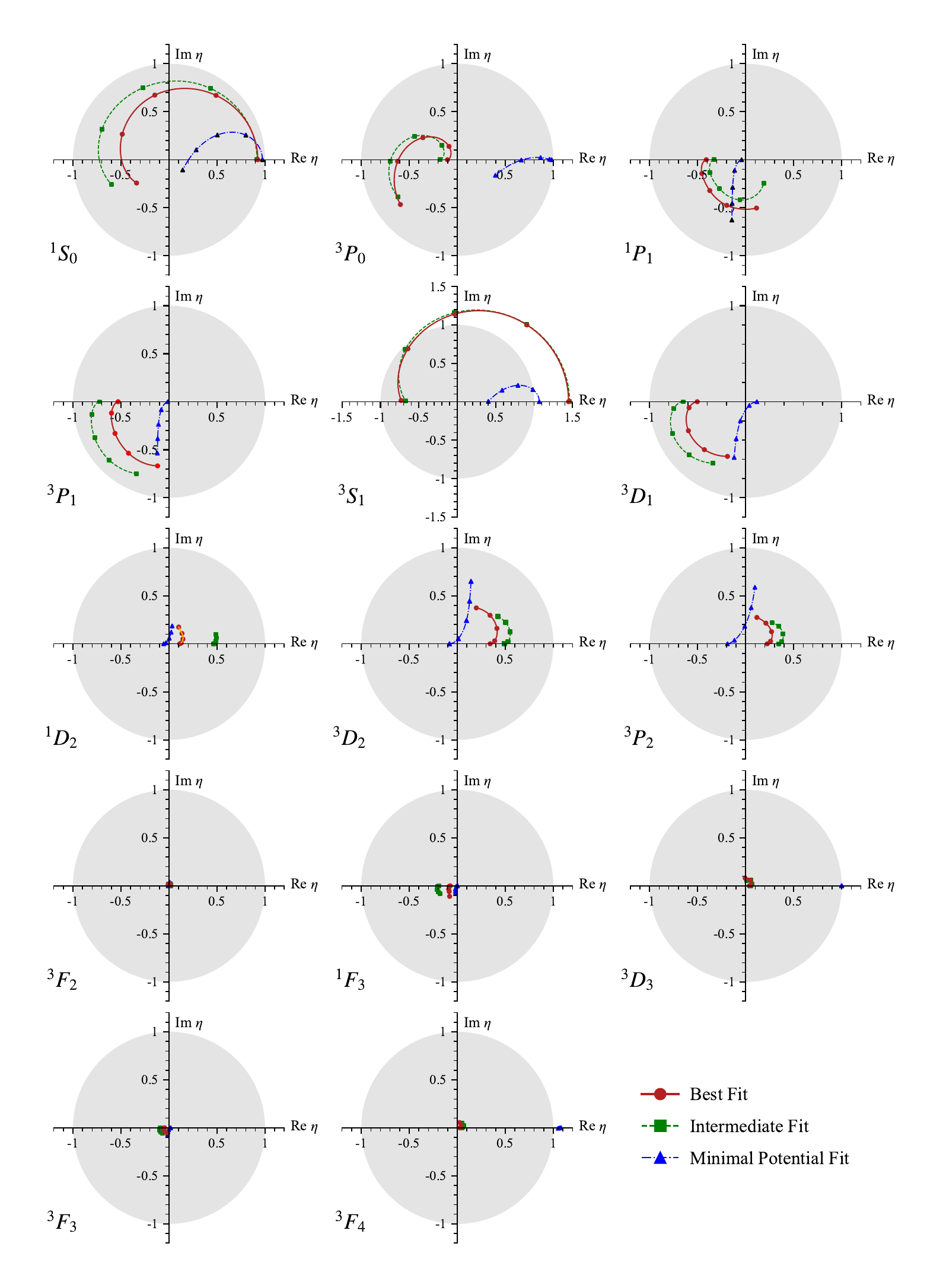} }
  \caption{\small Weinberg eigenvalues from EPT potentials fit up to a relative momentum of 405 MeV. These eigenvalues correspond to the phase shifts in Fig.~\ref{fig:400PhaseShifts}. The points denote the Weinberg eigenvalues at $k=0, 100, 200, 300, 405$ MeV. The Weinberg eigenvalues from the minimal, intermediate, and best-fit potentials are plotted in blue, green, and red, respectively. 
  \label{fig:400WeinbergEvals}}
\end{center}
\end{figure*}

The Lepage plots show that the fractional error can be made small at high energies without compromising the fractional error at low energies in virtually all channels. No clear obstruction is observed to extending $\nopi$ by introducing zeros of the scattering amplitude and adding higher-order shape parameters. Meanwhile, the behavior of the fractional error at low energies is similar to, though slightly worse than, that of the $\nopi$ fractional error. The difference is likely because the same channels were fit over a larger range of momenta with proportionally fewer parameters.

The Weinberg eigenvalues are plotted in Fig.~\ref{fig:400WeinbergEvals}. The potentials are beautifully perturbative up to $405$ MeV, except in the case of the $^3S_1$ channel, where the presence of the deuteron renders the potential inherently nonperturbative. The near-universal perturbativity of these potentials demonstrates one of the key benefits of EPT: a perturbative potential can be obtained in virtually any channel by fitting a sufficiently large range of momenta and by using subleading shape parameters to tame the asymptotic behavior of the phase shift.

The best-fit parameters and corresponding $\gamma$ values are listed in Tables~\ref{tab:ERE_400} and \ref{tab:poles400}, respectively, in Appendix~\ref{sec:par405MeV}.

\noindent 

\section{Conclusion}
\label{sec:conc}

\noindent This paper has introduced energy-independent NN potentials
which generate the effective-range expansion to a fixed order. These
potentials are equivalent to $\nopi$ for $k<M_\pi/2$ and are soft over phenomenologically relevant scales as
demonstrated by an analysis of Weinberg eigenvalues. 
This method is
aptly called Effective Pole Theory (EPT) since it is organized
according to the number of poles in a rational S-matrix, and exploits
the intrinsic S-matrix symmetry which permutes the poles.  As EPT
reproduces effective-range theory, which is a momentum expansion in
the inverse scattering amplitude, the potentials are most singular at
lowest order, in contrast with EFT calculations of the amplitude
itself, which become increasingly singular as fine-structure effects
are included.  This is because the natural organizational principle
consistent with a unitary S-matrix is an expansion in the inverse
scattering amplitude, while the EFT expansion of the amplitude
partitions unitarity into parts that are added at each order in the
expansion.

For $k>M_\pi/2$, 
EPT provides a smooth extension into
the region of pion physics. As, by construction, the sole
singularities of the S-matrix are simple poles, the branch-cut
singularities arising from pion physics are approximated by simple
poles. In the two-body sector, it is easy to check that this
approximation of short-distance physics can be made arbitrarily
precise, and it is not unreasonable to assume that the violence done
to short-distance physics using this method is no worse than the
mangling of the short-distance physics in chiral EFT introduced
through lack of control of cutoff effects.

The main conclusions of this paper are:

\begin{itemize}

\item It is straightforward to design momentum-dependent, separable and non-local potentials
which reproduce arbitrary orders---and nothing beyond---in the effective-range expansion for an
arbitrary partial wave. Construction of such potentials is greatly facilitated by exploiting the 
symmetry of the rational S-matrix with respect to permutation of all pole positions. Partial-wave mixing is
naturally accommodated.

\item EPT suggests a natural ordering in each partial wave beginning with a minimal potential that is singular and whose coupling strength exhibits universal RG flow.
This RG running regulates a linear ultraviolet 
divergence and, interestingly, the coupling strength always diverges at the RG scale $\mu= \sum_i \gamma_i$ where the $\{\gamma_i\}$ are the poles of the S-matrix.
The structure of the minimal potentials can be viewed as a consequence of causality as seen,
for instance, from a study of  Wigner bounds.

\item The EPT formalism admits refinements of the power counting according to the specifics of
the scattering data in a given partial wave. However, some refinements suggested by observation correspond to pathological potentials, e.g., non-Hermitian or acausal potentials,
and must be discarded.

\item The Weinberg eigenvalues show that the EPT potentials are---with few exceptions---perturbative over center-of-mass momenta up to pion-production threshold.

\end{itemize}

A subject of special interest which will be considered in future work is the calculation of three-body 
observables using the two-body potentials that have been considered here. It is well known that the 
three-body equations run into difficulty handling unphysical poles, and  therefore it is important to  
understand the role of three-body forces in the formalism with the momentum-dependent separable potentials. 
A primary motivation of this work is to use separable potentials to compute superfluid gaps in neutron matter~\cite{Ramanan:2020xrj}, building on the systematic weak-coupling expansion developed in Refs.~\cite{BCF2024,Beane:2026qim}.
This would enable meaningful comparison with Quantum Monte Carlo simulations~\cite{Gandolfi_2022} and provide a pathway to estimating induced gaps in higher partial waves~\cite{Fujimoto:2025zfa,Kumamoto:2024muw,Kohn:1965zz,Maiti_2013}.
Additionally, future work will focus on computing other many-body observables and systematically including three-body
forces.

\section*{Acknowledgments}

\begin{small}\sffamily
UvK would like to thank Jaber Balal Habashi, Mario Gattobigio, and Alejandro Kievsky for useful discussions.  This work was supported by 
the U.~S.~Department of Energy grants \textbf{DE-FG02-97ER-41014} (UW Nuclear Theory) and \textbf{DE-FG02-04ER41338} (UA). R.F. acknowledges support from the Institute for
Quantum Information and Matter, an NSF Physics Frontiers Center (PHY-2317110) and additionally from
a Burke Institute prize fellowship.   
\end{small}

\bibliography{bibi}

@article{Stoks:1994wp,
    author = "Stoks, V. G. J. and Klomp, R. A. M. and Terheggen, C. P. F. and de Swart, J. J.",
    title = "{Construction of high quality N N potential models}",
    eprint = "nucl-th/9406039",
    archivePrefix = "arXiv",
    reportNumber = "THEF-NYM-93-05",
    doi = "10.1103/PhysRevC.49.2950",
    journal = "Phys. Rev. C",
    volume = "49",
    pages = "2950--2962",
    year = "1994"
}

@article{Krivoruchenko:2010ft,
    author = "Krivoruchenko, M. I.",
    title = "{Remarks on the origin of Castillejo-Dalitz-Dyson poles}",
    eprint = "1001.1659",
    archivePrefix = "arXiv",
    primaryClass = "nucl-th",
    doi = "10.1103/PhysRevC.82.018201",
    journal = "Phys. Rev. C",
    volume = "82",
    pages = "018201",
    year = "2010"
}

@article{Li:2023hwe,
    author = "Li, Qingfeng and Lyu, Songlin and Ji, Chen and Long, Bingwei",
    title = "{Effective field theory with resonant P-wave interaction}",
    eprint = "2303.17292",
    archivePrefix = "arXiv",
    primaryClass = "nucl-th",
    doi = "10.1103/PhysRevC.108.024002",
    journal = "Phys. Rev. C",
    volume = "108",
    number = "2",
    pages = "024002",
    year = "2023"
}

@article{Rupak:2018gnc,
    author = "Rupak, Gautam and Vaghani, Akshay and Higa, Renato and van Kolck, Ubirajara",
    title = "{Fate of the neutron{\textendash}deuteron virtual state as an Efimov level}",
    eprint = "1806.01999",
    archivePrefix = "arXiv",
    primaryClass = "nucl-th",
    doi = "10.1016/j.physletb.2018.08.051",
    journal = "Phys. Lett. B",
    volume = "791",
    pages = "414--419",
    year = "2019"
}

@article{Kaplan:1996nv,
    author = "Kaplan, David B.",
    title = "{More effective field theory for nonrelativistic scattering}",
    eprint = "nucl-th/9610052",
    archivePrefix = "arXiv",
    reportNumber = "DOE-ER-40561-296, INT-96-00-153, UW-PT-96-31",
    doi = "10.1016/S0550-3213(97)00178-8",
    journal = "Nucl. Phys. B",
    volume = "494",
    pages = "471--484",
    year = "1997"
}

@article{Sanchez:2024xzl,
    author = "S{\'a}nchez, M. S. and Oller, J. A. and Entem, D. R.",
    title = "{Confronting the Lippmann{\textendash}Schwinger equation and the \(N/D\) method for coupled-wave separable potentials}",
    eprint = "2403.05486",
    archivePrefix = "arXiv",
    primaryClass = "nucl-th",
    doi = "10.1016/j.aop.2024.169767",
    journal = "Annals Phys.",
    volume = "469",
    pages = "169767",
    year = "2024"
}

@article{Behzadmoghaddam:2023gsh,
    author = "Behzadmoghaddam, B. and Radin, M. and Bayegan, S.",
    title = "{The $^{1}\!S_{0}$ channel of proton{\textendash}proton scattering in new chiral effective field theory power counting}",
    doi = "10.1038/s41598-023-33735-6",
    journal = "Sci. Rep.",
    volume = "13",
    number = "1",
    pages = "7041",
    year = "2023"
}

@article{SanchezSanchez:2021dki,
    author = "S{\'a}nchez S{\'a}nchez, M. and Duc, Dao Duy and Bonneau, L.",
    title = "{Cutoff effects in Hartree-Fock calculations at leading order of chiral effective field theory}",
    eprint = "2104.05474",
    archivePrefix = "arXiv",
    primaryClass = "nucl-th",
    doi = "10.1103/PhysRevC.107.064307",
    journal = "Phys. Rev. C",
    volume = "107",
    number = "6",
    pages = "064307",
    year = "2023"
}

@article{Yang:2020pgi,
    author = {Yang, C.-J. and Ekstr{\"o}m, A. and Forss{\'e}n, C. and Hagen, G.},
    title = "{Power counting in chiral effective field theory and nuclear binding}",
    eprint = "2011.11584",
    archivePrefix = "arXiv",
    primaryClass = "nucl-th",
    doi = "10.1103/PhysRevC.103.054304",
    journal = "Phys. Rev. C",
    volume = "103",
    number = "5",
    pages = "054304",
    year = "2021"
}

@article{Behzadmoghaddam:2020pqr,
    author = "Behzadmoghaddam, B. and Bayegan, S. and Arani, M. Moeini",
    title = "{Proton{\textendash}proton fusion in new pionless EFT power counting}",
    doi = "10.1140/epja/s10050-020-00166-0",
    journal = "Eur. Phys. J. A",
    volume = "56",
    number = "6",
    pages = "158",
    year = "2020"
}

@article{SanchezSanchez:2020kbx,
    author = "S{\'a}nchez S{\'a}nchez, M. and Smirnova, N. A. and Shirokov, A. M. and Maris, P. and Vary, J. P.",
    title = "{Improved description of light nuclei through chiral effective field theory at leading order}",
    eprint = "2002.12258",
    archivePrefix = "arXiv",
    primaryClass = "nucl-th",
    doi = "10.1103/PhysRevC.102.024324",
    journal = "Phys. Rev. C",
    volume = "102",
    number = "2",
    pages = "024324",
    year = "2020"
}

@article{Kohn:1965zz,
    author = "Kohn, W. and Luttinger, J. M.",
    title = "{New Mechanism for Superconductivity}",
    doi = "10.1103/PhysRevLett.15.524",
    journal = "Phys. Rev. Lett.",
    volume = "15",
    pages = "524--526",
    year = "1965"
}

@inproceedings{Maiti_2013,
	doi = {10.1063/1.4818400},
  
	url = {https://doi.org/10.1063%2F1.4818400},
  
	year = 2013,
	publisher = {{AIP}
},
  
	author = {Saurabh Maiti and Andrey V. Chubukov},
  
	title = {Superconductivity from repulsive interaction},
  
	booktitle = {{AIP} Conference Proceedings}
}

@article{Ramanan:2020xrj,
    author = "Ramanan, S. and Urban, M.",
    title = "{Pairing in pure neutron matter}",
    eprint = "2010.09632",
    archivePrefix = "arXiv",
    primaryClass = "nucl-th",
    doi = "10.1140/epjs/s11734-021-00008-0",
    journal = "Eur. Phys. J. ST",
    volume = "230",
    number = "2",
    pages = "567--577",
    year = "2021"
}

@article{Fujimoto:2025zfa,
    author = "Fujimoto, Yuki",
    title = "{Renormalization-group approach to Kohn-Luttinger superconductivity: Amplification of the pairing gap from {\ensuremath{\ell}}4 to {\ensuremath{\ell}}}",
    eprint = "2502.01169",
    archivePrefix = "arXiv",
    primaryClass = "cond-mat.supr-con",
    reportNumber = "N3AS-25-002, RIKEN-iTHEMS-Report-25",
    doi = "10.1103/PhysRevB.111.184510",
    journal = "Phys. Rev. B",
    volume = "111",
    number = "18",
    pages = "184510",
    year = "2025"
}

@article{Kumamoto:2024muw,
    author = "Kumamoto, Mia and Reddy, Sanjay",
    title = "{Kohn-Luttinger effect in dense matter and its implications for neutron stars}",
    eprint = "2405.12243",
    archivePrefix = "arXiv",
    primaryClass = "nucl-th",
    doi = "10.1103/PhysRevC.110.025804",
    journal = "Phys. Rev. C",
    volume = "110",
    number = "2",
    pages = "025804",
    year = "2024"
}

@misc{Beane:2026qim,
    author = "Beane, Silas R. and Farrell, Roland C. and Shotwell, Cameron P.",
    title = "{Universal enhancement of the pairing gap in a dilute Fermi superfluid}",
    eprint = "2610.20434",
    archivePrefix = "arXiv",
    primaryClass = "cond-mat.quant-gas",
    reportNumber = "NT@UW-26-23",
    month = "10",
    year = "2026"
}

@article{BCF2024,
    author = "Beane, Silas R. and Capatti, Zeno and Farrell, Roland C.",
    title = "{Universal corrections to the superfluid gap in a cold Fermi gas}",
    eprint = "2407.20168",
    archivePrefix = "arXiv",
    primaryClass = "nucl-th",
    reportNumber = "NT@UW-24-09, IQuS@UW-21-083",
    doi = "10.1103/PhysRevA.111.013309",
    journal = "Phys. Rev. A",
    volume = "111",
    number = "1",
    pages = "013309",
    year = "2025"
}

@article{Beane:2021bpr,
    author = "Beane, Silas R. and Farrell, Roland C.",
    title = "{Causality and dimensionality in geometric scattering}",
    eprint = "2112.02733",
    archivePrefix = "arXiv",
    primaryClass = "quant-ph",
    reportNumber = "NT@UW-21-14, IQuS@UW-21-015",
    doi = "10.1016/j.aop.2022.168841",
    journal = "Annals Phys.",
    volume = "440",
    pages = "168841",
    year = "2022"
}

@article{Tumino:2024qpk,
    author = "Tumino, Aurora and others",
    title = "{Test of the Charge Symmetry Hypothesis of NN Interaction from the Coulomb-Free p{\textendash}p Scattering Cross Section and Its Relation to Universality}",
    doi = "10.1007/s00601-024-01975-5",
    journal = "Few Body Syst.",
    volume = "66",
    number = "1",
    pages = "5",
    year = "2025"
}

@article{PavonValderrama:2005ku,
    author = "Pavon Valderrama, M. and Ruiz Arriola, E.",
    title = "{Low-energy NN scattering at next-to-next-to-next-to-next-to-leading order for partial waves with j $\le$ 5}",
    doi = "10.1103/PhysRevC.72.044007",
    journal = "Phys. Rev. C",
    volume = "72",
    pages = "044007",
    year = "2005"
}

@article{PhysRevC.41.898,
  title = {Asymptotic D-state to S-state ratio of the deuteron},
  author = {Rodning, N. L. and Knutson, L. D.},
  journal = {Phys. Rev. C},
  volume = {41},
  issue = {3},
  pages = {898--909},
  numpages = {0},
  year = {1990},
  month = {Mar},
  publisher = {American Physical Society},
  doi = {10.1103/PhysRevC.41.898},
  url = {https://link.aps.org/doi/10.1103/PhysRevC.41.898}
}

@article{Mohr:2024kco,
    author = "Mohr, Peter J. and Newell, David B. and Taylor, Barry N. and Tiesinga, Eite",
    title = "{CODATA recommended values of the fundamental physical constants: 2022*}",
    eprint = "2409.03787",
    archivePrefix = "arXiv",
    primaryClass = "hep-ph",
    doi = "10.1103/RevModPhys.97.025002",
    journal = "Rev. Mod. Phys.",
    volume = "97",
    number = "2",
    pages = "025002",
    year = "2025"
}

@book{Newton1982,
  author    = {Roger G. Newton},
  title     = {Scattering Theory of Waves and Particles},
  edition   = {2},
  publisher = {Springer},
  address   = {Berlin, Heidelberg},
  year      = {1982},
  series    = {Theoretical and Mathematical Physics},
  doi       = {10.1007/978-3-642-88128-2},
  isbn      = {978-3-642-88130-5},
}

@article{Eckart:1930zza,
    author = "Eckart, Carl",
    title = "{The Penetration of a Potential Barrier by Electrons}",
    doi = "10.1103/PhysRev.35.1303",
    journal = "Phys. Rev.",
    volume = "35",
    pages = "1303--1309",
    year = "1930"
}

@article{RevModPhys.21.488,
  title = {On the Connection between Phase Shifts and Scattering Potential},
  author = {Bargmann, V.},
  journal = {Rev. Mod. Phys.},
  volume = {21},
  issue = {3},
  pages = {488--493},
  numpages = {0},
  year = {1949},
  month = {Jul},
  publisher = {American Physical Society},
  doi = {10.1103/RevModPhys.21.488},
  url = {https://link.aps.org/doi/10.1103/RevModPhys.21.488}
}

@article{Kievsky:2023xgy,
    author = "Kievsky, Alejandro and Gattobigio, Mario and Girlanda, Luca and Viviani, Michele",
    title = "{From correlations to universal behavior in few-nucleon systems}",
    doi = "10.1051/epjconf/202329010008",
    journal = "EPJ Web Conf.",
    volume = "290",
    pages = "10008",
    year = "2023"
}

@article{Ekstrom:2024dqr,
    author = {Ekstr{\"o}m, Andreas and Platter, Lucas},
    title = "{Quantifying the breakdown scale of pionless effective field theory}",
    eprint = "2409.08197",
    archivePrefix = "arXiv",
    primaryClass = "nucl-th",
    doi = "10.1016/j.physletb.2024.139207",
    journal = "Phys. Lett. B",
    volume = "860",
    pages = "139207",
    year = "2025"
}

@article{Timoteo:2023dan,
    author = "Tim\'oteo, V. S. and van Kolck, U.",
    title = "{Two-body double pole and three-body bound states: Physical and unphysical quark masses}",
    eprint = "2304.01165",
    archivePrefix = "arXiv",
    primaryClass = "nucl-th",
    doi = "10.1016/j.physletb.2024.138619",
    journal = "Phys. Lett. B",
    volume = "852",
    pages = "138619",
    year = "2024"
}

@article{Konig:2019xxk,
    author = {K{\"o}nig, Sebastian},
    title = "{Energies and radii of light nuclei around unitarity}",
    eprint = "1910.12627",
    archivePrefix = "arXiv",
    primaryClass = "nucl-th",
    doi = "10.1140/epja/s10050-020-00098-9",
    journal = "Eur. Phys. J. A",
    volume = "56",
    number = "4",
    pages = "113",
    year = "2020"
}

@article{Dawkins:2019vcr,
    author = "Dawkins, William G. and Carlson, J. and van Kolck, U. and Gezerlis, Alexandros",
    title = "{Clustering of Four-Component Unitary Fermions}",
    eprint = "1908.04288",
    archivePrefix = "arXiv",
    primaryClass = "cond-mat.quant-gas",
    doi = "10.1103/PhysRevLett.124.143402",
    journal = "Phys. Rev. Lett.",
    volume = "124",
    number = "14",
    pages = "143402",
    year = "2020"
}

@article{vanKolck:2017jon,
    author = "van Kolck, U.",
    title = "{Unitarity and Discrete Scale Invariance}",
    doi = "10.1007/s00601-017-1271-9",
    journal = "Few Body Syst.",
    volume = "58",
    number = "3",
    pages = "112",
    year = "2017"
}

@article{Konig:2016iny,
    author = {K{\"o}nig, Sebastian},
    title = "{Second-order perturbation theory for ${}^{3}\mathrm{He}$ and pd scattering in pionless EFT}",
    eprint = "1609.03163",
    archivePrefix = "arXiv",
    primaryClass = "nucl-th",
    doi = "10.1088/1361-6471/aa60d6",
    journal = "J. Phys. G",
    volume = "44",
    number = "6",
    pages = "064007",
    year = "2017"
}

@article{Konig:2016utl,
    author = {K{\"o}nig, Sebastian and Grie{\ss}hammer, Harald W. and Hammer, H. W. and van Kolck, U.},
    title = "{Nuclear Physics Around the Unitarity Limit}",
    eprint = "1607.04623",
    archivePrefix = "arXiv",
    primaryClass = "nucl-th",
    reportNumber = "INT-PUB-16-019",
    doi = "10.1103/PhysRevLett.118.202501",
    journal = "Phys. Rev. Lett.",
    volume = "118",
    number = "20",
    pages = "202501",
    year = "2017"
}

@article{Phillips:1998uy,
    author = "Phillips, Daniel R. and Beane, S. R. and Birse, M. C.",
    title = "{Scheming in dimensional regularization}",
    eprint = "hep-th/9810049",
    archivePrefix = "arXiv",
    reportNumber = "UMD-PP-99-030, DOE-ER-40762-162",
    doi = "10.1088/0305-4470/32/18/313",
    journal = "J. Phys. A",
    volume = "32",
    pages = "3397--3407",
    year = "1999"
}

@misc{Thim:2026mzm,
    author = {Thim, Oliver and Ekstr{\"o}m, Andreas and Forss{\'e}n, Christian},
    title = "{Perturbative calculations of light nuclei up to N$^3$LO in chiral effective field theory}",
    eprint = "2604.14985",
    archivePrefix = "arXiv",
    primaryClass = "nucl-th",
    month = "4",
    year = "2026"
}

@article{vanKolck:2020llt,
    author = "van Kolck, U.",
    title = "{The Problem of Renormalization of Chiral Nuclear Forces}",
    eprint = "2003.06721",
    archivePrefix = "arXiv",
    primaryClass = "nucl-th",
    doi = "10.3389/fphy.2020.00079",
    journal = "Front. in Phys.",
    volume = "8",
    pages = "79",
    year = "2020"
}

@article{Beane:2000wh,
    author = "Beane, S. R. and Bedaque, Paulo F. and Childress, L. and Kryjevski, A. and McGuire, J. and van Kolck, U.",
    title = "{Singular potentials and limit cycles}",
    eprint = "quant-ph/0010073",
    archivePrefix = "arXiv",
    reportNumber = "NT-UW-00-023, DOE-ER-41132-102-INT-00, RBRC-140, KRL-MAP-271, NT@UW-00-023",
    doi = "10.1103/PhysRevA.64.042103",
    journal = "Phys. Rev. A",
    volume = "64",
    pages = "042103",
    year = "2001"
}

@article{Long:2007vp,
    author = "Long, B. and van Kolck, U.",
    title = "{Renormalization of Singular Potentials and Power Counting}",
    eprint = "0707.4325",
    archivePrefix = "arXiv",
    primaryClass = "quant-ph",
    reportNumber = "UA-ET-07-03, UA-ET-07-03",
    doi = "10.1016/j.aop.2008.01.003",
    journal = "Annals Phys.",
    volume = "323",
    pages = "1304--1323",
    year = "2008"
}

@article{Griesshammer:2023scn,
    author = "Grie{\ss}hammer, Harald W. and van Kolck, Ubirajara",
    title = "{Universality of three identical bosons with large, negative effective range}",
    eprint = "2308.01394",
    archivePrefix = "arXiv",
    primaryClass = "nucl-th",
    doi = "10.1140/epja/s10050-023-01196-0",
    journal = "Eur. Phys. J. A",
    volume = "59",
    number = "12",
    pages = "289",
    year = "2023"
}

@article{Contessi:2024vae,
    author = "Contessi, Lorenzo and Pavon Valderrama, Manuel and van Kolck, Ubirajara",
    title = "{Limits on an improved action for contact effective field theory in two-body systems}",
    eprint = "2403.16596",
    archivePrefix = "arXiv",
    primaryClass = "nucl-th",
    doi = "10.1016/j.physletb.2024.138903",
    journal = "Phys. Lett. B",
    volume = "856",
    pages = "138903",
    year = "2024"
}

@article{vanKolck:1997ut,
    author = "van Kolck, U.",
    editor = "Bernstein, A. M. and Drechsel, D. and Walcher, T.",
    title = "{Nucleon-nucleon interaction and isospin violation}",
    eprint = "hep-ph/9711222",
    archivePrefix = "arXiv",
    reportNumber = "DOE-ER-41014-44-N97",
    doi = "10.1007/BFb0104898",
    journal = "Lect. Notes Phys.",
    volume = "513",
    pages = "62",
    year = "1998"
}

@inproceedings{Lepage:1997cs,
    author = "Lepage, G. P.",
    title = "{How to renormalize the Schr\"odinger equation}",
    booktitle = "{8th Jorge Andr\'e Swieca Summer School on Nuclear Physics}",
    eprint = "nucl-th/9706029",
    archivePrefix = "arXiv",
    pages = "135--180",
    month = "2",
    year = "1997"
}

@misc{Gantenberg:2026xzy,
    author = "Gantenberg, Cullen and Kaplan, David B.",
    title = "{Dimer Effective Field Theory}",
    eprint = "2603.28842",
    archivePrefix = "arXiv",
    primaryClass = "nucl-th",
    reportNumber = "INT-PUB-26-010",
    month = "3",
    year = "2026"
}

@article{Hammer:2019poc,
    author = {Hammer, H.-W. and K{\"o}nig, S. and van Kolck, U.},
    title = "{Nuclear effective field theory: status and perspectives}",
    eprint = "1906.12122",
    archivePrefix = "arXiv",
    primaryClass = "nucl-th",
    doi = "10.1103/RevModPhys.92.025004",
    journal = "Rev. Mod. Phys.",
    volume = "92",
    number = "2",
    pages = "025004",
    year = "2020"
}

@article{SanchezSanchez:2017tws,
    author = "S{\'a}nchez S{\'a}nchez, M. and Yang, C.-J. and Long, Bingwei and van Kolck, U.",
    title = "{Two-nucleon $^1S_0$ amplitude zero in chiral effective field theory}",
    eprint = "1704.08524",
    archivePrefix = "arXiv",
    primaryClass = "nucl-th",
    reportNumber = "CTP-SCU-2017008",
    doi = "10.1103/PhysRevC.97.024001",
    journal = "Phys. Rev. C",
    volume = "97",
    number = "2",
    pages = "024001",
    year = "2018"
}

@article{Weinberg:1963zza,
    author = "Weinberg, Steven",
    title = "{Quasiparticles and the Born Series}",
    doi = "10.1103/PhysRev.131.440",
    journal = "Phys. Rev.",
    volume = "131",
    pages = "440--460",
    year = "1963"
}

@article{Bogner:2005sn,
    author = "Bogner, S. K. and Schwenk, A. and Furnstahl, R. J. and Nogga, A.",
    title = "{Is nuclear matter perturbative with low-momentum interactions?}",
    eprint = "nucl-th/0504043",
    archivePrefix = "arXiv",
    doi = "10.1016/j.nuclphysa.2005.08.024",
    journal = "Nucl. Phys. A",
    volume = "763",
    pages = "59--79",
    year = "2005"
}

@article{PhysRev.74.131,
  title = {On the Application of {H}eisenberg's Theory of ${S}$-Matrix to the Problems of Resonance Scattering and Reactions in Nuclear Physics},
  author = {Hu, Ning},
  journal = {Phys. Rev.},
  volume = {74},
  issue = {2},
  pages = {131--140},
  numpages = {0},
  year = {1948},
  month = {Jul},
  publisher = {American Physical Society},
  doi = {10.1103/PhysRev.74.131},
  url = {https://link.aps.org/doi/10.1103/PhysRev.74.131}
}

@article{PhysRev.83.249,
  title = {On the Connection of the Scattering and Derivative Matrices with Causality},
  author = {Sch\"utzer, Walter and Tiomno, J.},
  journal = {Phys. Rev.},
  volume = {83},
  issue = {2},
  pages = {249--251},
  numpages = {0},
  year = {1951},
  month = {Jul},
  publisher = {American Physical Society},
  doi = {10.1103/PhysRev.83.249},
  url = {https://link.aps.org/doi/10.1103/PhysRev.83.249}
}

@article{PhysRev.91.1267,
  title = {${S}$-Matrix and Causality Condition. {II.} Nonrelativistic Particles},
  author = {van Kampen, N. G.},
  journal = {Phys. Rev.},
  volume = {91},
  issue = {5},
  pages = {1267--1276},
  numpages = {0},
  year = {1953},
  month = {Sep},
  publisher = {American Physical Society},
  doi = {10.1103/PhysRev.91.1267},
  url = {https://link.aps.org/doi/10.1103/PhysRev.91.1267}
}

@article{Wigner:1955zz,
    author = "Wigner, Eugene P.",
    title = "{Lower Limit for the Energy Derivative of the Scattering Phase Shift}",
    doi = "10.1103/PhysRev.98.145",
    journal = "Phys. Rev.",
    volume = "98",
    pages = "145--147",
    year = "1955"
}

@article{Phillips:1996ae,
    author = "Phillips, Daniel R. and Cohen, Thomas D.",
    title = "{How short is too short? Constraining contact interactions in nucleon-nucleon scattering}",
    eprint = "nucl-th/9607048",
    archivePrefix = "arXiv",
    reportNumber = "DOE-ER-40762-092, UMD-PP-97-008",
    doi = "10.1016/S0370-2693(96)01411-6",
    journal = "Phys. Lett. B",
    volume = "390",
    pages = "7--12",
    year = "1997"
}

@article{Hammer:2010fw,
    author = "Hammer, H.-W. and Lee, Dean",
    title = "{Causality and the effective range expansion}",
    eprint = "1002.4603",
    archivePrefix = "arXiv",
    primaryClass = "nucl-th",
    reportNumber = "HISKP-TH-10-08",
    doi = "10.1016/j.aop.2010.06.006",
    journal = "Annals Phys.",
    volume = "325",
    pages = "2212--2233",
    year = "2010"
}

@article{Hammer:2009zh,
    author = "Hammer, H.-W. and Lee, Dean",
    title = "{Causality and universality in low-energy quantum scattering}",
    eprint = "0907.1763",
    archivePrefix = "arXiv",
    primaryClass = "nucl-th",
    reportNumber = "HISKP-TH-09-25",
    doi = "10.1016/j.physletb.2009.10.033",
    journal = "Phys. Lett. B",
    volume = "681",
    pages = "500--503",
    year = "2009"
}

@article{Braaten:2011vf,
    author = "Braaten, Eric and Hagen, P. and Hammer, H.-W. and Platter, L.",
    title = "{Renormalization in the Three-body Problem with Resonant P-wave Interactions}",
    eprint = "1110.6829",
    archivePrefix = "arXiv",
    primaryClass = "cond-mat.quant-gas",
    doi = "10.1103/PhysRevA.86.012711",
    journal = "Phys. Rev. A",
    volume = "86",
    pages = "012711",
    year = "2012"
}

@article{Nishida:2011np,
    author = "Nishida, Yusuke",
    title = "{Impossibility of the Efimov effect for p-wave interactions}",
    eprint = "1111.6961",
    archivePrefix = "arXiv",
    primaryClass = "cond-mat.quant-gas",
    reportNumber = "LA-UR-11-12008",
    doi = "10.1103/PhysRevA.86.012710",
    journal = "Phys. Rev. A",
    volume = "86",
    pages = "012710",
    year = "2012"
}

@article{Chowdhury:2023oas,
    author = "Chowdhury, Subham Dutta and Mishra, Ruchira and Son, Dam Thanh",
    title = "{Applied nonrelativistic conformal field theory: Scattering-length and effective-range corrections to rate of production of three neutrons at low relative momenta}",
    eprint = "2309.15177",
    archivePrefix = "arXiv",
    primaryClass = "hep-th",
    doi = "10.1103/PhysRevD.109.016001",
    journal = "Phys. Rev. D",
    volume = "109",
    number = "1",
    pages = "016001",
    year = "2024"
}

@article{Beane:2024kld,
    author = "Beane, Silas R. and Orlando, Domenico and Reffert, Susanne",
    title = "{Exact evaluation of large-charge correlation functions in nonrelativistic conformal field theory}",
    eprint = "2403.18898",
    archivePrefix = "arXiv",
    primaryClass = "hep-th",
    reportNumber = "YITP-24-48,NT@UW-24-05",
    doi = "10.1103/PhysRevD.110.025011",
    journal = "Phys. Rev. D",
    volume = "110",
    number = "2",
    pages = "025011",
    year = "2024"
}

@article{Beane:2025tum,
    author = "Beane, Silas R. and Orlando, Domenico and Reffert, Susanne",
    title = "{Unnuclear matter at large charge}",
    eprint = "2501.10505",
    archivePrefix = "arXiv",
    primaryClass = "nucl-th",
    reportNumber = "NT@UW-24-12,YITP-24-173",
    doi = "10.1103/5mvy-4p7q",
    journal = "Phys. Rev. D",
    volume = "112",
    number = "1",
    pages = "014028",
    year = "2025"
}

@article{Teng:2024exc,
    author = "Teng, Yu Ping and Grie{\ss}hammer, Harald W.",
    title = "{On two nucleons near unitarity with perturbative pions}",
    eprint = "2410.09653",
    archivePrefix = "arXiv",
    primaryClass = "nucl-th",
    doi = "10.1140/epja/s10050-025-01675-6",
    journal = "Eur. Phys. J. A",
    volume = "61",
    number = "9",
    pages = "211",
    year = "2025"
}

@misc{Griesshammer:2025els,
    author = "Grie{\ss}hammer, Harald W.",
    title = "{The Unitarity-Limit Expansion for Two Nucleons with Perturbative Pions: Digest and Ideas}",
    eprint = "2504.13353",
    archivePrefix = "arXiv",
    primaryClass = "nucl-th",
    month = "4",
    year = "2025"
}

@article{Bertulani:2002sz,
    author = "Bertulani, C. A. and Hammer, H.-W. and van Kolck, U.",
    title = "{Effective field theory for halo nuclei}",
    eprint = "nucl-th/0205063",
    archivePrefix = "arXiv",
    doi = "10.1016/S0375-9474(02)01270-8",
    journal = "Nucl. Phys. A",
    volume = "712",
    pages = "37--58",
    year = "2002"
}

@article{Kaplan:1999qa,
    author = "Kaplan, David B. and Steele, James V.",
    title = "{The Long and short of nuclear effective field theory expansions}",
    eprint = "nucl-th/9905027",
    archivePrefix = "arXiv",
    reportNumber = "CERN-TH-99-138, DOE-ER-40561-55-INT99",
    doi = "10.1103/PhysRevC.60.064002",
    journal = "Phys. Rev. C",
    volume = "60",
    pages = "064002",
    year = "1999"
}

@article{Beane:2000fi,
    author = "Beane, Silas R. and Savage, Martin J.",
    title = "{Rearranging pionless effective field theory}",
    eprint = "nucl-th/0011067",
    archivePrefix = "arXiv",
    reportNumber = "NT-UW-00-028, JLAB-THY-00-63, NT@UW-00-028",
    doi = "10.1016/S0375-9474(01)01088-0",
    journal = "Nucl. Phys. A",
    volume = "694",
    pages = "511--524",
    year = "2001"
}

@article{Peng:2021pvo,
    author = {Peng, Rui and Lyu, Songlin and K\"onig, Sebastian and Long, Bingwei},
    title = "{Constructing chiral effective field theory around unnatural leading-order interactions}",
    eprint = "2112.00947",
    archivePrefix = "arXiv",
    primaryClass = "nucl-th",
    doi = "10.1103/PhysRevC.105.054002",
    journal = "Phys. Rev. C",
    volume = "105",
    number = "5",
    pages = "054002",
    year = "2022"
}

@article{vanKolck:1998bw,
    author = "van Kolck, U.",
    title = "{Effective field theory of short-range forces}",
    eprint = "nucl-th/9808007",
    archivePrefix = "arXiv",
    reportNumber = "KRL-MAP-230, NT-UW-98-01",
    doi = "10.1016/S0375-9474(98)00612-5",
    journal = "Nucl. Phys. A",
    volume = "645",
    pages = "273--302",
    year = "1999"
}

@article{Kaplan:1998tg,
    author = "Kaplan, David B. and Savage, Martin J. and Wise, Mark B.",
    title = "{A New expansion for nucleon-nucleon interactions}",
    eprint = "nucl-th/9801034",
    archivePrefix = "arXiv",
    reportNumber = "DOE-ER-40561-352, INT-97-00-189, NT-UW-98-05, CALT-68-2155",
    doi = "10.1016/S0370-2693(98)00210-X",
    journal = "Phys. Lett. B",
    volume = "424",
    pages = "390--396",
    year = "1998"
}

@article{Kaplan:1998we,
    author = "Kaplan, David B. and Savage, Martin J. and Wise, Mark B.",
    title = "{Two nucleon systems from effective field theory}",
    eprint = "nucl-th/9802075",
    archivePrefix = "arXiv",
    reportNumber = "DOE-ER-40561-357, INT-98-00-5, NT-UW-98-08, CALT-68-2161",
    doi = "10.1016/S0550-3213(98)00440-4",
    journal = "Nucl. Phys. B",
    volume = "534",
    pages = "329--355",
    year = "1998"
}

@article{Gandolfi_2022,
	doi = {10.3390/condmat7010019},
  
	url = {https://doi.org/10.3390%2Fcondmat7010019},
  
	year = 2022,
	month = {feb},
  
	publisher = {{MDPI} {AG}
},
  
	volume = {7},
  
	number = {1},
  
	pages = {19},
  
	author = {Stefano Gandolfi and Georgios Palkanoglou and Joseph Carlson and Alexandros Gezerlis and Kevin E. Schmidt},
  
	title = {The 1S0 Pairing Gap in Neutron Matter},
  
	journal = {Condensed Matter}
}

@article{Beane:2020wjl,
    author = "Beane, Silas R. and Farrell, Roland C.",
    title = "{Geometry and entanglement in the scattering matrix}",
    eprint = "2011.01278",
    archivePrefix = "arXiv",
    primaryClass = "hep-th",
    reportNumber = "NT@UW-20-08, IQuS@UW-20-01",
    doi = "10.1016/j.aop.2021.168581",
    journal = "Annals Phys.",
    volume = "433",
    pages = "168581",
    year = "2021"
}

@article{Beane:2021xrk,
    author = "Beane, Silas R. and Farrell, Roland C.",
    title = "{UV/IR symmetries of the S-matrix and RG flow}",
    eprint = "2112.03472",
    archivePrefix = "arXiv",
    primaryClass = "hep-ph",
    reportNumber = "NT@UW-21-15, IQuS@UW-21-016",
    doi = "10.1016/j.nuclphysa.2022.122478",
    journal = "Nucl. Phys. A",
    volume = "1024",
    pages = "122478",
    year = "2022"
}

@article{Beane:2021dab,
    author = "Beane, Silas R. and Farrell, Roland C.",
    title = "{Symmetries of the Nucleon\textendash{}Nucleon S-Matrix and Effective Field Theory Expansions}",
    eprint = "2112.05800",
    archivePrefix = "arXiv",
    primaryClass = "nucl-th",
    reportNumber = "NT@UW-21-16, IQuS@UW-21-017",
    doi = "10.1007/s00601-022-01748-y",
    journal = "Few Body Syst.",
    volume = "63",
    number = "2",
    pages = "45",
    year = "2022"
}

@misc{NNOnline,
  title = {{NN}-Online},
  author = {Radboud University Nijmegen},
  year = {2005},
  howpublished = {\url{http://nn-online.org/}},
  note = {Accessed: 2018-12-01}
}

@inproceedings{deSwart:1995ui,
	title        = {The Low-energy n p scattering parameters and the deuteron},
	author       = {de Swart, J. J. and Terheggen, C. P. F. and Stoks, V. G. J.},
	year         = 1995,
	month        = 9,
	booktitle    = {{3rd International Symposium on Dubna Deuteron 95}},
	archiveprefix = {arXiv},
	eprint       = {nucl-th/9509032},
	primaryclass = {nucl-th},
	reportnumber = {THEF-NYM-95-11}
}

\clearpage

\appendix

\section{Best-fit parameters for 
EPT potentials fit in \texorpdfstring{$\nopi$}{pionless EFT}}
\label{sec:parPiSlash}

This appendix collects the values for the EPT potential parameters in $\nopi$ corresponding to the best fits to the Nijmegen PWA~\cite{NNonline} described in Section~\ref{sec:NNcomptodata}. The ERE and pole parameters are given in Tables~\ref{tab:ERE_EFT} and \ref{tab:polesEFT}, respectively.

\begin{table}[H]
\caption{The best-fit ERE parameters in all channels up through the f-wave for the EPT potentials fit in $\nopi$. (See Section~\ref{sec:NNcomptodata} for details.) $(a)_{\ell\ell'}$, $(r)_{\ell\ell'}$, and $(v_n)_{\ell\ell'}$ have units of fm$^{\ell+\ell'+1}$, fm$^{1-\ell-\ell'}$, and fm$^{2n-\ell-\ell'-1}$, respectively. All effective-range parameters marked by an asterisk are not fit to the data, but are fixed to guarantee the phase shift has the appropriate asymptotic behavior or because they are auxiliary parameters. \label{tab:ERE_EFT}\\}
\begin{minipage}{\linewidth}
\begin{center}
\begin{tabular}{ccccccccc} \hline \hline
$^{2S+1}L_J$ & Order & $a$ & $r$ & $v_2$ & $v_3$ & $v_4$ & $v_5$ & $k_{\max}$'s (MeV)\\ 
\hline\hline
$^1S_0$ & LO & $-23.86(13)$ & \text{—} & \text{—} & \text{—} & \text{—} & \text{—} & 3{\bf--}12 \\
$^1S_0$ & NLO & $-23.728(5)$ & $2.657(3)$ & \text{—} & \text{—}&  \text{—} & \text{—} & 35{\bf--}54 \\
$^1S_0$ & NNLO & $-23.724(6)$ & $2.658(4)$ & $-0.13(3)$ & $0.1$* & \text{—} & \text{—} & 65{\bf--}94 \\
$^1S_0$ & N$^3$LO & $-23.735(4)$ & $2.668(3)$ & $-0.27(3)$ & $1.04(12)$ & \text{—} & \text{—} & 90{\bf--}109 \\
\hline
$^3P_0$ & NLO & $-2.431(17)$ & \text{—} & \text{—} & \text{—} & \text{—} & \text{—} & 15{\bf--}34 \\
$^3P_0$ & NNLO & $-2.543(10)$ & $4.26(6)$ & \text{—} & \text{—} & \text{—} & \text{—} & 50{\bf--}79 \\
$^3P_0$ & N$^3$LO & $-2.485(6)$ & $3.69(4)$ & $2.14(7)$ & \text{—} & \text{—} & \text{—} & 90{\bf--}109 \\
\hline
$^1P_1$ & NLO & $2.57(3)$ & \text{—} & $-2$* & \text{—} & \text{—} & \text{—} & 15{\bf--}34\\
$^1P_1$ & NNLO & $2.83(2)$ & $-7.21(7)$ & \text{—} & \text{—} & \text{—} & \text{—} & 65{\bf--}94\\
$^1P_1$ & N$^3$LO & $2.779(9)$ & $-6.83(4)$ & $-1.19(7)$ & \text{—} & \text{—} & \text{—} & 100{\bf--}124\\
\hline
$^3P_1$ & NLO & $1.429(18)$ & \text{—} & $-2$* & \text{—} & \text{—} & \text{—} & 20{\bf--}39\\
$^3P_1$ & NNLO & $1.5450(8)$ & $-8.538(9)$ & \text{—} & \text{—} & \text{—} & \text{—} & 60{\bf--}99\\
$^3P_1$ & N$^3$LO & $1.5446(3)$ & $-8.517(4)$ & $-0.090(6)$ & \text{—} & \text{—} & \text{—} & 105{\bf--}129\\
\hline
$^3S_1$ & LO & $5.4631(5)$ & \text{—} & \text{—} & \text{—} & \text{—} & \text{—} & 3{\bf--}22\\
$E_1$ & LO & $0$* & \text{—} & \text{—} & \text{—} & \text{—} & \text{—} & \text{—}\\
$^3D_1$ & LO & $0$* & \text{—} & \text{—} & \text{—} & \text{—} & \text{—} & \text{—}\\
$^3S_1$ & NLO & $5.4167(8)$ & $1.8162(12)$ & \text{—} & \text{—} & \text{—} & \text{—} & 50{\bf--}69\\
$E_1$ & NLO & $1.599(11)$ & \text{—} & \text{—} & \text{—} & \text{—} & \text{—} & 15{\bf--}29\\
$^3D_1$ & NLO & $6.5$* & \text{—} & \text{—} & \text{—} & \text{—} & \text{—} & \text{—}\\
$^3S_1$ & NNLO & $5.395(4)$ & $1.952(6)$ & $-0.256(11)$ & $0.1$* & \text{—} & \text{—} & 75{\bf--}104\\
$E_1$ & NNLO & $1.645(3)$ & \text{—} & \text{—} & \text{—} & \text{—} & \text{—} & 30{\bf--}49\\
$^3D_1$ & NNLO & $6.30(3)$ & $-2.31(8)$ & \text{—} & \text{—} & \text{—} & \text{—} & 20{\bf--}49\\
$^3S_1$ & N$^3$LO & $5.4110(18)$ & $1.9269(13)$ & $-0.50(3)$ & $2.4(2)$ & $-3.3(3)$ & $0.1$* & 110{\bf--}129\\
$E_1$ & N$^3$LO & $1.678(5)$ & \text{—} & \text{—} & \text{—} & \text{—} & \text{—} & 55{\bf--}74\\
$^3D_1$ & N$^3$LO & $6.127(3)$ & $-3.22(4)$ & $-2.176(9)$ & \text{—} & \text{—} & \text{—} & 50{\bf--}79\\
\hline
$^1D_2$ & NNLO & $-1.21(3)$ & \text{—} & \text{—} & \text{—} & \text{—} & \text{—} & 20{\bf--}39 \\
$^1D_2$ & N$^3$LO & $-1.55(4)$ & $19.3(6)$ & \text{—} & \text{—} & \text{—} & \text{—} & 50{\bf--}84 \\
$^1D_2$ & N$^4$LO & $-1.47(2)$ & $16.6(3)$ & $11.3(4)$ & \text{—} & \text{—} & \text{—} & 95{\bf--}124 \\
\hline
$^3D_2$ & NNLO & $-6.80(10)$ & \text{—} & \text{—} & \text{—} & \text{—} & \text{—} & 15{\bf--}34 \\
$^3D_2$ & N$^3$LO & $-7.70(8)$ & $3.33(6)$ & \text{—} & \text{—} & \text{—} & \text{—} & 45{\bf--}64 \\
$^3D_2$ & N$^4$LO & $-7.428(12)$ & $2.909(10)$ & $2.18(2)$ & \text{—} & \text{—} & \text{—} & 70{\bf--}119 \\
\hline
$^3P_2$ & NLO & $-0.2826(3)$ & \text{—} & \text{—} & \text{—} & \text{—} & \text{—} & 10{\bf--}75 \\
$E_2$ & NLO & $0$* & \text{—} & \text{—} & \text{—} & \text{—} & \text{—} & \text{—} \\
$^3F_2$ & NLO & $0$* & \text{—} & \text{—} & \text{—} & $-2$* & \text{—} & \text{—} \\
$^3P_2$ & NNLO & $-0.2872(7)$ & $-0.73(8)$ & $-2$* & \text{—} & \text{—} & \text{—} & 55{\bf--}84 \\
$E_2$ & NNLO & $1.30(2)$ & \text{—} & \text{—} & \text{—} & \text{—} & \text{—} & 35{\bf--}49 \\
$^3F_2$ & NNLO & $-1$* & \text{—} & \text{—} & \text{—} & $-2$* & \text{—} & \text{—} \\
$^3P_2$ & N$^3$LO & $-0.2944(18)$ & $-1.23(12)$ & $-$0.13(9) & \text{—} & \text{—} & \text{—} & 85{\bf--}114 \\
$E_2$ & N$^3$LO & $1.91(6)$ & \text{—} & \text{—} & \text{—} & \text{—} & \text{—} & 50{\bf--}69 \\
$^3F_2$ & N$^3$LO & $-$1.56(11) & $-$9.0(4) & \text{—} & \text{—} & $-2$* & \text{—} & 35{\bf--}84\\
\hline\hline
\end{tabular}
\end{center}
\end{minipage}
\end{table}

\begin{table}
\addtocounter{table}{-1}
\caption{The best-fit ERE parameters in $\nopi$ (continued).}
\begin{minipage}{\linewidth}
\begin{center}
\begin{tabular}{ccccccccc} \hline \hline
$^{2S+1}L_J$ & Order & $a$ & $r$ & $v_2$ & $v_3$ & $v_4$ & $v_5$ & $k_{\max}$'s (MeV)\\ 
\hline\hline
$^3P_2$ & N$^4$LO & $-0.286(5)$ & $-12(5)$ & $-$3.0(12) & $-$2(2) & 8.6(12) & $-$2* & 115{\bf--}129 \\
$E_2$ & N$^4$LO & $1.48(2)$ & \text{—} & \text{—} & \text{—} & \text{—} & \text{—} & 75{\bf--}104 \\
$^3F_2$ & N$^4$LO & $-$1.433(7) & $-$3.0(3) & $-$40.7(12) & \text{—} & $-2$* & \text{—} & 85{\bf--}104\\
\hline
$^1F_3$ & N$^3$LO & $8.3(11)$ & \text{—} & \text{—} & \text{—} & $-2$* & \text{—} & 15{\bf--}34 \\
$^1F_3$ & N$^4$LO & $10.8(7)$ & $-6.1(3)$ & \text{—} & \text{—} & \text{—} & \text{—} & 45{\bf--}59 \\
$^1F_3$ & N$^5$LO & $7.5(3)$ & $-3.0(2)$ & $-14.7(7)$ & \text{—} & $-2$* & \text{—} & 65{\bf--}94 \\
$^1F_3$ & N$^6$LO & $8.58(8)$ & $-4.11(4)$ & $-9.21(14)$ & $-18.4(3)$ & \text{—} & \text{—} & 100{\bf--}124 \\
\hline  
$^3D_3$ & NNLO & $-0.113(2)$ & \text{—} & \text{—} & \text{—} & \text{—} & \text{—} & 15{\bf--}29 \\
$^3D_3$ & N$^3$LO & $-0.158(8)$ & $460(30)$ & \text{—} & \text{—} & \text{—} & \text{—} & 30{\bf--}49\\
$^3D_3$ & N$^4$LO & $-0.142(5)$ & $320(20)$ & $1620(120)$ & \text{—} & \text{—} & \text{—} & 50{\bf--}69\\
\hline
$^3F_3$ & N$^3$LO & $1.95(8)$ & \text{—} & \text{—} & \text{—} & $-2$* & \text{—} & 30{\bf--}44\\
$^3F_3$ & N$^4$LO & $4.5(5)$ & $-17.1(7)$ & \text{—} & \text{—} & \text{—} & \text{—} & 55{\bf--}74\\
$^3F_3$ & N$^5$LO & $2.42(9)$ & $-7.6(5)$ & $-29(1)$ & \text{—} & $-2$* & \text{—} & 80{\bf--}104\\
$^3F_3$ & N$^6$LO & $2.726(15)$ & $-9.85(8)$ & $-20.7(2)$ & $-21.6(5)$ & \text{—} & \text{—} & 105{\bf--}124\\
\hline
$^3F_4$ & N$^3$LO & $-0.03057(18)$ & \text{—} & \text{—} & \text{—} & \text{—} & \text{—} & 50{\bf--}69\\
$^3F_4$ & N$^4$LO & $-0.0349(14)$ & $80(20)$ & \text{—} & \text{—} & \text{—} & \text{—} & 70{\bf--}84\\
$^3F_4$ & N$^5$LO & $-0.0309(12)$ & $-70(30)$ & $360(60)$ & \text{—} & \text{—} & \text{—} & 85{\bf--}99\\
$^3F_4$ & N$^6$LO & $-0.0299(13)$ & $-140(50)$ & $630(150)$ & $-700(300)$ & \text{—} & \text{—} & 100{\bf--}119\\
\hline\hline
\end{tabular}
\end{center}
\end{minipage}
\end{table}

\begin{table}
\caption{The best-fit S-matrix poles in all channels up through the f-wave for the EPT potentials fit in $\nopi$. (See Section~\ref{sec:NNcomptodata} for details.) All poles are in units of MeV. Because the number of S-matrix poles increases with the order of the ERE, the number of entries in each row of the table varies accordingly. Redundant S-matrix poles are listed first. Complex-conjugate poles are listed together. For channels with more than eight poles at a given order, the remaining poles are listed in the following row. \label{tab:polesEFT}\\}
\begin{minipage}{\linewidth}
\begin{center}
\begin{tabular}{cccccccccc} \hline \hline
$^{2S+1}L_J$ & Order & $\gamma_1$ & $\gamma_2$ & $\gamma_3$ & $\gamma_4$ & $\gamma_5$ & $\gamma_6$ & $\gamma_7$ & $\gamma_8$\\ 
\hline\hline
$^1S_0$ & LO & $-$8.27 & \text{—} & \text{—} & \text{—}&  \text{—} & \text{—} & \text{—} & \text{—}\\
$^1S_0$ & NLO & 156.44 & $-$7.90 & \text{—} & \text{—}&  \text{—} & \text{—} & \text{—} & \text{—}\\
$^1S_0$ & NNLO & \multicolumn{2}{c}{$205.07 \pm 307.03\, i$} & 145.99 & \multicolumn{2}{c}{$-274.11\pm288.43\, i$} & $-$7.90 & \text{—} & \text{—}\\
$^1S_0$ & N$^3$LO & \multicolumn{2}{c}{$107.30 \pm 180.11\, i$} & 128.46 & \multicolumn{2}{c}{$-167.58 \pm 159.89\, i$} & $-$7.89 & \text{—} & \text{—} \\
\hline
$^3P_0$ & NLO & 146.76 & \multicolumn{2}{c}{$-73.38\pm127.10\, i$} & \text{—} & \text{—} & \text{—} & \text{—} & \text{—}\\
$^3P_0$ & NNLO & 77.86 & $-$401.90 & $-$96.57 & \text{—} & \text{—} & \text{—} & \text{—} & \text{—}\\
$^3P_0$ & N$^3$LO & 218.74 & 92.16 & \multicolumn{2}{c}{$-109.26\pm47.24\, i$} & \text{—} & \text{—} & \text{—} & \text{—}\\
\hline
$^1P_1$ & NLO & \multicolumn{2}{c}{$72.70\pm 89.23\, i$} & \multicolumn{2}{c}{$-122.03 \pm 85.71\, i$} & \text{—} & \text{—} & \text{—} & \text{—}\\
$^1P_1$ & NNLO & 64.79 & 706.17 & $-$59.34 & \text{—} & \text{—} & \text{—} & \text{—} & \text{—}\\
$^1P_1$ & N$^3$LO & 250.61 & 69.30 & $-$423.14 & $-$62.28 & \text{—} & \text{—} & \text{—} & \text{—}\\
\hline
$^3P_1$ & NLO & \multicolumn{2}{c}{$86.77\pm 104.23\, i$} & \multicolumn{2}{c}{$-136.11 \pm 101.61\, i$} & \text{—} & \text{—} & \text{—} & \text{—}\\
$^3P_1$ & NNLO & 80.81 & 835.21 & $-$73.68 & \text{—} & \text{—} & \text{—} & \text{—} & \text{—}\\
$^3P_1$ & N$^3$LO & 640.31 & 81.11 & $-$2829.69 & $-$73.87 & \text{—} & \text{—} & \text{—} & \text{—}\\
\hline
$^3S_1$ & LO & 36.12 & \text{—} & \text{—} & \text{—} & \text{—} & \text{—} & \text{—} & \text{—}\\
$^3S_1$ & NLO & 165.82 & 51.47 & \text{—} & \text{—} & \text{—} & \text{—} & \text{—} & \text{—}\\
$^3D_1$ & NLO & \multicolumn{2}{c}{$42.57\pm131.03\, i$} & 137.77 & \multicolumn{2}{c}{$-111.46\pm80.98\, i$} & \text{—} & \text{—} & \text{—}\\
$^3S_1$ & NNLO & \multicolumn{2}{c}{$148.96\pm327.41\, i$} & 122.37 & 55.03 & \multicolumn{2}{c}{$-237.66\pm282.81\, i$} & \text{—} & \text{—}\\
$^3D_1$ & NNLO & \multicolumn{2}{c}{$99.93\pm187.20\, i$} & 74.54 & $-$195.98 & $-$78.42 & \text{—} & \text{—} & \text{—}\\
$^3S_1$ & N$^3$LO & \multicolumn{2}{c}{$91.00\pm128.47\, i$} & \multicolumn{2}{c}{$9.83\times10^{-4}\pm1113.37\, i$} & 91.85 & 56.20 & \multicolumn{2}{c}{$-143.32\pm93.03\, i$}\\
\multicolumn{2}{c}{(cont.)} & \multicolumn{2}{c}{$-21.71\pm196.38\, i$} & \text{—} & \text{—} & \text{—} & \text{—} & \text{—} & \text{—} \\
$^3D_1$ & N$^3$LO & \multicolumn{2}{c}{$280.47\pm0.03\, i$} & 71.02 & $-$128.00 & $-$74.63 & \text{—} & \text{—} & \text{—}\\
\hline\hline
\end{tabular}
\end{center}
\end{minipage}
\end{table}

\begin{table}
\addtocounter{table}{-1}
\caption{The best-fit S-matrix poles in $\nopi$ (continued). \\}
\begin{minipage}{\linewidth}
\begin{center}
\begin{tabular}{cccccccccc} \hline \hline
$^{2S+1}L_J$ & Order & $\gamma_1$ & $\gamma_2$ & $\gamma_3$ & $\gamma_4$ & $\gamma_5$ & $\gamma_6$ & $\gamma_7$ & $\gamma_8$\\ 
\hline\hline
$^1D_2$ & NNLO & \multicolumn{2}{c}{$153.62 \pm 111.61\, i$} & $-$189.88 & \multicolumn{2}{c}{$-58.68\pm180.59\, i$} & \text{—} & \text{—} & \text{—} \\
$^1D_2$ & N$^3$LO & 418.01 & 51.12 & \multicolumn{2}{c}{$-209.05\pm365.62\, i$} & $-$51.03 & \text{—} & \text{—} & \text{—} \\
$^1D_2$ & N$^4$LO & 152.17 & 60.56 & $-$2211.33 & $-$165.49 & $-$60.28 & \text{—} & \text{—} & \text{—} \\
\hline
$^3D_2$ & NNLO & \multicolumn{2}{c}{$108.81 \pm 79.06\, i$} & $-$134.50 & \multicolumn{2}{c}{$-41.56\pm127.92\, i$} & \text{—} & \text{—} & \text{—} \\
$^3D_2$ & N$^3$LO & 229.23 & 55.51 & \multicolumn{2}{c}{$-114.98\pm206.24\, i$} & $-$54.79 & \text{—} & \text{—} & \text{—} \\
$^3D_2$ & N$^4$LO & 124.31 & 67.15 & $-$332.54 & $-$224.63 & $-$64.60 & \text{—} & \text{—} & \text{—} \\
\hline
$^3P_2$ & NLO & 300.69 & \multicolumn{2}{c}{$-150.35\pm260.41\, i$} & \text{—} & \text{—} & \text{—} & \text{—} & \text{—} \\
$^3P_2$ & NNLO & \multicolumn{2}{c}{$94.89\pm99.45\, i$} & \multicolumn{2}{c}{$-144.22\pm88.98\, i$} & \text{—} & \text{—} & \text{—} & \text{—} \\
$^3F_2$ & NNLO & \multicolumn{2}{c}{$127.46\pm56.02\, i$} & \multicolumn{2}{c}{$46.36\pm134.75\, i$} & \multicolumn{2}{c}{$-153.80\pm54.06\, i$} & \multicolumn{2}{c}{$-69.36\pm133.32\, i$} \\
$^3P_2$ & N$^3$LO & \multicolumn{2}{c}{$118.86\pm117.03\, i$} & $-$1640.14 & $-$126.04 & \text{—} & \text{—} & \text{—} & \text{—} \\
$^3F_2$ & N$^3$LO & \multicolumn{2}{c}{$97.96\pm194.15\, i$} & 28.29 & 211.33 & $-$245.59 & \multicolumn{2}{c}{$-130.16\pm193.04\, i$} & $-$28.29 \\
$^3P_2$ & N$^4$LO & \multicolumn{2}{c}{$84.86\pm87.82\, i$} & \multicolumn{2}{c}{$8.74\pm200.11\, i$}  & 156.08 & \multicolumn{2}{c}{$-0.97\pm389.15\, i$} & $-$139.66\\
\multicolumn{2}{c}{(cont.)} & \multicolumn{2}{c}{$-100.84\pm89.91\, i$} & \text{—} & \text{—} & \text{—} & \text{—} & \text{—} & \text{—}\\
$^3F_2$ & N$^4$LO & \multicolumn{2}{c}{$272.55\pm295.65\, i$} & \multicolumn{2}{c}{$39.86\pm29.30\, i$} & \multicolumn{2}{c}{$-321.89\pm295.10\, i$} & \multicolumn{2}{c}{$-39.86\pm29.31\, i$} \\
\hline
$^1F_3$ & N$^3$LO & \multicolumn{2}{c}{$42.81\pm125.72\, i$} & \multicolumn{2}{c}{$118.68\pm52.30\, i$} & \multicolumn{2}{c}{$-145.25\pm50.00\, i$} & \multicolumn{2}{c}{$-65.57\pm124.12\, i$}\\
$^1F_3$ & N$^4$LO & \multicolumn{2}{c}{$75.98\pm235.62\, i$} & 245.83 & 34.34 & \multicolumn{2}{c}{$-198.90\pm145.64\, i$} & $-$34.33 & \text{—}\\
$^1F_3$ & N$^5$LO & \multicolumn{2}{c}{$204.93\pm230.74\, i$} & \multicolumn{2}{c}{$53.43\pm29.09\, i$} & \multicolumn{2}{c}{$-254.26\pm229.23\, i$} & \multicolumn{2}{c}{$-53.44\pm29.16\, i$}\\
$^1F_3$ & N$^6$LO & \multicolumn{2}{c}{$98.79\pm36.09\, i$} & 56.50 & 3626.25 & \multicolumn{2}{c}{$-96.16\pm37.13\, i$} & $-$56.39 & \text{—}\\
\hline
$^3D_3$ & NNLO & \multicolumn{2}{c}{$247.11 \pm 179.54\, i$} & $-$305.45 & \multicolumn{2}{c}{$-94.39\pm290.50\, i$} & \text{—} & \text{—} & \text{—} \\
$^3D_3$ & N$^3$LO & 1208.34 & 32.80 & \multicolumn{2}{c}{$-604.17\pm1046.96\, i$} & $-$32.79 & \text{—} & \text{—} & \text{—} \\
$^3D_3$ & N$^4$LO & \multicolumn{2}{c}{$47.25\pm18.38\, i$} & $-$319900.46 & \multicolumn{2}{c}{$-47.25\pm18.38\, i$} & \text{—} & \text{—} & \text{—} \\
\hline
$^3F_3$ & N$^3$LO & \multicolumn{2}{c}{$53.02\pm151.55\, i$} & \multicolumn{2}{c}{$143.85\pm62.95\, i$} & \multicolumn{2}{c}{$-169.85\pm61.44\, i$} & \multicolumn{2}{c}{$-76.35\pm150.37\, i$}\\
$^3F_3$ & N$^4$LO & \multicolumn{2}{c}{$94.44\pm288.85\, i$} & 302.37 & 31.87 & \multicolumn{2}{c}{$-244.63\pm178.52\, i$} & $-$31.87 & \text{—}\\
$^3F_3$ & N$^5$LO & \multicolumn{2}{c}{$247.90\pm274.43\, i$} & \multicolumn{2}{c}{$59.69\pm32.44\, i$} & \multicolumn{2}{c}{$-297.22\pm273.20\, i$} & \multicolumn{2}{c}{$-59.70\pm32.49\, i$}\\
$^3F_3$ & N$^6$LO & 163.80 & \multicolumn{2}{c}{$78.35\pm11.16\, i$} & 4260.72 & $-$153.62 & \multicolumn{2}{c}{$-79.07\pm10.61\, i$} & \text{—}\\
\hline
$^3F_4$ & N$^3$LO & \multicolumn{2}{c}{$202.49\pm253.91\, i$} & 324.76 & \multicolumn{2}{c}{$-292.60\pm140.91\, i$} & \multicolumn{2}{c}{$-72.27\pm316.62\, i$} & \text{—}\\
$^3F_4$ & N$^4$LO & \multicolumn{2}{c}{$322.19\pm251.34\, i$} & 167.78 & \multicolumn{2}{c}{$-123.88\pm402.91\, i$} & $-$394.65 & $-$169.76 & \text{—}\\
$^3F_4$ & N$^5$LO & 1409.95 & \multicolumn{2}{c}{$-69.42\pm82.44\, i$} & \multicolumn{2}{c}{$-704.95\pm1219.42\, i$} & \multicolumn{2}{c}{$69.45\pm82.45\, i$} & \text{—}\\
$^3F_4$ & N$^6$LO & 136804.73 & \multicolumn{2}{c}{$64.29\pm71.33\, i$} & \multicolumn{2}{c}{$-0.12\pm183.02\, i$} & \multicolumn{2}{c}{$-64.30\pm71.33\, i$} & \text{—}\\
\hline\hline
\end{tabular}
\end{center}
\end{minipage}
\end{table}

\clearpage

\section{Best-fit parameters for potentials fit up to 405 MeV}

\label{sec:par405MeV}

This appendix collects the values for the EPT potential parameters corresponding to the best fits to the Nijmegen PWA~\cite{NNonline} described in Section~\ref{sec:NNtopiprod}, where amplitude zeros are taken into account. The ERE and pole parameters are given in Tables~\ref{tab:ERE_400} and \ref{tab:poles400}, respectively.

\begin{table}[H]
\caption{The best-fit ERE parameters in all channels up through the f-wave for EPT potentials fit up to $405$ MeV. (See Section~\ref{sec:NNtopiprod} for details.) $(a)_{\ell\ell'}$, $(r)_{\ell\ell'}$, and $(v_n)_{\ell\ell'}$ have units of fm$^{\ell+\ell'+1}$, fm$^{1-\ell-\ell'}$, and fm$^{2n-\ell-\ell'-1}$, respectively. The $a$ values are fixed from fits to the phase shift data near threshold. $k_*$ is fixed by observing when the phase-shift data crosses zero. \label{tab:ERE_400}\\}
\begin{minipage}{\linewidth}
\begin{center}
\resizebox{0.99\columnwidth}{!}{%
\begin{tabular}{ccccccccccc} \hline \hline
$^{2S+1}L_J$ & Label & $a$ & $r$ & $v_2$ & $v_3$ & $v_4$ & $v_5$ & $v_6$ & $k_*$ (MeV) & $k_{\max}$ (MeV)\\ 
\hline\hline
$^1S_0$ & Minimal & $-23.735$ & $2.52$ & \text{—} & \text{—} & \text{—} & \text{—} & \text{—} & $352.6$ & 100\\
$^1S_0$ & Intermediate & $-23.735$ & $2.58$ & $0.195$ & $0.0924$ & \text{—} & \text{—} & \text{—} & $352.6$ & 405 \\
$^1S_0$ & Best-Fit & $-23.735$ & $2.60$ & $0.0864$ & $0.220$ & $0.00568$ & $0.00881$ & \text{—} & $352.6$ & 405 \\
\hline
$^3P_0$ & Minimal & $-2.48$ & $3.62$ & $1.99$ & \text{—} & \text{—} & \text{—} & \text{—} & $295.4$ & 200 \\
$^3P_0$ & Intermediate & $-2.48$ & $3.62$ & $1.98$ & $0.890$ & $0.406$ & \text{—} & \text{—} & $295.4$ & 300 \\
$^3P_0$ & Best-Fit & $-2.48$ & $3.70$ & $1.84$ & $1.02$ & $0.387$ & $0.178$ & \text{—} & $295.4$ & 405 \\
\hline
$^1P_1$ & Minimal & $2.78$ & $-7.30$ & \text{—} & \text{—} & \text{—} & \text{—} & \text{—} & \text{—} & 100\\
$^1P_1$ & Intermediate & $2.78$ & $-6.96$ & $-1.30$ & $1.00$ & $-0.209$ & \text{—} & \text{—} & \text{—} & 300\\
$^1P_1$ & Best-Fit & $2.78$ & $-6.77$ & $-1.78$ & $1.78$ & $-0.717$ & $0.134$ & $-0.00955$ & \text{—} & 405\\
\hline
$^3P_1$ & Minimal & $1.545$ & $-8.30$ & \text{—} & \text{—} & \text{—} & \text{—} & \text{—} & \text{—} & 250\\
$^3P_1$ & Intermediate & $1.545$ & $-9.17$ & $0.463$ & $-0.050$ & \text{—} & \text{—} & \text{—} & \text{—} & 300\\
$^3P_1$ & Best-Fit & $1.545$ & $-8.82$ & $0.157$ & $0.0892$ & $-0.0150$ & \text{—} & \text{—} & \text{—} & 405\\
\hline
$^3S_1$ & Minimal & $5.420$ & $1.75$ & \text{—} & \text{—} & \text{—} & \text{—} & \text{—} & $407.8$ & 150\\
$^3S_1$ & Intermediate & $5.420$ & $1.76$ & $0.109$ & $0.0490$ & \text{—} & \text{—} & \text{—} & $407.8$ & 300\\
$^3S_1$ & Best-Fit & $5.420$ & $1.72$ & $0.167$ & $0.0292$ & $0.00917$ & \text{—} & \text{—} & $407.8$ & 405\\
\hline
$^3D_1$ & Minimal & 6.5 & $-$3.49 & $-$3.54 & \text{—} & \text{—} & \text{—} & \text{—} & \text{—} & 300\\
$^3D_1$ & Intermediate & 6.5 & $-$3.70 & $-$3.39 & $-$0.00494 & $-0.0244$ & \text{—} & \text{—} & \text{—} & 405\\
$^3D_1$ & Best-Fit & 6.5 & $-$3.74 & $-$3.31 & $-$0.0764 & 0.000723 & $-0.00293$ & \text{—} & \text{—} & 405\\
\hline
$^1D_2$ & Minimal & $-$1.5 & 16.9 & 8.65 & \text{—} & \text{—} & \text{—} & \text{—} & \text{—} & 405\\
$^1D_2$ & Intermediate & $-$1.5 & 18.9 & 7.57 & 0.232 & \text{—} & \text{—} & \text{—} & \text{—} & 405\\
$^1D_2$ & Best-Fit & $-$1.5 & 17.0 & 12.4 & $-6.09$ & $3.13$ & $-0.640$ & $0.0447$ & \text{—} & 405\\
\hline
$^3D_2$ & Minimal & $-$7.4 & 2.58 & 2.39 & \text{—} & \text{—} & \text{—} & \text{—} & \text{—} & 250\\
$^3D_2$ & Intermediate & $-$7.4 & 3.60 & 1.20 & 0.602 & \text{—} & \text{—} & \text{—} & \text{—} & 350\\
$^3D_2$ & Best-Fit & $-$7.4 & 3.21 & 1.69 & 0.256 & 0.0702 & \text{—} & \text{—} & \text{—} & 405\\
\hline
$^3P_2$ & Minimal & $-$0.30 & 4.72 & \text{—} & \text{—} & \text{—} & \text{—} & \text{—} & \text{—} & 200\\
$^3P_2$ & Intermediate & $-$0.30 & 3.52 & 0.996 & \text{—} & \text{—} & \text{—} & \text{—} & \text{—} & 405\\
$^3P_2$ & Best-Fit & $-$0.30 & 1.97 & 1.90 & $-0.295$ & $0.0275$ & \text{—} & \text{—} & \text{—} & 405\\
\hline
$^3F_2$ & Minimal & $-$1.0 & 33.6 & 28.5 & 47.6 & \text{—} & \text{—} & \text{—} & \text{—} & 275\\
$^3F_2$ & Intermediate & $-$1.0 & 33.1 & 20.0 & 76.3 & $-27.2$ & 7.78 & \text{—} & \text{—} & 350\\
$^3F_2$ & Best-Fit & $-$1.0 & 16.4 & 64.9 & $-4.94$ & 36.7 & $-14.6$ & $2.87$ & \text{—} & 375\\
\hline
$^3D_3$ & Minimal & $-$0.14 & 215 & $-$60.7 & \text{—} & \text{—} & \text{—} & \text{—} & \text{—} & 200\\
$^3D_3$ & Intermediate & $-$0.14 & 169 & $-$49.5 & 15.7 & \text{—} & \text{—} & \text{—} & \text{—} & 300\\
$^3D_3$ & Best-Fit & $-$0.14 & 218 & $-$107 & 60.8 & $-$13.8 & 1.27 & \text{—} & \text{—} & 405\\
\hline
$^1F_3$ & Minimal & 8 & $-$6.66 & $-$2.99 & $-25.4$ & \text{—} & \text{—} & \text{—} & \text{—} & 250\\
$^1F_3$ & Intermediate & 8 & $-$5.85 & $-$4.93 & $-$22.7 & $-$1.12 & \text{—} & \text{—} & \text{—} & 300\\
$^1F_3$ & Best-Fit & 8 & $-$1.56 & $-$18.1 & 2.46 & $-$20.1 & 5.55 & $-0.430$ & \text{—} & 405\\
\hline
$^3F_3$ & Minimal & 2.7 & $-$10.9 & $-$17.7 & $-$25.7 & \text{—} & \text{—} & \text{—} & \text{—} & 200\\
$^3F_3$ & Intermediate & 2.7 & $-$9.29 & $-$22.2 & $-$18.1 & $-$3.92 & \text{—} & \text{—} & \text{—} & 300\\
$^3F_3$ & Best-Fit & 2.7 & $-$9.45 & $-$22.0 & $-$18.0 & $-$4.63 & 0.550 & $-0.125$ & \text{—} & 405\\
\hline 
$^3F_4$ & Minimal & $-$0.03 & 73.5 & 128 & $-$11.4 & \text{—} & \text{—} & \text{—} & \text{—} & 250\\
$^3F_4$ & Intermediate & $-$0.03 & 54.2 & 160 & $-$43.4 & 10.3 & \text{—} & \text{—} & \text{—} & 300\\
$^3F_4$ & Best-Fit & $-$0.03 & 37.7 & 190 & $-$82.8 & 33.5 & $-6.02$ & 0.513 & \text{—} & 405\\
\hline\hline
\end{tabular}
}
\end{center}
\end{minipage}
\end{table}

\begin{table}
\caption{The best-fit S-matrix poles in all channels up through the f-wave for EPT potentials fit up to 405 MeV. (See Section~\ref{sec:NNtopiprod} for details.) Because the number of S-matrix poles increases with the order of the ERE, the number of entries in each row of the table varies accordingly. All poles are in units of MeV. Redundant S-matrix poles are listed first. Complex-conjugate poles are listed together. For channels with more than eight poles at a given order, the remaining poles are listed in the following row.  \label{tab:poles400}\\}
\begin{minipage}{\linewidth}
\begin{center}
\begin{tabular}{cccccccccc} \hline \hline
$^{2S+1}L_J$ & Label & $\gamma_1$ & $\gamma_2$ & $\gamma_3$ & $\gamma_4$ & $\gamma_5$ & $\gamma_6$ & $\gamma_7$ & $\gamma_8$\\ 
\hline\hline
$^1S_0$ & Minimal & 232.75 & 561.16 & $-$7.91 & \text{—}&  \text{—} & \text{—} & \text{—} & \text{—}\\
$^1S_0$ & Intermediate & \multicolumn{2}{c}{$192.94 \pm 372.67\, i$} & 175.04 & \multicolumn{2}{c}{$-276.50 \pm 499.44\, i$} & $-$7.91 &\text{—} & \text{—}\\
$^1S_0$ & Best-Fit & \multicolumn{2}{c}{$192.89 \pm 228.68\, i$} & \multicolumn{2}{c}{$90.51 \pm 427.36\, i$} & 155.97 & \multicolumn{2}{c}{$-86.88 \pm 457.23\, i$} & $-7.90$ \\
\multicolumn{2}{c}{(cont.)} & \multicolumn{2}{c}{$-270.56\pm210.64\, i$} & \text{—} & \text{—} & \text{—} & \text{—} & \text{—} & \text{—}\\
\hline
$^3P_0$ & Minimal & \multicolumn{2}{c}{$331.82\pm274.92\, i$} & 91.28 & \multicolumn{2}{c}{$-115.87\pm50.24\, i$} & \text{—} & \text{—} & \text{—} \\
$^3P_0$ & Intermediate & \multicolumn{2}{c}{$290.57\pm298.30\, i$} & 91.20 & 591.28 & \multicolumn{2}{c}{$-515.73\pm 654.82\, i$} & \multicolumn{2}{c}{$-116.08\pm 50.11\, i$}\\
$^3P_0$ & Best-Fit & \multicolumn{2}{c}{$241.73\pm186.49\, i$} & \multicolumn{2}{c}{$170.53\pm511.19\, i$} & $89.44$ & \multicolumn{2}{c}{$-146.00\pm 624.08\, i$} & $-378.24$\\
\multicolumn{2}{c}{(cont.)} & \multicolumn{2}{c}{$-121.85\pm48.26\, i$} & \text{—} & \text{—} & \text{—} & \text{—} & \text{—} & \text{—}\\
\hline
$^1P_1$ & Minimal & 64.96 & 714.49 & $-$59.55 & \text{—} & \text{—} & \text{—} & \text{—} & \text{—}\\
$^1P_1$ & Intermediate & \multicolumn{2}{c}{$127.78\pm343.02\, i$} & 68.83 & 187.08 & \multicolumn{2}{c}{$-107.64\pm383.04\, i$} & $-234.29$ & $-61.91$\\
$^1P_1$ & Best-Fit & \multicolumn{2}{c}{$172.48\pm303.77\, i$} & \multicolumn{2}{c}{$92.30\pm464.67\, i$} & 71.22 & 156.47 & \multicolumn{2}{c}{$-65.86\pm484.55\, i$} \\
\multicolumn{2}{c}{(cont.)} & \multicolumn{2}{c}{$-185.31\pm337.78\, i$} & $-191.69$ & $-63.23$ & \text{—} & \text{—} & \text{—} & \text{—}\\
\hline
$^3P_1$ & Minimal & 82.16 & 811.37 & $-$74.60 & \text{—} & \text{—} & \text{—} & \text{—} & \text{—}\\
$^3P_1$ & Intermediate & \multicolumn{2}{c}{$322.27\pm412.46\, i$} & 76.85 & \multicolumn{2}{c}{$-325.22\pm636.98\, i$} & $-70.96$& \text{—} & \text{—}\\
$^3P_1$ & Best-Fit & \multicolumn{2}{c}{$206.51\pm454.06\, i$} & 79.07 & 409.80 & $-$503.11 & \multicolumn{2}{c}{$-163.15\pm558.14\, i$} & $-72.49$\\
\hline
$^3S_1$ & Minimal & 364.84 & 364.85 & 45.48 & \text{—} & \text{—} & \text{—} & \text{—} & \text{—}\\
$^3S_1$ & Intermediate & \multicolumn{2}{c}{$205.07\pm342.89\, i$} & 198.88 & 45.64 & \multicolumn{2}{c}{$-327.33\pm462.50\, i$} & \text{—} & \text{—}\\
$^3S_1$ & Best-Fit & \multicolumn{2}{c}{$289.28\pm55.14\, i$} & \multicolumn{2}{c}{$188.82\pm443.00\, i$} & 45.21 & $-579.42$ & \multicolumn{2}{c}{$-210.99\pm547.12\, i$}\\
\hline
$^3D_1$ & Minimal & $65.58$ & $141.39$ & $667.69$ & $-108.69$ & $-68.41$ & $\text{—}$ & $\text{—}$ & $\text{—}$ \\
$^3D_1$ & Intermediate & \multicolumn{2}{c}{$ 476.37\pm284.67\,i $} & $62.34$ & $154.31$ & \multicolumn{2}{c}{$ -493.82\pm636.15\,i $} & $-117.59$ & $-64.16$ \\
$^3D_1$ & Best-Fit & \multicolumn{2}{c}{$ 417.39\pm482.71\,i $} & $61.56$ & $162.14$ & $435.19$ & $-695.74$ & \multicolumn{2}{c}{$ -306.91\pm637.25\,i $} \\
\multicolumn{2}{c}{(cont.)} & $-120.93$ & $-63.18$ & $\text{—}$ & $\text{—}$ & $\text{—}$ & $\text{—}$ & $\text{—}$ & $\text{—}$ \\
\hline
$^1D_2$ & Minimal & $58.24$ & $176.05$ & $-1684.09$ & $-199.08$ & $-58.02$ & $\text{—}$ & $\text{—}$ & $\text{—}$ \\
$^1D_2$ & Intermediate & $54.13$ & $202.74$ & $1619.74$ & $-730.55$ & $-241.82$ & $-54.00$ & $\text{—}$ & $\text{—}$ \\
$^1D_2$ & Best-Fit & \multicolumn{2}{c}{$ 201.63\pm266.65\,i $} & \multicolumn{2}{c}{$ 82.98\pm524.21\,i $} & $59.71$ & $128.72$ & \multicolumn{2}{c}{$ -94.49\pm508.55\,i $} \\
\multicolumn{2}{c}{(cont.)} & \multicolumn{2}{c}{$ -187.61\pm256.12\,i $} & $-134.02$ & $-59.44$ & $\text{—}$ & $\text{—}$ & $\text{—}$ & $\text{—}$ \\
\hline
$^3D_2$ & Minimal & $78.91$ & $103.20$ & $-416.49$ & $-166.29$ & $-71.69$ & $\text{—}$ & $\text{—}$ & $\text{—}$ \\
$^3D_2$ & Intermediate & $55.89$ & $202.12$ & $451.18$ & \multicolumn{2}{c}{$ -163.12\pm143.39\,i $} & $-55.13$ & $\text{—}$ & $\text{—}$ \\
$^3D_2$ & Best-Fit & \multicolumn{2}{c}{$ 366.36\pm382.73\,i $} & $60.91$ & $151.62$ & $-486.05$ & \multicolumn{2}{c}{$ -199.81\pm137.78\,i $} & $-59.58$ \\
\hline
$^3P_2$ & Minimal & $196.64$ & \multicolumn{2}{c}{$ -331.19\pm143.40\,i $} & \text{—} & \text{—} & \text{—} & \text{—} & \text{—} \\
$^3P_2$ & Intermediate & \multicolumn{2}{c}{$ 286.37\pm77.96\,i $} & \multicolumn{2}{c}{$ -187.30\pm150.11\,i $} & \text{—} & \text{—} & \text{—} & \text{—} \\
$^3P_2$ & Best-Fit & \multicolumn{2}{c}{$ 223.45\pm553.08\,i $} & \multicolumn{2}{c}{$ 198.04\pm118.85\,i $} & \multicolumn{2}{c}{$ -264.38\pm511.62\,i $} & \multicolumn{2}{c}{$ -157.10\pm140.20\,i $} \\
\hline
$^3F_2$ & Minimal & \multicolumn{2}{c}{$ 126.48\pm75.75\,i $} & $50.74$ & $-9397.85$ & \multicolumn{2}{c}{$ -127.71\pm76.39\,i $} & $-50.75$ & $\text{—}$ \\
$^3F_2$ & Intermediate & \multicolumn{2}{c}{$ 164.19\pm320.00\,i $} & \multicolumn{2}{c}{$ 103.55\pm77.38\,i $} & $50.00$ & \multicolumn{2}{c}{$ -163.79\pm315.52\,i $} & $-50.00$ \\
\multicolumn{2}{c}{(cont.)} & \multicolumn{2}{c}{$ -103.96\pm77.72\,i $} & $\text{—}$ & $\text{—}$ & $\text{—}$ & $\text{—}$ & $\text{—}$ & $\text{—}$ \\
$^3F_2$ & Best-Fit & \multicolumn{2}{c}{$ 170.36\pm143.38\,i $} & \multicolumn{2}{c}{$ 134.38\pm359.15\,i $} & \multicolumn{2}{c}{$ 60.35\pm34.18\,i $} & \multicolumn{2}{c}{$ -136.02\pm356.36\,i $} \\
\multicolumn{2}{c}{(cont.)} & \multicolumn{2}{c}{$ -168.73\pm144.95\,i $} & \multicolumn{2}{c}{$ -60.35\pm34.15\,i $} & $\text{—}$ & $\text{—}$ & $\text{—}$ & $\text{—}$ \\
\hline
$^1F_3$ & Minimal & \multicolumn{2}{c}{$ 88.21\pm79.38\,i $} & $38.70$ & $5008.41$ & \multicolumn{2}{c}{$ -87.75\pm78.04\,i $} & $-38.69$ & $\text{—}$ \\
$^1F_3$ & Intermediate & \multicolumn{2}{c}{$ 92.35\pm72.65\,i $} & $42.09$ & $798.97$ & $-977.06$ & \multicolumn{2}{c}{$ -91.44\pm71.30\,i $} & $-42.08$ \\
$^1F_3$ & Best-Fit & \multicolumn{2}{c}{$ 135.95\pm133.19\,i $} & \multicolumn{2}{c}{$ 87.72\pm504.29\,i $} & \multicolumn{2}{c}{$ 45.30\pm34.42\,i $} & \multicolumn{2}{c}{$ -83.28\pm517.95\,i $} \\
\multicolumn{2}{c}{(cont.)} & \multicolumn{2}{c}{$ -140.42\pm131.58\,i $} & \multicolumn{2}{c}{$ -45.28\pm34.45\,i $} & $\text{—}$ & $\text{—}$ & $\text{—}$ & $\text{—}$ \\
\hline
$^3D_3$ & Minimal & $49.94$ & $11983.89$ & \multicolumn{2}{c}{$ -2.88\pm267.39\,i $} & $-49.94$ & $\text{—}$ & $\text{—}$ & $\text{—}$ \\
$^3D_3$ & Intermediate & \multicolumn{2}{c}{$ 124.73\pm281.49\,i $} & $56.05$ & \multicolumn{2}{c}{$ -118.45\pm275.95\,i $} & $-56.04$ & $\text{—}$ & $\text{—}$ \\
$^3D_3$ & Best-Fit & \multicolumn{2}{c}{$ 126.23\pm239.99\,i $} & \multicolumn{2}{c}{$ 117.00\pm432.65\,i $} & $48.97$ & \multicolumn{2}{c}{$ -122.09\pm426.36\,i $} & $-48.96$ \\
\multicolumn{2}{c}{(cont.)} & \multicolumn{2}{c}{$ -121.14\pm239.91\,i $} & $\text{—}$ & $\text{—}$ & $\text{—}$ & $\text{—}$ & $\text{—}$ & $\text{—}$ \\
\hline\hline
\end{tabular}
\end{center}
\end{minipage}
\end{table}

\begin{table}
\addtocounter{table}{-1}
\caption{The best-fit S-matrix poles in all channels up through the f-wave for EPT potentials fit up to 405 MeV (continued).\\}
\begin{minipage}{\linewidth}
\begin{center}
\begin{tabular}{cccccccccc} \hline \hline
$^{2S+1}L_J$ & Label & $\gamma_1$ & $\gamma_2$ & $\gamma_3$ & $\gamma_4$ & $\gamma_5$ & $\gamma_6$ & $\gamma_7$ & $\gamma_8$\\ 
\hline\hline
$^3F_3$ & Minimal & \multicolumn{2}{c}{$ 117.67\pm43.58\,i $} & $59.64$ & $5070.51$ & \multicolumn{2}{c}{$ -115.06\pm44.41\,i $} & $-59.58$ & $\text{—}$ \\
$^3F_3$ & Intermediate & \multicolumn{2}{c}{$ 278.29\pm54.71\,i $} & \multicolumn{2}{c}{$ 71.34\pm21.73\,i $} & $-375.76$ & $-230.67$ & \multicolumn{2}{c}{$ -71.56\pm21.79\,i $} \\
$^3F_3$ & Best-Fit & \multicolumn{2}{c}{$ 275.81\pm461.25\,i $} & \multicolumn{2}{c}{$ 235.88\pm77.12\,i $} & \multicolumn{2}{c}{$ 72.08\pm20.30\,i $} & \multicolumn{2}{c}{$ -267.06\pm471.10\,i $} \\
\multicolumn{2}{c}{(cont.)} & \multicolumn{2}{c}{$ -244.38\pm51.91\,i $} & \multicolumn{2}{c}{$ -72.33\pm20.34\,i $} & $\text{—}$ & $\text{—}$ & $\text{—}$ & $\text{—}$ \\
\hline
$^3F_4$ & Minimal & \multicolumn{2}{c}{$ 112.70\pm82.91\,i $} & $2416.40$ & \multicolumn{2}{c}{$ -84.26\pm642.13\,i $} & \multicolumn{2}{c}{$ -112.81\pm82.75\,i $} & $\text{—}$ \\
$^3F_4$ & Intermediate & \multicolumn{2}{c}{$ 193.28\pm351.54\,i $} & \multicolumn{2}{c}{$ 103.79\pm81.17\,i $} & \multicolumn{2}{c}{$ -183.62\pm345.94\,i $} & \multicolumn{2}{c}{$ -103.87\pm81.08\,i $} \\
$^3F_4$ & Best-Fit & \multicolumn{2}{c}{$ 211.54\pm279.18\,i $} & \multicolumn{2}{c}{$ 150.80\pm473.17\,i $} & \multicolumn{2}{c}{$ 98.00\pm79.59\,i $} & \multicolumn{2}{c}{$ -155.65\pm464.86\,i $} \\
\multicolumn{2}{c}{(cont.)} & \multicolumn{2}{c}{$ -206.64\pm282.72\,i $} & \multicolumn{2}{c}{$ -98.06\pm79.53\,i $} & $\text{—}$ & $\text{—}$ & $\text{—}$ & $\text{—}$ \\
\hline\hline
\end{tabular}
\end{center}
\end{minipage}
\end{table}

\end{document}